\documentclass{aastex61}

\shorttitle{CIII] emission in Green Pea Galaxies}
\shortauthors{Ravindranath et al.}

\begin{document}

\title{The Semi-forbidden CIII]$\lambda$1909\AA~ Emission in the Rest-Ultraviolet Spectra of Green Pea Galaxies}

\correspondingauthor{Swara Ravindranath}
\email{swara@stsci.edu}

\author[0000-0002-5269-6527]{Swara Ravindranath}
\affil{Space Telescope Science Institute, 3700 San Martin Dr., Baltimore, MD 21218-2410, USA}

\author{TalaWanda Monroe}
\affiliation{Space Telescope Science Institute, 3700 San Martin Dr., Baltimore, MD 21218-2410, USA}

\author{Anne Jaskot}
\affiliation{Astronomy Department, Williams College, Williamstown, MA 01267, USA}
\affiliation{Hubble Fellow}

\author{Henry C. Ferguson}
\affiliation{Space Telescope Science Institute, 3700 San Martin Dr., Baltimore, MD 21218-2410, USA}

\author{Jason Tumlinson}
\affiliation{Space Telescope Science Institute, 3700 San Martin Dr., Baltimore, MD 21218-2410, USA}




\begin{abstract}
We used the Space Telescope Imaging Spectrograph (STIS) on the {\it Hubble Space Telescope (HST)} to observe the semi-forbidden CIII] $\lambda\lambda$1907,1909\AA~ 
doublet emission in Green Pea galaxies at $0.13 \leq z \leq 0.3$. We detect CIII] emission in 7/10 galaxies with CIII] equivalent widths that range from 2-10\AA~, confirming 
that CIII] emission is almost ubiquitous in low-mass, low-metallicity (12+log(O/H) $<$ 8.4) galaxies which are characterized by strong optical [OIII] $\lambda$5007\AA~ emission. 
The composite UV spectrum shows evidence for the HeII $\lambda$1640\AA~ emission line and interstellar absorption features (eg; CIV $\lambda\lambda$1548,1550\AA~, 
AlIII $\lambda\lambda$1854,1862\AA~). We do not detect the OIII] $\lambda\lambda$1661,1666\AA~ emission with $>$ 3-$\sigma$ significance. The observed CIII] emission line 
strengths are consistent with the predictions from photoionization models which incorporate the effects of binary stellar evolution with young stellar ages $\leq$ $3-5$ Myrs, and high 
ionization parameters (log$U$ $>$ -2). The hard ionizing radiation from young massive stars, and high nebular temperatures at low-metallicities can account for the observed 
high equivalent widths of CIII] $\lambda$1909\AA~ and [OIII] $\lambda$5007\AA~ emission lines. Some of the star-forming galaxies at high redshift, and local blue compact dwarf 
galaxies show offsets from the EW(CIII]) versus EW([OIII]) model grids, indicating an additional contribution to the continuum emission from composite stellar populations, or different 
C/O abundances, nebular temperatures and electron densities than assumed in the photoionization models. The Green Pea galaxies do not show a significant correlation between 
the Ly$\alpha$ and CIII] equivalent widths, and the observed scatter is likely due to the variations in the optical depth of Ly$\alpha$ to the neutral gas. Green Pea galaxies are likely 
to be density-bounded, and we examined the dependence of CIII] emission on the Lyman continuum optical depth.  The potential LyC leaker galaxies in our sample have high 
CIII] equivalent widths that can only be reproduced by starburst ages as young as $<$ 3 Myrs and harder ionizing spectra than the non-leakers. Among the galaxies with similar 
metallicities and ionization parameters, the CIII] equivalent width appears to be stronger for those with higher optical depth to LyC, as expected from the photoionization models. 
There are various factors that affect the CIII] emission line strengths and further investigation of a larger sample of CIII]-emitters is necessary to calibrate the dependence of CIII] 
emission on the escape of LyC radiation, and to enable application of the CIII] diagnostics to galaxies in the reionization epoch. 

\end{abstract}




\section{Introduction} \label{sec:intro}

The semi-forbidden [CIII]$\lambda$1907\AA~ + CIII]$\lambda$1909\AA~ doublet (hereafter, CIII]$\lambda$1909 or CIII]) is one of the strongest nebular emission line 
features observed in the rest-frame ultraviolet (rest-UV)  spectrum of low-metallicity, 12+log(O/H) $\leq$ 8.4 ($Z < 0.5 Z_{\odot}$) star-forming galaxies (SFGs) both 
locally (Garnett et al. 1995; Rigby et al. 2015; Berg et al. 2016, 2019; Senchyna et al. 2017, 2019) and at high redshifts (Fosbury et al. 2003, Erb et al. 2010, Bayliss 
et al. 2014; Stark et al. 2014, 2015a, 2017; Vanzella et al. 2016, de Barros et al. 2016, Maseda et al. 2017, Amo\'{r}in et al. 2017; Laporte et al. 2017; Berg et al. 2018; 
Le F\`{e}vre et al. 2019, Hutchison et al. 2019). The CIII] emission line is frequently observed in gravitationally-lensed SFGs at $z>2-6$ as the strongest line after 
Ly$\alpha$ $\lambda$1216\AA~ emission. Since the Ly$\alpha$ emission at $z>6$ is expected to be severely quenched by the increasingly neutral intergalactic 
medium (IGM), CIII] is emerging as an alternative redshift indicator for galaxies in the reionization era (Stark et al. 2015a, 2017; Ding et al. 2017). At $z>6$, the 
rest-UV spectrum of galaxies is redshifted to the Near-InfraRed (NIR) wavelengths, and is accessible to the spectrographs on upcoming facilities (such as, the 
{\it James Webb Space Telescope}, and the $>$20m-class Extremely Large Telescopes), which will address the key goal of identifying the sources of reionization. 
In recent years, there has been considerable effort to develop spectral diagnostics involving the CIII] emission line and other UV spectral features (e.g., 
CIV$\lambda\lambda$1548, 1550\AA~, HeII $\lambda$1640\AA~, OIII]$\lambda\lambda$1661, 1666\AA~; hereafter, OIII]$\lambda$1663 or OIII]) that can be used 
to understand the nature of the reionizers, their hard ionizing continua, and the physical conditions in their interstellar media (Jaskot \& Ravindranath 2016, hereafter 
JR16; Feltre et al. 2016; Gutkin et al. 2016; Nakajima et al. 2018a; Byler et al. 2018). The CIII] emission observed at high redshifts is mostly from strongly-lensed 
low-metallicity ($<$ 0.5 $Z_{\odot}$) , low-mass ($<$ 10$^{10}$ M$_{\odot}$) galaxies with high specific star formation rates (SSFRs $\gtrsim$ 2 Gyr$^{-1}$), and 
their CIII] equivalent widths show a broad range EW(CIII]) $\sim$ 3-25\AA~(Fosbury et al. 2003, Erb et al. 2010, Bayliss et al. 2014; Stark et al. 2014; Vanzella et 
al. 2016, de Barros et al. 2016; Berg et al. 2018). The composite rest-frame UV spectra of Lyman-break galaxies at $z\sim 3$ (Shapley et al. 2003), and of SFGs 
at $z=1-2.5$ (Steidel et al. 2016, Amor{\'i}n et al. 2017) also reveal the CIII]$\lambda$1909 and OIII]$\lambda$1663 nebular lines at sub-solar metallicities. 

In the local Universe (z$\sim$0), the UV spectra of star-forming galaxies obtained with the Faint Object Spectrograph, and Goddard High Resolution Spectrograph 
on the {\it Hubble Space Telescope (HST)} revealed CIII]$\lambda$1909\AA~ and OIII]$\lambda$1663\AA~ in metal-poor Blue Compact Dwarfs (BCDs) and 
Wolf-Rayet galaxies (Garnett et al. 1995; Leitherer et al. 2011). These UV lines, although measured only in a small sample of low-metallicity galaxies, were used to 
determine C/O ratios and infer the elemental carbon abundances (Garnett et al. 1995). Based on the compilation of all the available CIII] equivalent width (EW) 
measurements for star-forming galaxies at low and high redshifts, Rigby et al. (2015) found that low-metallicity galaxies tend to have a stronger CIII] emission, and 
EW(CIII]) $\geq$ 5\AA~ is only found in galaxies with $Z < 0.5 Z_{\odot}$. Only three galaxies in their compilation of 46 local SFGs showed EW(CIII]) $\geq$ 14\AA~, 
comparable to the strongest CIII] emitters at $z > 2$, and all three are Wolf-Rayet (WR) galaxies with high H$\alpha$ equivalent widths. More recently, UV spectroscopy 
of BCDs (Berg et al. 2016, 2019), He II-emitters (Senchyna et al. 2017), and extremely metal-poor galaxies (Senchyna et al. 2019) with the {\it HST}/Cosmic Origins 
Spectrograph (COS), has increased the number of CIII] emission line measurements in the local Universe. Most of these SFGs have EW(CIII]) in the range of values 
seen for the $z\sim2$ galaxies, and only few of them with very low metallicities reach high EWs $\sim$ 12-14\AA~. The distribution of EW(CIII]) as a function of 
metallicity from these new studies also confirms that CIII] emission EWs $>$ 5\AA~ only occurs at low-metallicities, with a clear transition occurring at 12+ log(O/H) 
$<$ 8.4 or $Z < 0.5 Z_{\odot}$ (Senchyna et al. 2017, 2019). 

The physical properties of the CIII]-emitters at $z\sim 3$ inferred from the UV-optical spectra reveals that the ISM conditions in these low-metallicity SFGs are different 
from the high-metallicity SFGs, as evidenced by the presence of extreme optical emission lines with high EWs. Most of the CIII]-emitters for which optical spectra are 
available, show high EWs ($\sim$ 1000\AA~) for the [OIII]$\lambda$5007\AA~ (hereafter, [OIII]$\lambda$5007 or [OIII]) emission line (Stark et al. 2014; Vanzella et al. 
2016, Maseda et al. 2017), and would be identified as Extreme Emission Line Galaxies (EELGs; van der Wel et al. 2011) based on the contribution of the [OIII] emission 
to their broad-band colors. The star-forming galaxies at $z>2$ with strong UV-optical emission lines in their spectra, commonly show high ionization parameters (log $U$ 
$\geq$ -2), blue UV continuum slopes ($\beta \sim -2$), low dust content, and often show evidence for strong outflows (Steidel et al. 2014, Shapley et al. 2015, Maseda 
et al. 2014). The best fit photoionization models based on the photometric Spectral Energy Distributions (SEDs) suggest low metallicities and large EW([OIII] +H$\beta$) 
($>$ 700\AA) for the CIII] emitters at $z > 6$ (Stark et al. 2015a, 2017). The contribution of strong nebular [OIII]$\lambda$5007+H$\beta$ to the Spitzer IRAC [3.6]-[4.5] 
colors is being increasingly used to identify $z\sim 6.6 - 9.0$ galaxies (Smit et al. 2014, 2015; Roberts-Borsani et al. 2016; Stark et al. 2016), and their follow-up 
spectroscopy have revealed some of the strongest CIII] emitters with EW(CIII]) $>$ 20\AA~ (Stark et al. 2017, Hutchison et al. 2019). Many of these high redshift sources 
are also Lyman Alpha Emitters (LAEs), and the EWs of the CIII] and Ly$\alpha$ emission lines are found to be correlated (Shapley et al. 2003; Stark et al, 2014, 2015a; 
Rigby et al. 2015). LAEs are also actively star-forming with high SSFRs, low dust content, strong [OIII]$\lambda$5007\AA~ emission, and their 
[OIII]$\lambda$5007\AA~/[OII]$\lambda$3727\AA~ ratios (hereafter, O$_{32}$) imply high ionization parameters (Nakajima \& Ouchi 2014; Nakajima et al. 2016).

Although low metallicities and young stellar ages are favorable for CIII], the strength of the emission is determined by many other factors which are not yet fully understood.  
CIII] is a collisionally-excited emission line, and high nebular temperatures and densities are expected to enhance the emission due to the increased collisional rates. 
Photoionization codes are being used increasingly to try and understand the role of age, ionization parameter, metallicity, and dust on the emergent CIII] nebular line flux 
(JR16; Nakajima et al. 2018a). The models are able to reproduce the observed EW(CIII]) values including the very high values $>$ 10-15\AA~, but require extreme values 
of ionization parameter (logU $>$ -2), and very young stellar populations with ages $<$ 3 Myrs at $Z < 0.5 Z_\odot$. Only models that incorporate the effects of binary stellar 
evolution (Eldridge \& Stanway 2009;  Eldridge et al. 2017) are able to provide the hard ionizing continuum required to produce the strong CIII] emission with EW(CIII]) $>$ 
10\AA~ over long timescales ($>$ 3 Myrs) as compared to the single-star models (JR16). The stellar population models including massive star binaries are also known to 
consistently account for the observed nebular emission line ratios in the rest-frame UV and optical spectra of $z\sim 2.4$ galaxies (Steidel et al. 2016). Photoionization models 
predict that high optical depths, high C/O ratios, and the presence of shocks can enhance the CIII] emission for a given age, metallicity, and logU (JR16). In recent years, UV 
spectral diagnostics involving CIII], CIV, and He II have been identified (Feltre et al. 2016; Gutkin et al. 2016), that have the ability to distinguish between an ionizing continuum 
powered by star-forming galaxies versus Active Galactic Nuclei (AGNs). However, observations of low-metallicity galaxies that span the range of physical properties that determine 
the CIII] strengths are limited, offering few constraints to disentangle the effect of various model parameters on the emergent nebular line fluxes. 

The Green Pea Galaxies (GPs) are among the closest low redshift analogs, to the low-mass, low-metallicity SFGs in the reionization era. The GPs at $z=0.1-0.3$ were originally 
identified in the Sloan Digital Sky Survey (SDSS) DR7 data, through a $gri$ color selection based on the unusually strong [OIII]$\lambda$5007 emission (EW([OIII] $>$ 300\AA~) in 
the $r$-band (Cardamone et al. 2009). Their optical emission line spectra imply high star formation rates (SFRs $>$10 - 60 M$_{\odot}$ yr$^{-1}$), high specific SFRs (10$^{-7}$ 
to 10$^{-9}$ yr$^{-1}$), low metallicities ($Z \leq 0.5 Z_{\odot}$), low extinction ($E(B-V) \leq 0.25$), and the presence of hard ionizing spectra produced by very hot stars (Cardamone 
et al. 2009; Izotov et al. 2011, Amor{\'i}n et al. 2012). GPs have high electron temperatures ($>$ 15,000K) derived from the optical nebular lines, and the high O$_{32}$ ratios suggest 
that they have  high ionization parameters (Kewley \& Dopita 2002; Jaskot \& Oey 2013). The strong Balmer lines, and detection of the HeII $\lambda$4686\AA~ line indicate very 
young ages of $<$ 3-5 Myrs for the dominant stellar population (Jaskot \& Oey 2013). Although the UV-optical spectra of GPs are dominated by the recently formed young stars, 
modeling their star formation history shows that they host an older population ($>$ 1 Gyr), which contributes most of the stellar mass (Amor{\'i}n et al. 2012). GPs have high EW([OIII]) 
($>$ 500-1000\AA~), similar to the CIII] emitters at $z\sim 2$, and their O$_{32}$ ratios, ionization parameters, metallicities, and sSFRs are more extreme than the normal, low-$z$ 
star-forming galaxies, but are comparable to LAEs at $z>2$ (Nakajima \& Ouchi 2014). The GPs have low stellar masses (M$\sim$ 10$^{8}$ - 10$^{10}$ M$_{\odot}$), are 
UV-luminous ($L_{FUV} \sim 3\times10^{10} L_{\odot}$), and have compact sizes $\leq$ 5 kpc. In the {\it HST} images, some of the GPs appear as clumpy galaxies, with one 
or few bright super-star clusters that dominate the morphology, giving them a close resemblance to star-forming galaxies at $z>2$ (Izotov et al. 2018b, Henry et al. 2015).  

The GPs are extremely valuable for exploring the nebular line diagnostics, because they are at low redshifts, $z<0.3$, and the entire rest UV-optical spectra are accessible to 
determine their physical properties, ISM conditions, and escape fractions in much more detail than can be done for the high-redshift galaxies. The rest-UV spectra of GPs 
taken with the {\it HST}/COS covering the rest-wavelengths 950-1450\AA~ show strong Ly$\alpha$ emission with a variety of profile shapes and escape fractions, 
$f_{esc}$(Ly$\alpha$) = 0 - 98\% (Jaskot \& Oey 2014; Henry et al. 2015; Izotov et al. 2018a). Their Ly$\alpha$ profiles are most often double-peaked, and the separation 
between the peaks is an indicator of the column density of neutral hydrogen and possible leakage of ionizing radiation (Verhamme et al. 2015, Verhamme et al. 2017). 
Direct measurements of the Lyman continuum (LyC) escape are only available for a handful of galaxies at low redshifts, and GPs have been the most promising candidates  
with high observed escape fractions, $f_{esc}(LyC)$ = 3\% - 72\% (Izotov et al. 2016a,b; Izotov et al. 2018a,b). Recently, Schaerer et al. (2018) reported the detection of intense 
CIII] $\lambda$ 1909 emission with EW(CIII]) = 11.7$\pm$ 2.9\AA~ in GP J1154+2443, a galaxy that has $f_{esc}$(Ly$\alpha$) = 98\%, and is a confirmed Lyman continuum 
emitter with $f_{esc}(LyC)$ = 46\%. In JR16, we presented various diagnostics involving the CIII] emission line, and offered specific predictions for the dependence of EW(CIII]) 
on the optical depth in GPs. Galaxies that have weaker EW(CIII]) than predicted for their O$_{32}$ ratio are likely to be density-bounded or optically thin systems with a high 
fraction of Lyman continuum escape. Thus, EW (CIII]) can be an indicator of the optical depth, if the ionization parameter, age, and metallicity can be constrained using other 
emission lines in the spectra. According to the JR16 models with $Z\sim 0.003$ and age $<$ 10 Myr, for the GPs with O$_{32}$ ratios $\sim 1-10$, the CIII] EWs are $\leq 1-4$\AA~ 
at low optical depths, and EW (C III]) $>$ 6\AA~ would require extremely young stellar ages, $\leq$ 2 Myrs. The photoionization models predict CIII] EWs for most of the GPs 
to be in the range of $2-10\AA~$ depending on the age of the starburst. Further, the models predict CIV $\lambda$1549\AA~ emission in GPs to be $\lesssim$ 50\% of the CIII] 
emission and He II $\lambda$ 1640\AA~ $<$ 10\% of CIII], although the latter may be affected by contributions from other sources, such as, a significant Wolf-Rayet population 
or shocks.

In this paper, we present  new rest-UV spectroscopy of a sample of ten Green Pea Galaxies, and compare it with the JR16 photoionization model predictions to 
understand the conditions that give rise to the CIII] nebular emission and its dependence on metallicity ($Z$) and ionization parameters ($U$). Since optical emission line 
ratios from SDSS spectra  offer independent constraints on the metallicity ($Z$) and ionization parameter ($U$), and the Ly$\alpha$ lines offer information about the optical 
depth, these galaxies are excellent candidates to test the model predictions for CIII] emission. The details of observations and analysis are presented in section 2. The results 
on the correlations between CIII] emission and other UV-optical lines are presented in section 3, followed by the comparison with photoionization models in section 4. We 
discuss the results from GPs in the context of low-metallicity galaxies at high redshifts and implications for galaxies in the reionization era in section 5. Throughout this paper 
we assume 12+log(O/H) = 8.69 for Solar metallicity (Asplund et al. 2009).

\section{HST/STIS Observations and Data Analysis} \label{sec:style}

$HST$ observations for the sample of GP galaxies were obtained with the Space Telescope Imaging Spectrograph (STIS) through the program GO-14134 (PI: Ravindranath), 
which was awarded 18 orbits. All the galaxies in our sample have archival $HST$/COS spectra covering the wavelengths that include the Ly$\alpha$ emission. The STIS 
observations were designed to use the HST UV spectra to measure the CIII]1909\AA~ line fluxes and equivalent widths, investigate the correlations between CIII] emission and 
Ly$\alpha$ emission, and examine the behavior of CIII] emission relative to the rest-optical emission-line diagnostics derived from the SDSS spectra. 

\subsection{The Sample and Observations} \label{subsec:sample and observations}

We selected all the GP galaxies that had existing archival $HST$/COS spectra covering the wavelengths that include the Ly$\alpha$ emission, 
from two previous HST GO programs that observed GPs.
The sample consists of ten GPs (Table 1) at redshifts, $z=0.1-0.3$, with low metallicities ($<$ 0.4 Z$_{\odot}$; Izotov et al. 2011), of which four are classified as ``{\it extreme}'' 
GP galaxies based on their high optical emission line ratios, with [OIII]$\lambda\lambda$4959,5007/[OII]$\lambda$3727 $>$ 9 (Jaskot \& Oey 2013).  {\it HST}/COS G130M, G160M 
spectra are available in the archive for eight of the objects, covering the rest-wavelengths $\lambda$ $\sim$ 950-1450\AA~  from GO-12928 (PI: Henry), and two of the remaining 
galaxies have COS G160M spectra from GO-13293 (PI: Jaskot). The COS G160M spectra from both these programs provide Ly$\alpha$ emission line detections with signal-to-noise 
(S/N) $>$ 10. The sample of GPs represent a range of physical properties with EW (Ly$\alpha$) $\sim 0 -170\AA~$, nebular oxygen abundances with 7.8$<$12+log(O/H)$<$8.3 
($0.15-0.40$ $Z_{\odot}$), and stellar masses ($< 3\times 10^{9}$ M$_{\odot}$).  The GPs have high UV surface-brightnesses ($<$ 20 mag/arcsec$^{-2}$) in the {\it GALEX} 
NUV images, and compact sizes ($\sim$ 1 arcsec) which corresponds to physical sizes of 2.5-3.5 kiloparsecs for their redshifts.

\begin{deluxetable*}{ccccccccc}[b!]
\tablecaption{Properties of the Green Pea Galaxies and $HST$/STIS Observations Summary \label{tab:mathmode}}
\tablecolumns{9}
\tablenum{1}
\tablewidth{0pt}
\tablehead{
\colhead{Name} & \colhead{R.A.} & \colhead{Dec.} &
\colhead{Redshift} &
\colhead{{\it g}} & \colhead{NUV} & \colhead{Exposure Times} & \colhead{$\lambda_{rest frame}$} \\
\colhead{} & \colhead{(J2000)} & \colhead{(J2000)} & \colhead{} &
\colhead{(AB mag)} &\colhead{(AB mag)} & \colhead{(s)} & \colhead{($\AA$~)}
}
\startdata
J030321$-$075923 & 03:03:21.41 & $-$07:59:23.2 & 0.165 &19.4 & 19.6 & 4924 & 1345$-$2733 \\
J081552$+$215623 & 08:15:52.00 & $+$21:56:23.6 & 0.141 & 20.1 & 20.1 & 7499 &  1374$-$2790\\
J091113$+$183108 & 09:11:13.34 & $+$18:31:08.1 & 0.262 &19.5 & 19.8 & 4896 &  1242$-$2522\\
J105330$+$523752 & 10:53:30.82 & $+$52:37:52.8 & 0.253 &18.8 & 19.2 & 2300 &  1251$-$2541\\
J113303$+$651341 & 11:33:03.79 & $+$65:13:41.3 & 0.241 & 20.1 & 19.7 & 5282 & 1263$-$2565 \\
J113722$+$352426 & 11:37:22.14 & $+$35:24:26.6 & 0.194 & 18.9 & 19.3 & 2163 &  1313$-$2666\\
J121903$+$152608 & 12:19:03.98 & $+$15:26:08.5 & 0.196 & 19.5 & 19.3 &1909 &  1311$-$2662\\
J124423$+$021540 & 12:44:23.37 & $+$02:15:40.4 & 0.239 & 19.2 & 19.9 & 4974 &  1265$-$2569\\
J124834$+$123402 & 12:48:34.63 & $+$12:34:02.9 & 0.263 & 19.9 & 19.9 & 4698 &  1241$-$2520\\
J145735$+$223201 & 14:57:35.13 & $+$ 22:32:01.7 & 0.149 & 19.4 & 19.9 & 4926 &  1364$-$2771\\
\enddata

\tablecomments{The coordinates, redshifts, and {\it g} magnitudes are from SDSS DR7, and  the $NUV$ magnitudes are taken from $GALEX$. 
The last two columns provide the total exposure times for the $HST$/STIS observations, and the rest-frame UV wavelength coverage in 
the G230L grating for each galaxy based on its redshift.
}
\end{deluxetable*}

The $HST$/STIS observations were carried out using the G230L grating that covers the wavelength range 1700-3200\AA~ at a spectral resolution of R $\sim$ 300 - 600 across 
these wavelengths. The average dispersion with the G230L grating is 1.58\AA~ per pixel, and the 2-pixel resolution element yields a 3.16\AA~ wavelength resolution for the 
spectrum. STIS is equipped with the NUV-MAMA detector, with a pixel scale of 0.025 arcsecs/pixel (or 0.05 arcsecs per 2-pixel resolution element). The GP galaxies are barely 
resolved in SDSS images, with sizes $\sim$ 1 arcsec. We used the 52$^{\prime\prime}$ $\times$ 0.5$^{\prime\prime}$ slit to optimize between slit loss and spectral resolution, resulting 
in R$\sim$ 500 over most of the wavelength range. The target acquisition images were taken using short exposures ($\sim$ 100s) on the STIS CCD detector. The exposure times 
for the STIS NUV spectra were estimated based on the {\it GALEX} NUV flux for the continuum and using the predicted CIII]$\lambda$1909\AA~ and [OIII]$\lambda$1663\AA~ 
emission line fluxes from CLOUDY photoionization models (Ferland et al. 2013). The total exposure times varied from 2500s to 7500s based on the target brightness and predicted 
emission line fluxes. 

\begin{figure}
\plottwo{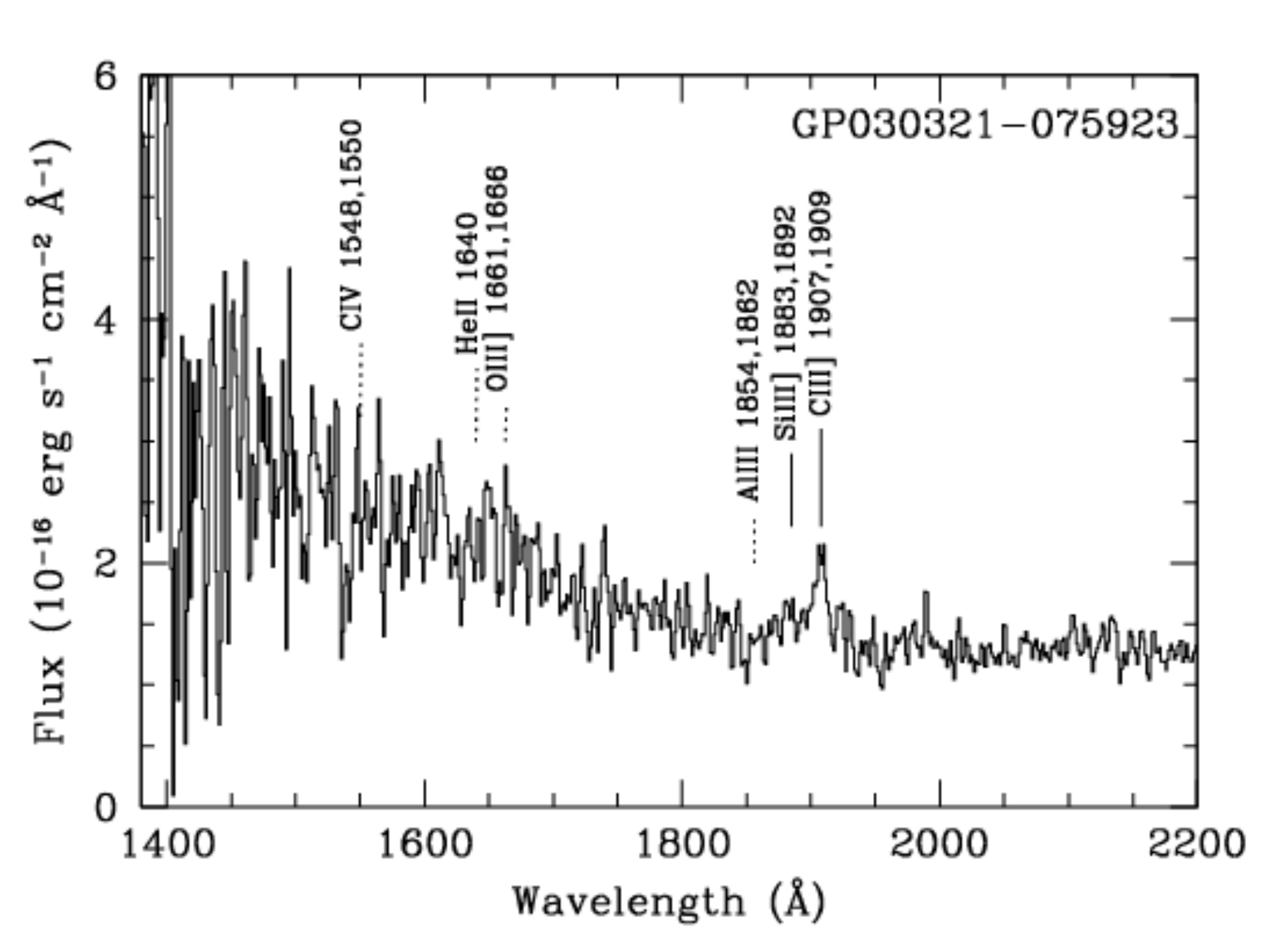}{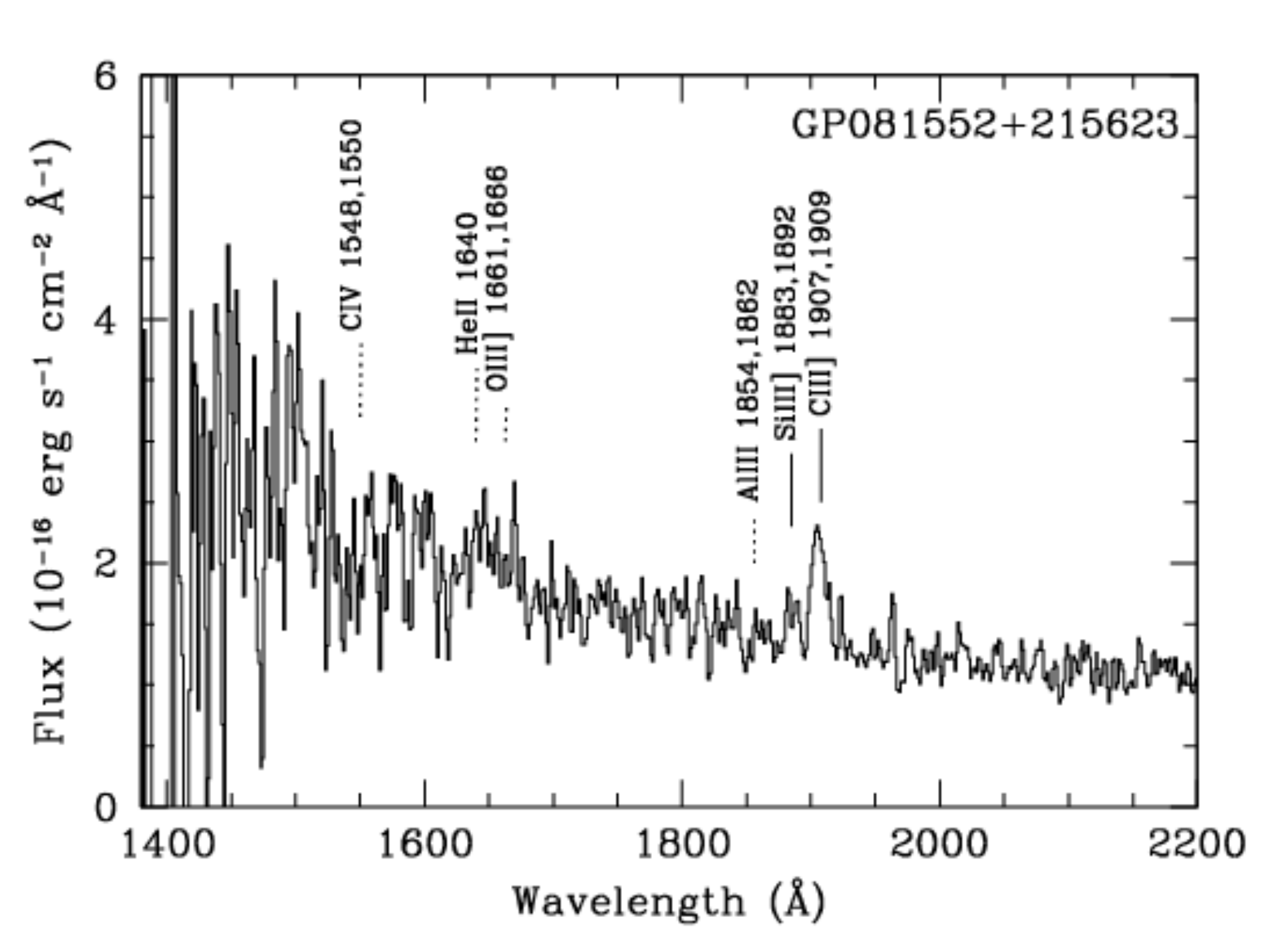}
\plottwo{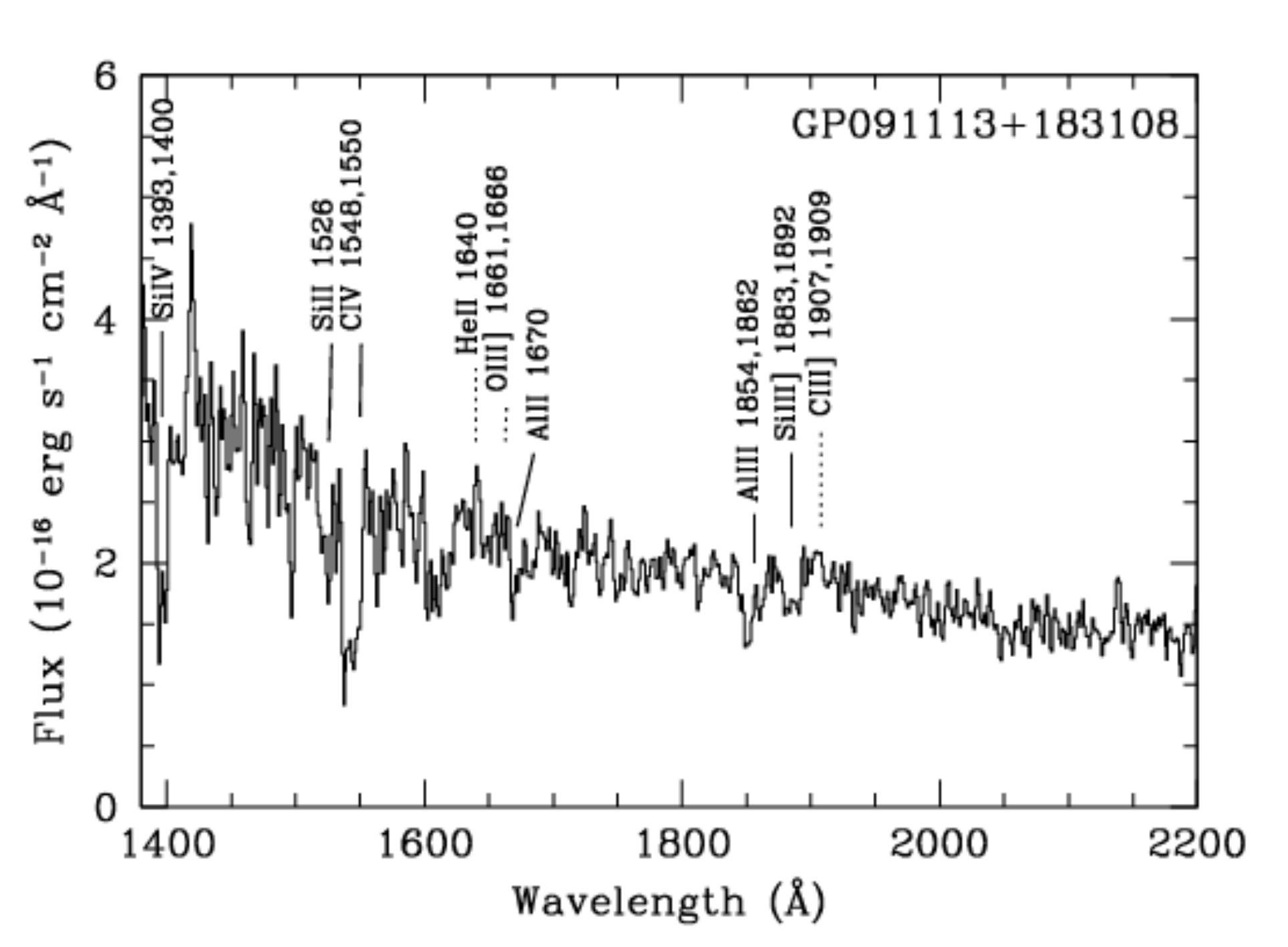}{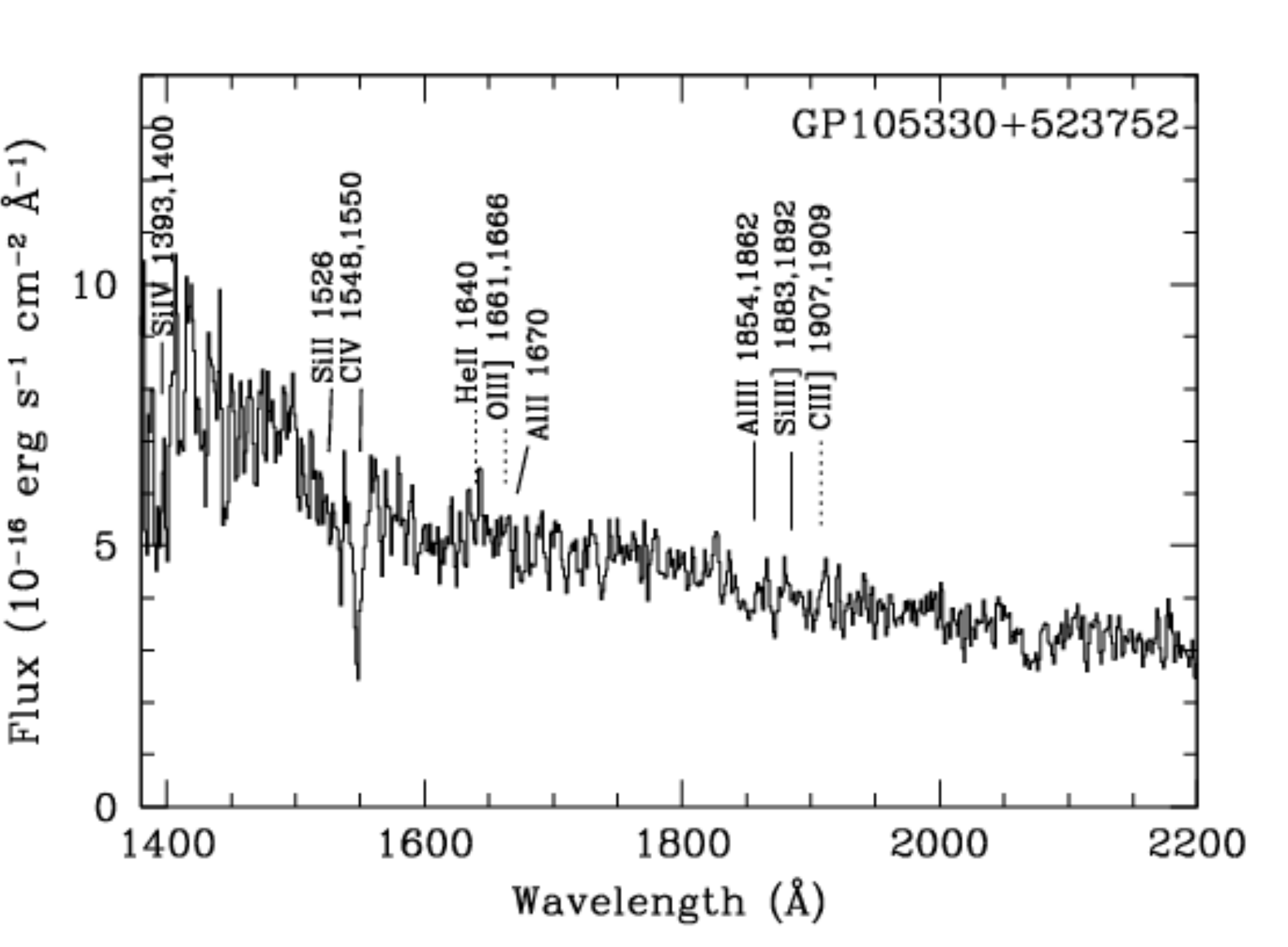}
\plottwo{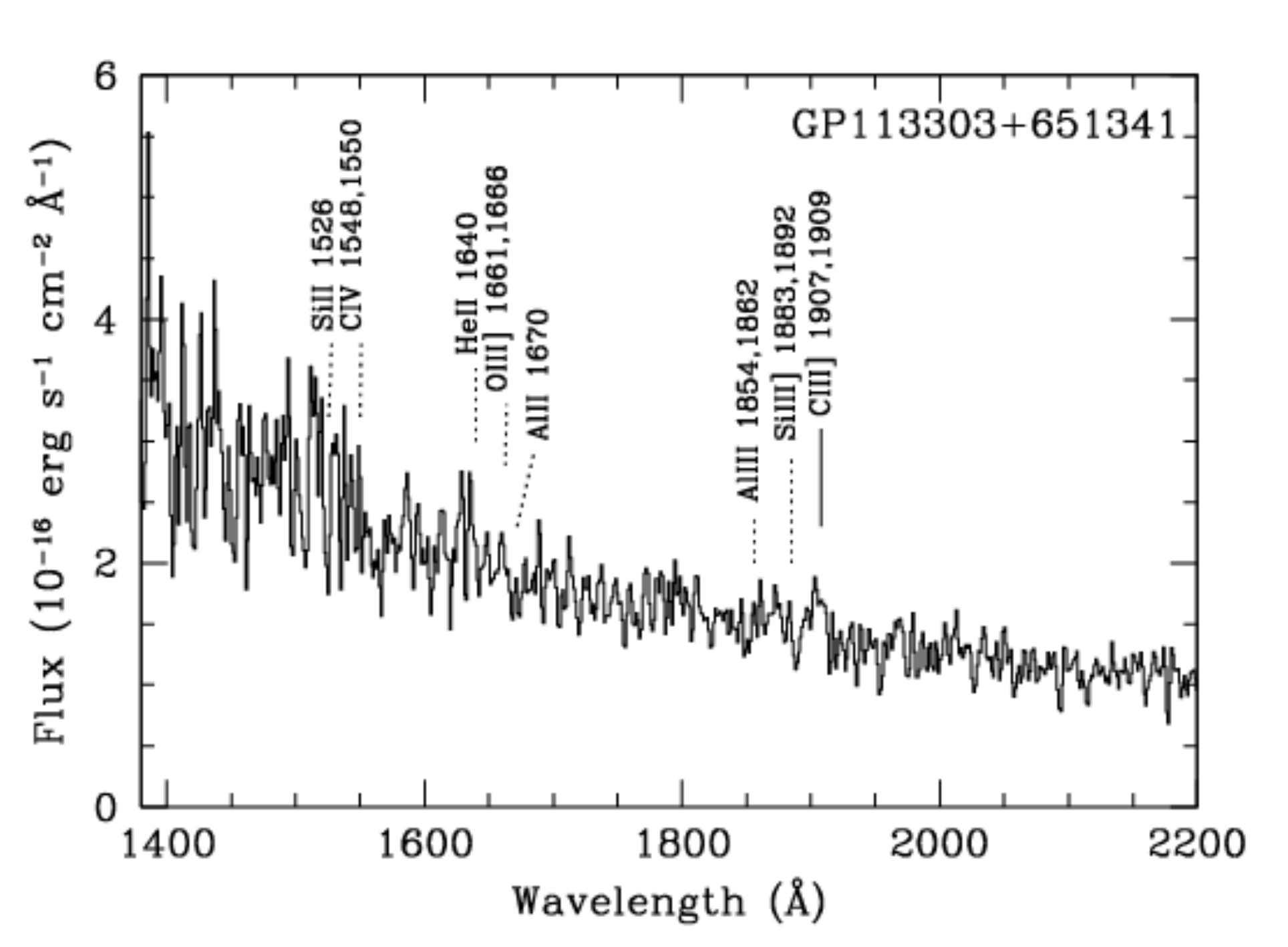}{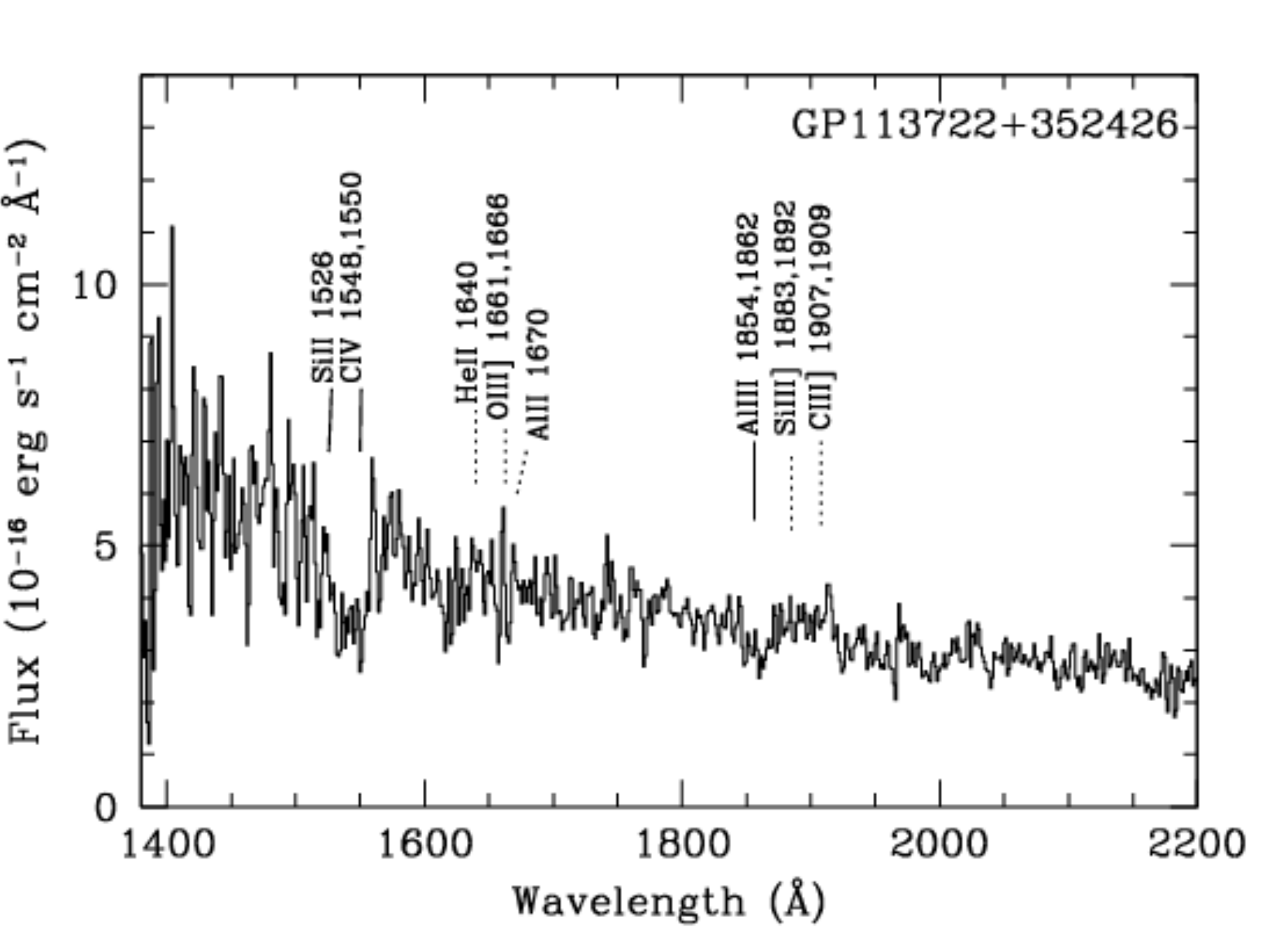}
\caption{The HST STIS spectra covering the rest-UV wavelengths from $\sim$ 1400 $-$ 2200\AA~ for the Green Pea galaxies J030321$-$075923, J081552$+$215623, 
J091113$+$183108, J105330$+$523752, J113303$+$651341, and J113722$+$352426. The locations of the detected spectral features are marked with solid lines, and 
the positions of other expected spectral lines are marked using dotted lines.}
\end{figure}

\begin{figure}
\plottwo{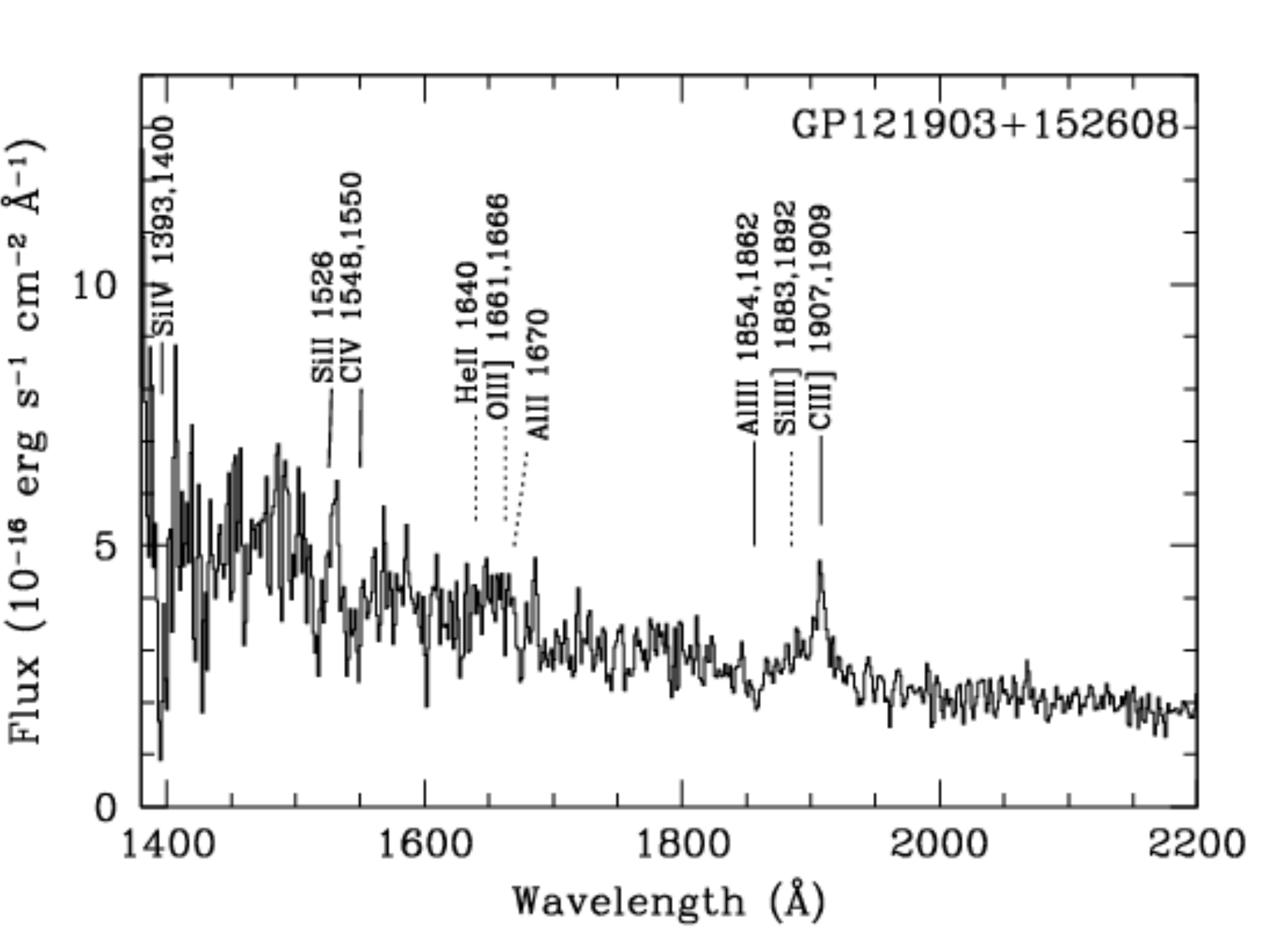}{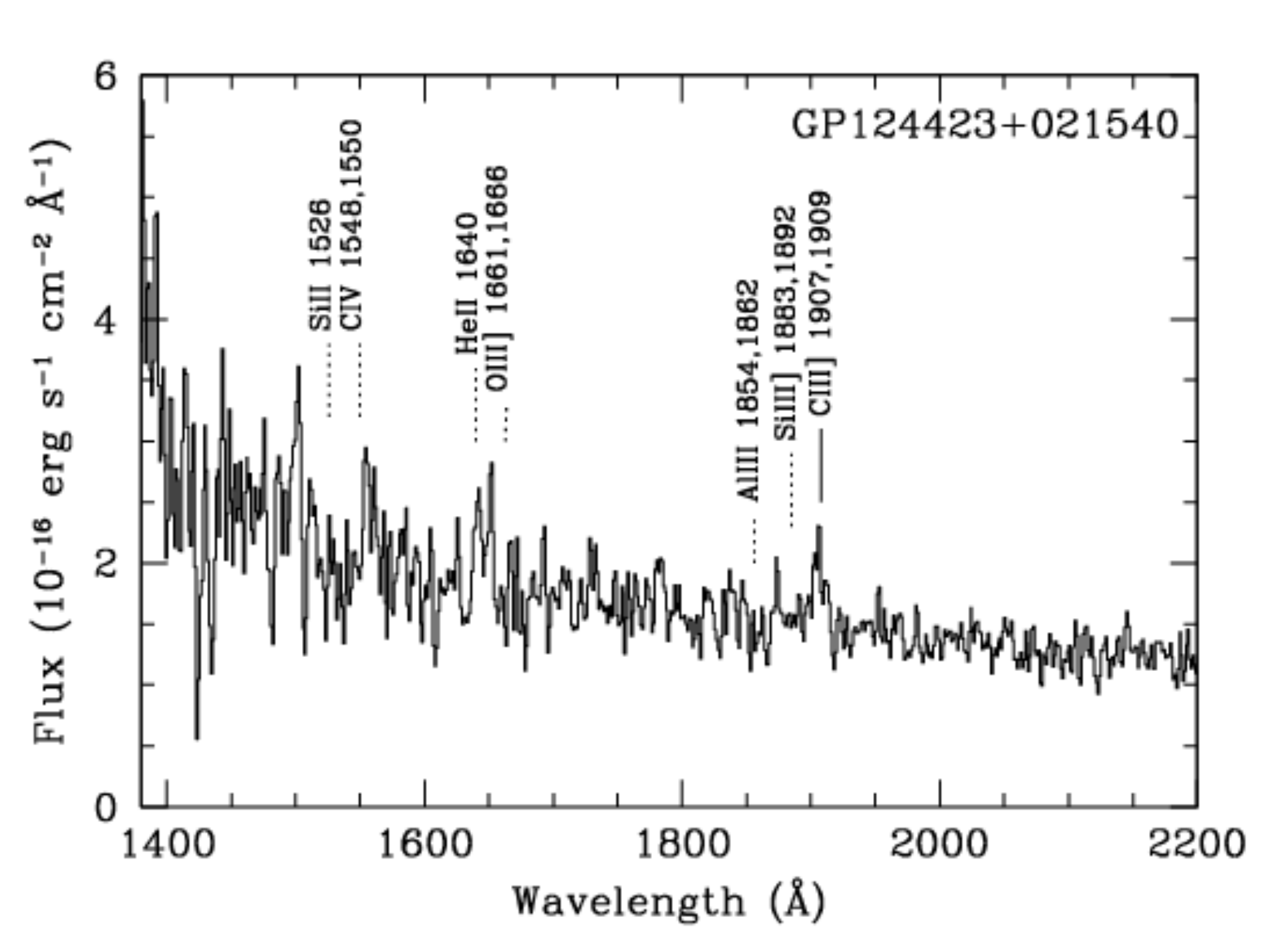}
\plottwo{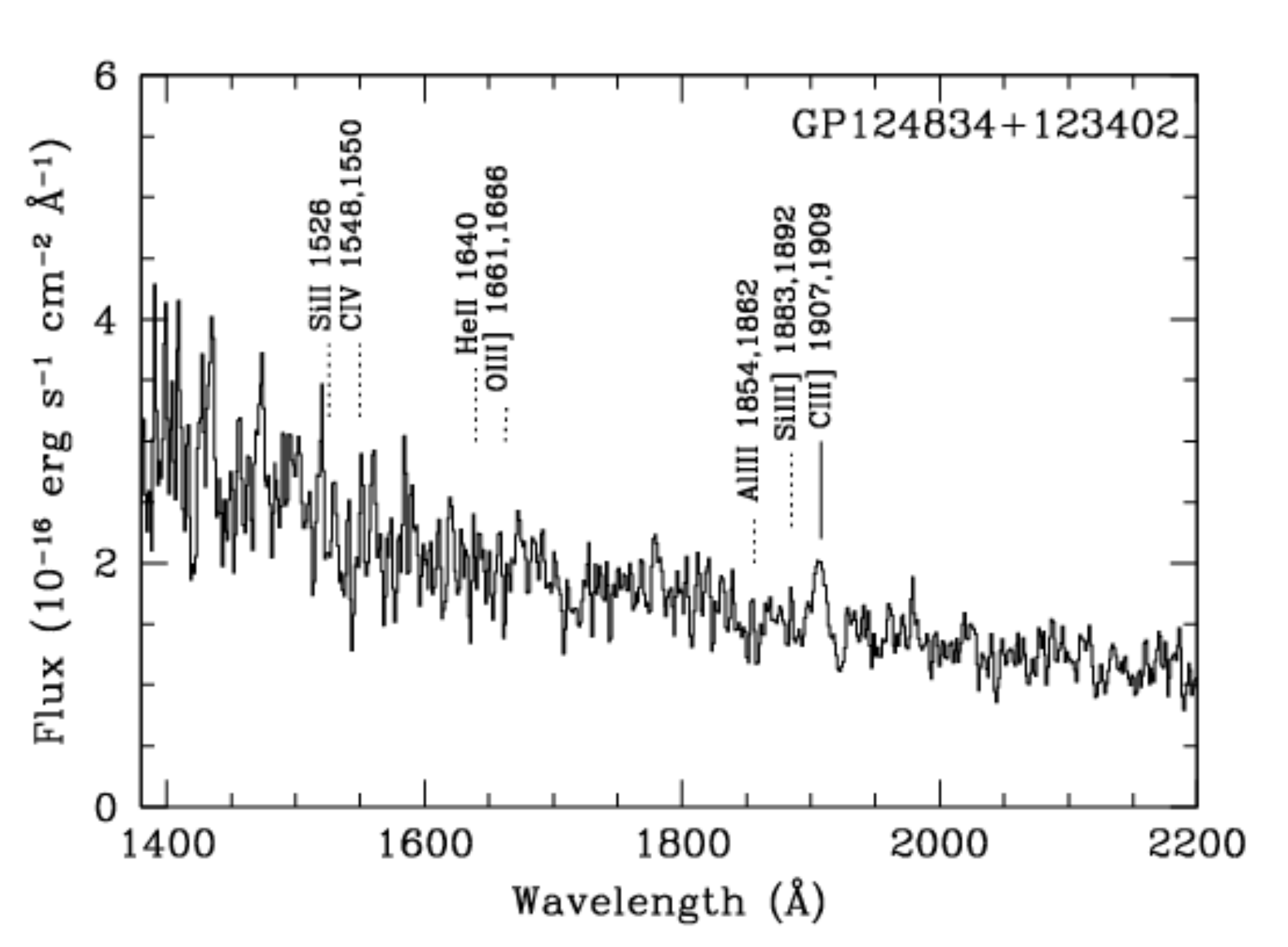}{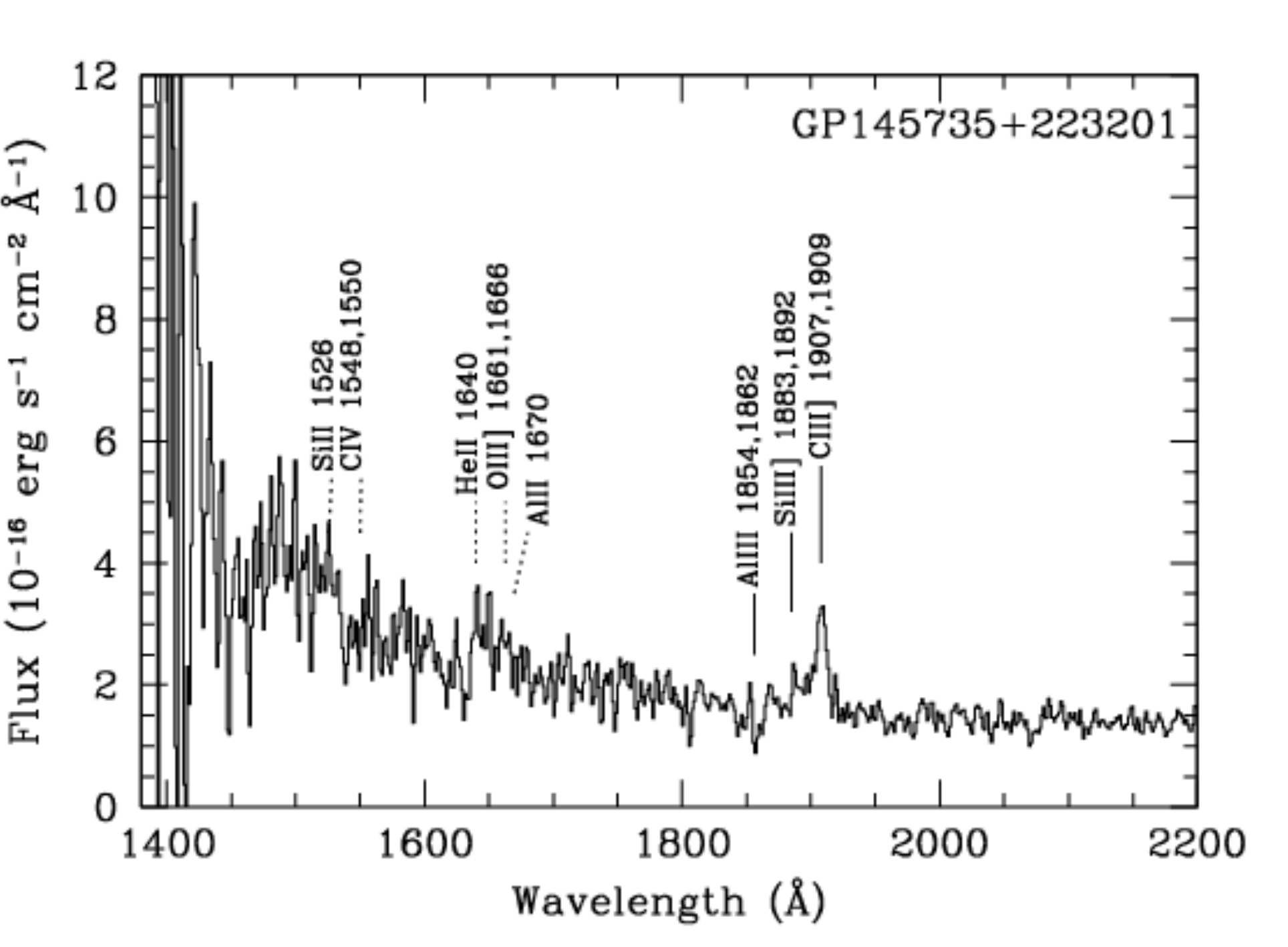}
\caption{ Same as Figure 1, for the galaxies J121903+152608, J124423+021540, J124834+123402, J145735+223201.}
\end{figure}


\subsection{Analysis} \label{subsec:analysis}

We retrieved STIS spectra  from the MAST archive, which are pipeline-processed and calibrated using CALSTIS. The pipeline produces flux-calibrated, 
rectified two-dimensional spectroscopic  images, and one-dimensional spectra of flux versus wavelength for each target. The default aperture used for 
spectral extraction in the pipeline is 11 pixels (0.275 arcsecs) along the cross-dispersion axis. However, some of the GPs are clearly more extended in the 
spatial direction based on the 2-dimensional spectroscopic image and the acquisition images. We re-processed the raw data files using updated reference 
files, and performed the spectral extraction using optimized extraction boxes. The standard extraction height of 11 pixels was used for the more compact 
GPs, and the spectra of the more extended sources were extracted using 21 pixels in the spatial direction. The extracted 1-D spectra were smoothed by 2 
pixels in the wavelength direction to a final spectral resolution of 3.2\AA~ per pixel. In the cases where there were multiple exposures from 2 or more 
orbits, the one-dimensional spectra from multi-orbit visits were combined using {\it scombine} in pyRAF. The final combined spectra have S/N $\geq$ 4 per 
pixel in the continuum {\bf at $\sim$ $1900-2000$\AA~} for all the GPs. 

The {\it HST}/STIS spectra covering rest-UV wavelengths from 1400-2200\AA~ for the sample are presented in Figure 1 and Figure 2.  The CIII] emission 
line is detected with greater than 3-$\sigma$ significance in 7 of the 10 GPs. We measured the fluxes and equivalent widths of the CIII] emission line using 
the {\it splot} task within the {\it specred} package in PyRAF. We used the HST/COS acquisition images to measure the extent of the UV continuum and to
correct for the slit losses due to the 0.5$^{\prime\prime}$ $\times$ 0.525$^{\prime\prime}$ aperture used for HST/STIS spectral extraction for the GPs. In 
most cases the UV continuum is compact and the extraction box includes $75-80$\% of the flux, except for two galaxies which have multiple UV-bright knots 
and only 62\% of the flux is within the extraction box. There is a possibility that the nebular emission may be more extended than the continuum. We used
the HST/ACS images taken in the narrow-band ramp filter tuned to [OIII]$\lambda$5007\AA~ from HST-GO-13293 (PI: Jaskot) to measure the slit losses for 
4 galaxies (J0303-0759, J0815+2156, J1219+1526, and J1457+2232) in our sample. The extent of the nebular emission and measured slit losses were 
found to be comparable to that measured from the UV continuum in all cases. The EW(CIII]) measurements were performed by using direct integration of the 
area under the line, and also by using the Gaussian profile fitting method. The measurements from the two methods are in close agreement and the average 
of the two is used for further analysis. The continuum flux ($F_{\nu}$) is assumed to be flat in the vicinity of the CIII] line. The noise in the continuum spectrum 
on either side of the emission line contributes to the uncertainty in the measured EWs. We include this in the errors by using repeated measurements with 
different continuum levels to determine the r.m.s uncertainty arising from the continuum noise. 

We do not detect any measurable CIII] emission in 3 GPs, J0911+1831, J1053+5237, and J1137+3524, and provide upper limits on their CIII] fluxes in 
Table 2. These GPs with no CIII] emission line in their rest-UV spectrum have dominant interstellar (and possibly stellar) absorption features, in particular, 
Si IV $\lambda\lambda$1393, 1402\AA~ and C IV $\lambda\lambda$1548, 1550\AA~. We detect strong CIV $\lambda\lambda$1548,1550\AA~ absorption lines 
with EW(C IV) = $-10.53\pm$ 0.3\AA~, $-4.55\pm$ 0.02\AA~, and $-14.8\pm$ 1.91 in J0911+1831, J1053+5237, and J1137+3524 respectively. In the case 
of J0911+1831 and J1053+5237, the spectra also show SiIV $\lambda\lambda$1393, 1402\AA~ absorption. GP1137+3524 has a lower redshift (z=0.194) 
and only a broad CIV absorption is detected. SiIV absorption cannot be identified for this galaxy because it falls in the low S/N blue end of the spectrum.

We do not detect the OIII]$\lambda$1663\AA~ doublet in any of the GPs, although it is one of the most prominent nebular lines seen in the rest-UV 
spectra of low-metallicity galaxies. We provide 3-$\sigma$ upper limits for the OIII]$\lambda$1663\AA~ doublet in Table 2. The photoionization models from 
JR16 show a tight correlation between the [OIII]$\lambda$4363\AA~ optical line emission and the OIII]$\lambda$1663\AA~ emission, and this allows us to
predict the expected OIII] fluxes. We estimate a value for the ionization parameter ($U$) using the O$_{32}$ vs. C III] EW diagram from JR16. Then, keeping 
$U$ fixed, we interpolate the predicted OIII]1663/[OIII] 4363 ratios as a function of metallicity ($Z$), given the GPs' calculated metallicities. The uncertainty in 
$Z$ dominates over the uncertainty in our $U$ estimate, and the errors provided in Table 2 show the uncertainty in the predictions given the GPs' metallicity 
uncertainties. The predicted OIII] $\lambda$1663\AA~ fluxes are below or close to the detection limits in most cases. In the LyC leaker GP galaxy, J1154+2443, 
Schaerer et al. (2018) detect OIII] $\lambda$1663\AA~ with EW = 5.8$\pm$2.9\AA~, which is $\sim$ 0.5 EW(CIII]). However, that galaxy has a much lower 
metallicity than our sample with 12+log(O/H) $\sim$ 7.65, closer to the metallicities of BCDs (Berg et al. 2016, 2019) that also show significant OIII]$\lambda$1663\AA~ 
detections. There are 5 galaxies in our sample with $Z \leq 0.2 Z_{\odot}$, but even in such low-metallicity galaxies, the  OIII]$\lambda$ 1663\AA~ can be 
weak with median EW $<$ 2.4\AA~  (Senchyna et al. 2017). 

We measured the fluxes and EWs of the Ly$\alpha$ emission from their $HST$/COS FUV spectra, and the [OIII]$\lambda$4363\AA~, [OIII]$\lambda$5007\AA~, 
and H${\alpha}$ emission lines from the SDSS spectra for all of the GPs. The line fluxes were corrected for Milky Way extinction using the Fitzpatrick (1999) 
extinction law and the Schlafly \& Finkbeiner (2011) dust map. The fluxes were corrected for internal reddening using the Balmer decrement derived from the observed 
H${\alpha}$/H${\beta}$ ratio and the Cardelli, Clayton, \& Mathias (1989)  reddening law, taking into account the variation of  R$_{V}$ in the presence of intense 
UV radiation as outlined in Izotov et al. (2017). All of the GPs exhibit very low internal reddening with E(B-V) = 0.03 - 0.2. The extinction-corrected fluxes for 
rest-UV and rest-optical emission lines are provided in Table 2, and the EWs are presented in Table 3.

\begin{figure}
\plottwo{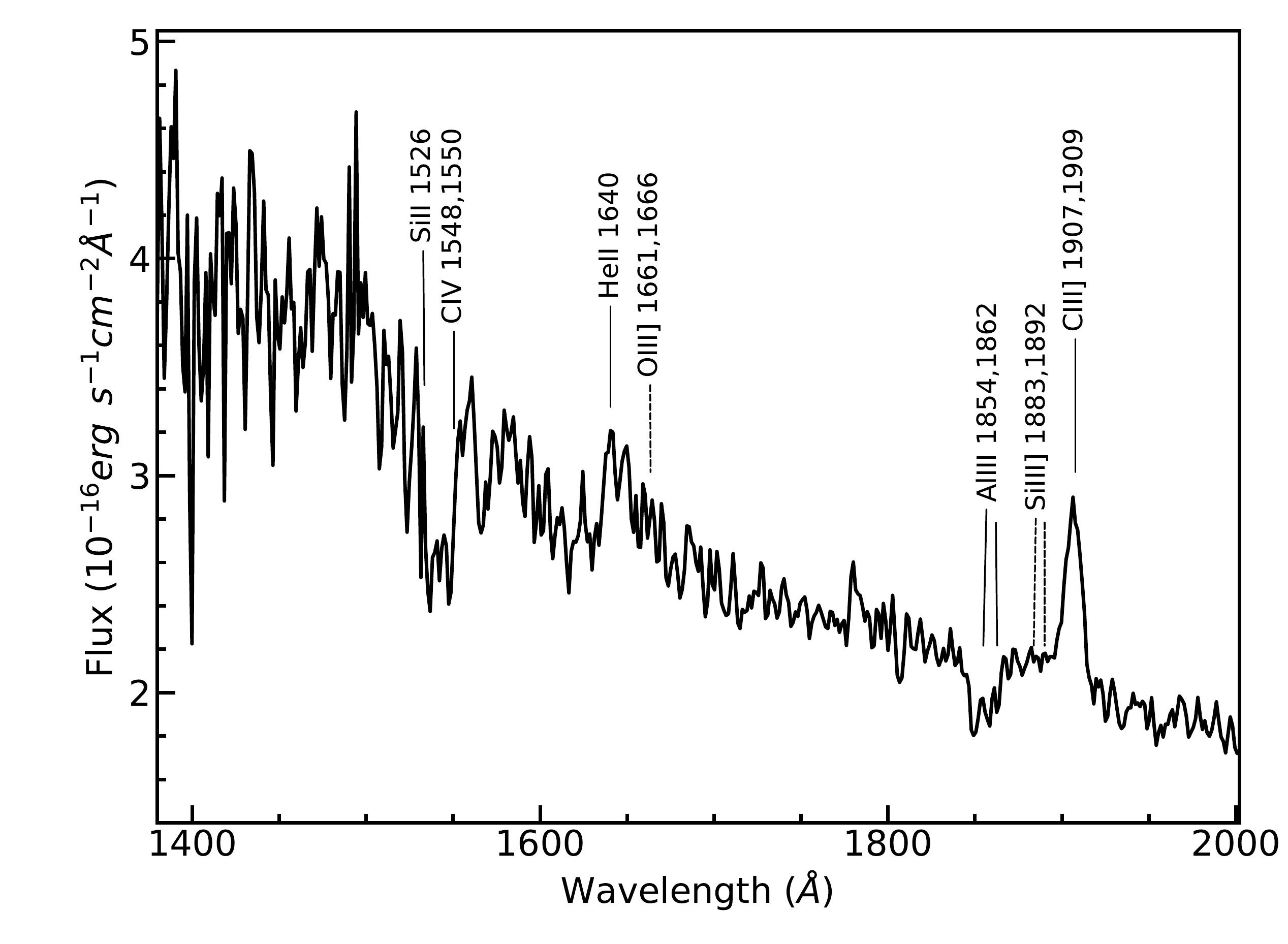}{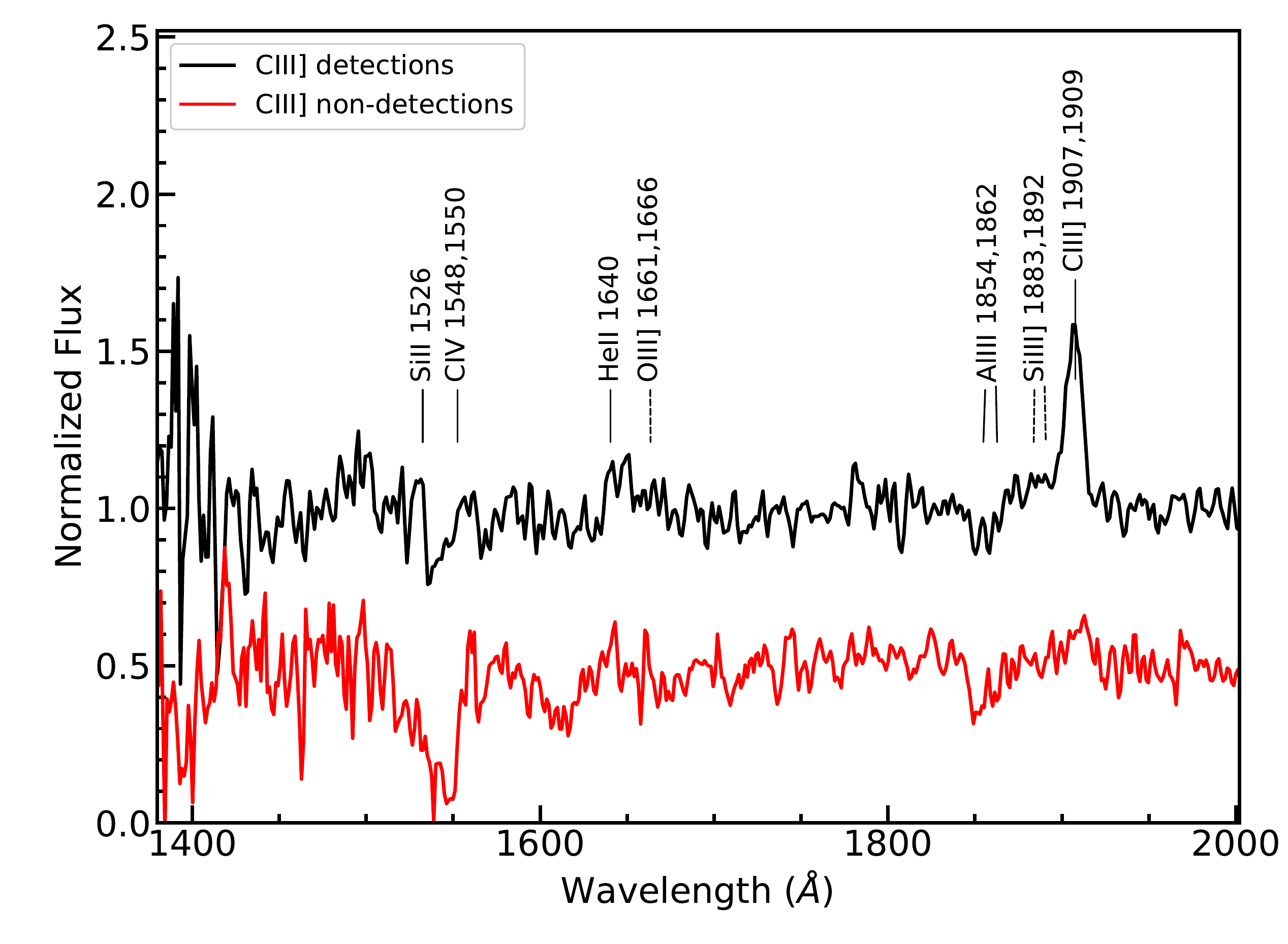}
\caption{: (left) The composite spectrum covering the rest-UV wavelengths 1400-2000\AA~ from an averaged stack of the sample of 10 Green Pea galaxies. 
In addition to the dominant CIII] emission, the weak HeII $\lambda$1640\AA~ emission feature is revealed along with the CIV $\lambda\lambda$1548,1550\AA~ and AlIII 
$\lambda\lambda$1854,1862\AA~ absorption features in the composite spectrum. The OIII]$\lambda\lambda$ 1661,1666\AA~ lines are not detected at $>$ 1-$\sigma$ 
significance  in the stacked spectrum. (Right) The normalized composite UV spectrum obtained by stacking the spectra of the Green Pea galaxies which show CIII]
emission (black), and those that do not (red). The composite spectrum of the non-detections is offset by -0.5 for visualization. The CIV wind features and interstellar 
absorption features are stronger when the CIII] emission is weak. }
\end{figure}

\subsection{The Composite Rest-UV Spectrum} \label{subsec:composite}

The stacked composite rest-UV spectrum of the GPs was created after applying a doppler correction to bring each individual spectrum to the rest-frame wavelengths
using the redshifts inferred from the [OIII]$\lambda$5007\AA~ emission line in the SDSS spectrum. Each GP spectrum was scaled to match the flux density between
2000-2100\AA~, which is redward of the CIII] doublet and is a continuum region that has high S/N and is free of absorption lines. The stacking was done using the 
{\it scombine} task in the {\it PyRAF/Specred} package, by averaging at each dispersion point after applying a 3-sigma clipping factor. The composite spectrum created
from all 10 GPs is shown in Figure 3. The composite spectrum reveals weak spectral features, particularly the broad emission feature at the location of HeII 
$\lambda$1640\AA~ line. Only 3 GPs in the sample show weak He II emission feature in their individual spectra, with EW (HeII) = 0.77$\pm$0.05\AA~ for J0815+2156, 
2.32$\pm$0.02\AA~ for J1244+0215, and 1.02$\pm$0.03\AA~ for J1457+2232. The EW(He II) as measured in the composite spectrum is $\sim$ 20\% of the CIII] emission flux. 
As noted in JR16, the photoionization models for GPs predict much weaker nebular HeII $\lambda$1640\AA~, $<$ 10\% of CIII]. However, it is known that the models tend to 
underpredict the HeII emission strengths, as evidenced also by the observed higher optical HeII $\lambda$4686\AA/H$\beta$ line ratios for GPs. Additional sources, such as 
Wolf-Rayet stars, shocks, or high-mass X-ray binaries have  been invoked to explain the higher than predicted HeII strengths (Shirazi \& Brinchmann 2012; Jaskot \& Oey 
2013). The composite spectrum also reveals stellar photospheric and interstellar absorption features, including SiII$\lambda$1526\AA~, CIV$\lambda$1548,1550\AA~, 
and AlIII$\lambda$1854,1862\AA~. The CIV absorption does not show the characteristic P-Cygni profile shape in the composite spectrum, because the stacking is 
performed using a small sample of 10 galaxies, of which only three have the strong CIV absorption feature and with widely different profile shapes. 

In Figure 3, the normalized composite spectrum is shown separately for the GPs with CIII] detections and non-detections. The interstellar absorption features are more 
prominent in the composite spectra of galaxies with CIII] non-detections compared to the stack of the entire sample. The CIV $\lambda\lambda$ 1548,1550\AA~ absorption 
profiles are stronger in the weak CIII]-emitters and those with non-detections. The very broad wing of the CIV feature in the composite spectrum is mainly contributed by the 
absorption profile of GP1137+3524 (Figure 1). The SiIV$\lambda$1393,1402\AA~ interstellar absorption lines are also more easily identified in the composite for galaxies with 
CIII] non-detections. In the case of the CIII]-emitters, there may be many factors that make the absorption lines less prominent, such as interstellar emission filling 
in the absorption features, the stronger nebular continuum from the ionized gas, or a difference in the age of the stellar population. Notably, the HeII$\lambda$ 1640\AA~ 
emission feature is prominent in the stack created using only the CIII]-emitters, and is not present in the composite from the CIII] non-detections. The OIII]$\lambda$1663\AA~ 
doublet remains undetected even in the stack created using only the CIII] detections. 

\section{Semi-Forbidden CIII] Emission in Green Pea Galaxies} \label{sec:style}

The CIII] nebular emission line is clearly the most prominent spectral feature in the HST/STIS rest-UV 1400-2200\AA~ spectra of the GPs. 
We detect CIII] emission in 7/10 GPs in our sample, and the measured EWs are provided in Table 3 along with upper limits for the three non-detections.
The rest-frame CIII] EWs for the GPs span values from as low as 1.67\AA~ to as high as 9.35\AA~, comparable to the strengths of CIII] emission seen in CIII] 
emitters at high redshifts ($z\sim 2-6$). When detected, the EW(CIII]) is $\geq$ 3\AA~ in most cases and is consistent with the predictions for 
CIII] emission lines in GPs from the photoionization models of JR16. Based on the GPs' metallicities, derived using the direct abundance method, and their observed 
[OIII] $\lambda$5007\AA~, [OII] $\lambda$3727\AA~, and [Ne III] $\lambda$3869\AA~  line strengths, these models predict C III] EWs in the range of 2-10\AA~, which 
is in close agreement with what we observe in the HST STIS spectrum. None of the GPs in our sample have the very high ($>$15\AA~) EWs seen in high redshifts 
(z$\sim$ 3-6) galaxies (Stark et al. 2017; Ding et al. 2017; de Barros et al. 2016). However, GPs with lower metallicities than this sample, such as, J1154+2443 from 
Schaerer et al. (2018), do show high EW(CIII]).

\subsection{Frequency of CIII]-emitters}

GP galaxies are, by definition, a class of strong [OIII] $\lambda$5007\AA~ emitters with EW([OIII]) $>$ 300\AA~. The frequency of CIII]-emitters ($\sim$ 70\%) we 
find among the GPs suggests that extreme emission-line galaxies selected by the presence of strong high ionization lines (e.g; [OIII]$\lambda$5007) are very likely 
to also be strong CIII]-emitters. It is interesting, therefore, to see how this fraction compares to star-forming galaxies selected using different criteria. In the sample 
of HeII emitters in the local Universe, CIII] doublet emission is detected in 7/10 galaxies, with EW(CIII]) $\sim 3-14.8$\AA~ (Senchyna et al. 2017). The HeII emitters 
span a wider range in metallicities (7.81 $<$ 12+log(O/H) $<$ 8.48), particularly towards lower metallicities than our GP sample. Local blue compact dwarf galaxies 
with low-metallicities (7.2 $<$ 12+log(O/H) $<$ 8.2) have a higher fraction of strong CIII]-emitters (Berg et al. 2016, 2019) with EW(CIII]) reaching as high as $>$ 
15\AA~ at the lowest metallicities. 


\begin{deluxetable*}{ccCcccccc}[b!]
\tablecaption{Emission-line Fluxes of UV and Optical nebular lines \label{tab:mathmode}}
\tablecolumns{9}
\tablenum{2}
\tablewidth{0pt}
\tablehead{
\colhead{Name} & F$_{\lambda}$(CIII]1909) & F$_{\lambda}^{\dagger}$(OIII]1666) &
\colhead{I$_{\lambda}$(CIII]1909)}&
\colhead{I$_{\lambda}^{\dagger}$(OIII]1666)} & \colhead{I$_{\lambda}$([OIII]5007)} & \colhead{$\frac{OIII]\lambda1666}{[OIII]\lambda4363}$$^{e}$} & \colhead{I$_{\lambda}^{e}$(OIII]1666)}
}
\startdata
J030321$-$075923   &  0.849$\pm$0.09  & 0.714  & 1.918$\pm$0.21     &1.428   & 52.88   & 1.709$^{+0.086}_{-0.254}$  & 1.76$^{+0.15}_{-0.28}$\\
J081552$+$215623  & 1.617$\pm$0.01   & 0.460  & 3.210$\pm$0.01     & 0.853  & 42.45   & 1.456$^{+0.114}_{-0.075}$  & 1.08$^{+0.12}_{-0.10}$\\
J091113$+$183108  & 0.446$^{\dagger}$  & 0.605  &2.318$^{\dagger}$   & 2.991  &  27.84  & 0.928$^{+0.384}_{-0.256}$  & 0.30$^{+0.15}_{-0.11}$\\
J105330$+$523752  & 0.987$^{\dagger}$  & 2.505  & 2.262$^{\dagger}$  & 5.579  &  43.77  & 0.849$^{+0.179}_{-0.256}$  & 0.35$^{+0.09}_{-0.12}$\\
J113303$+$651341  & 0.368$\pm$0.08   & 0.556  & 0.865$\pm$0.18     & 1.278  & 16.32    & 1.185$^{+0.252}_{-0.146}$  & 0.33$^{+0.08}_{-0.06}$\\
J113722$+$352426  & 0.856$^{\dagger}$  & 1.479  & 1.666$^{\dagger}$  & 2.791  &  65.06  & 0.898$^{+0.115}_{-0.256}$  & 0.61$^{+0.10}_{-0.18}$\\
J121903$+$152608  & 1.992$\pm$0.01   & 1.141  & 3.069$\pm$0.02     & 1.708 &  61.37  & 1.768$^{+0.062}_{-0.254}$  & 2.21$^{+0.16}_{-0.34}$\\
J124423$+$021540  & 0.656$\pm$0.06   & 0.742  & 1.794$\pm$0.18      & 1.971  &  76.83  & 1.013$^{+0.064}_{-0.258}$  & 0.91$^{+0.09}_{-0.24}$\\
J124834$+$123402  & 0.998$\pm$0.09   & 0.727  & 2.439$\pm$0.25     & 1.748  &  33.13  & 1.055$^{+0.132}_{-0.258}$  & 0.44$^{+0.08}_{-0.12}$\\
J145735$+$223201  & 2.080$\pm$0.02   & 0.622  & 5.846$\pm$0.06     & 1.605  &  95.61  & 1.400$^{+0.077}_{-0.064}$  & 2.34$^{+0.19}_{-0.18}$\\
\enddata

\tablecomments{The observed emission line fluxes (columns 2 and 3) and fluxes corrected for internal reddening and Milky Way extinction (columns 4, 5, and 6) in units of 10$^{-15}$ erg/s/cm$^{2}$. 
The 3-$\sigma$ upper limits are indicated by the $^{\dagger}$ symbol. The OIII]$\lambda$1666 fluxes are all upper limits ($^{\dagger}$) from $HST$/STIS spectra, and the expected fluxes ($^{e}$) 
(column 8) are based on the OIII]$\lambda$1666/[OIII]$\lambda$4363 ratio (column 7) from CLOUDY models. The [OIII]$\lambda$4363 and [OIII]$\lambda$5007 fluxes are from the SDSS spectra. 
}
\end{deluxetable*}

At higher redshifts, CIII]-emitters have been identified in surveys of star-forming galaxies that employ a more broader range of selection criteria (e.g., UV luminosity 
or color selection). Using MUSE observations, Maseda et al. (2017) find 17 CIII]-emitters which represents only $\sim$ 3\% of their photometric sample of star-forming 
galaxies between $1.5\leq z \leq 4$ in the Hubble Deep Field South and Ultra Deep Field. The CIII] EWs of these galaxies are in the range of $2-10$\AA~, similar to the 
GPs in most cases. However, a larger fraction (5 out of 17) among these high redshift star-forming galaxies are strong CIII]-emitters with EW(CIII]) $>$ 10\AA~, and 3 of 
them have EW(CIII]) $>$ 11.7\AA~ which is the highest value observed for the GPs (Schaerer et al. 2018). Among the UV-selected star-forming population at $2<z<3.8$ 
in the VUDS survey (Le F\`{e}vre et al. 2019), 24\% of the SFGs have EW(CIII])$>$ 3\AA~ and of these 4\% of have $>$ 10\AA~. While the number of known high redshift 
CIII]-emitters at $z>6$ is small, in almost all cases they have high EWs for CIII], and their broad-band colors suggest the presence of strong optical emission lines with 
EW([OIII]) $>$ 500\AA~ (Stark et al. 2014; 2017). In summary, the frequency of CIII]-emitters among GPs and other low-metallicity galaxies with strong optical 
emission-lines is higher, compared to populations of star-forming galaxies selected based on UV luminosity or colors which likely span a wider range of metallicities.

\subsection{CIII] and Lyman-$\alpha$ Emission}

Previous studies have highlighted the empirical relation between the CIII] and Ly$\alpha$ emission line equivalent widths (Shapley et al. 2003; Stark et al. 2014, 2015a; 
Rigby et al. 2015; Nakajima et al. 2018a). Both CIII] and Ly$\alpha$ emission lines are produced in the ionized gas and are powered by the ionizing radiation from the 
young, massive stars. Ly$\alpha$ is a resonant line and is sensitive to neutral gas, while CIII] emission escapes freely. The relation between CIII] and Ly$\alpha$ EWs 
is important in the context of the potential use of CIII] emission as a redshift indicator, and a tracer of the ionizing populations during the reionization epoch.  At $z>6$, 
the Ly$\alpha$ photons are resonantly scattered by the neutral Hydrogen in the intergalactic medium (IGM), making the detection of Ly$\alpha$ emission difficult (Tilvi 
et al. 2014; Konno et al. 2014; Mason et al. 2018).  The CIII] emission from actively star-forming galaxies at these redshifts, however, is often strong and easily observable 
(Stark et al. 2015a, 2017; Ding et al. 2017, Hutchison et al. 2019) making it a useful diagnostic emission line at high redshifts. 
 
 \begin{figure}
\plottwo{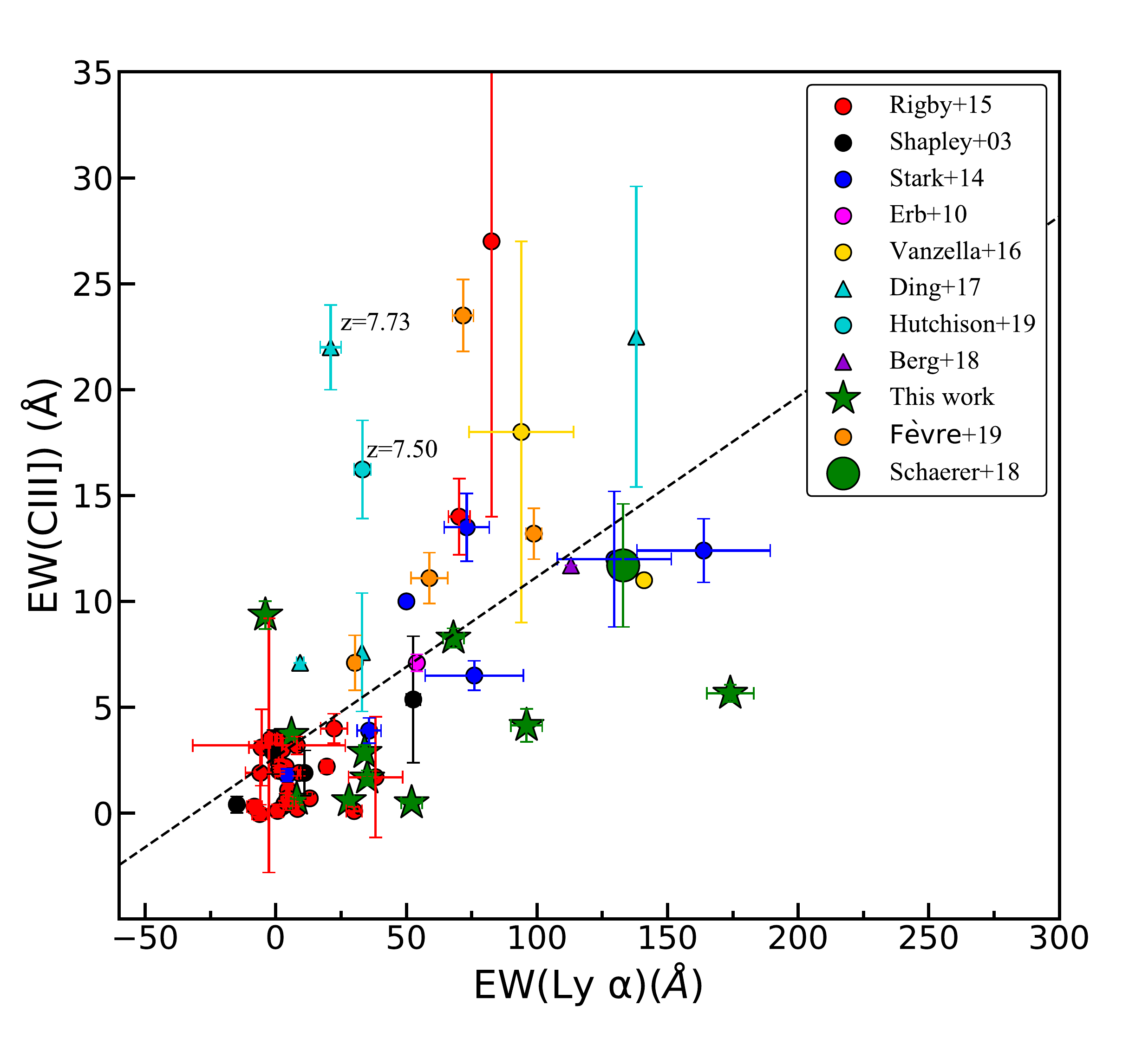}{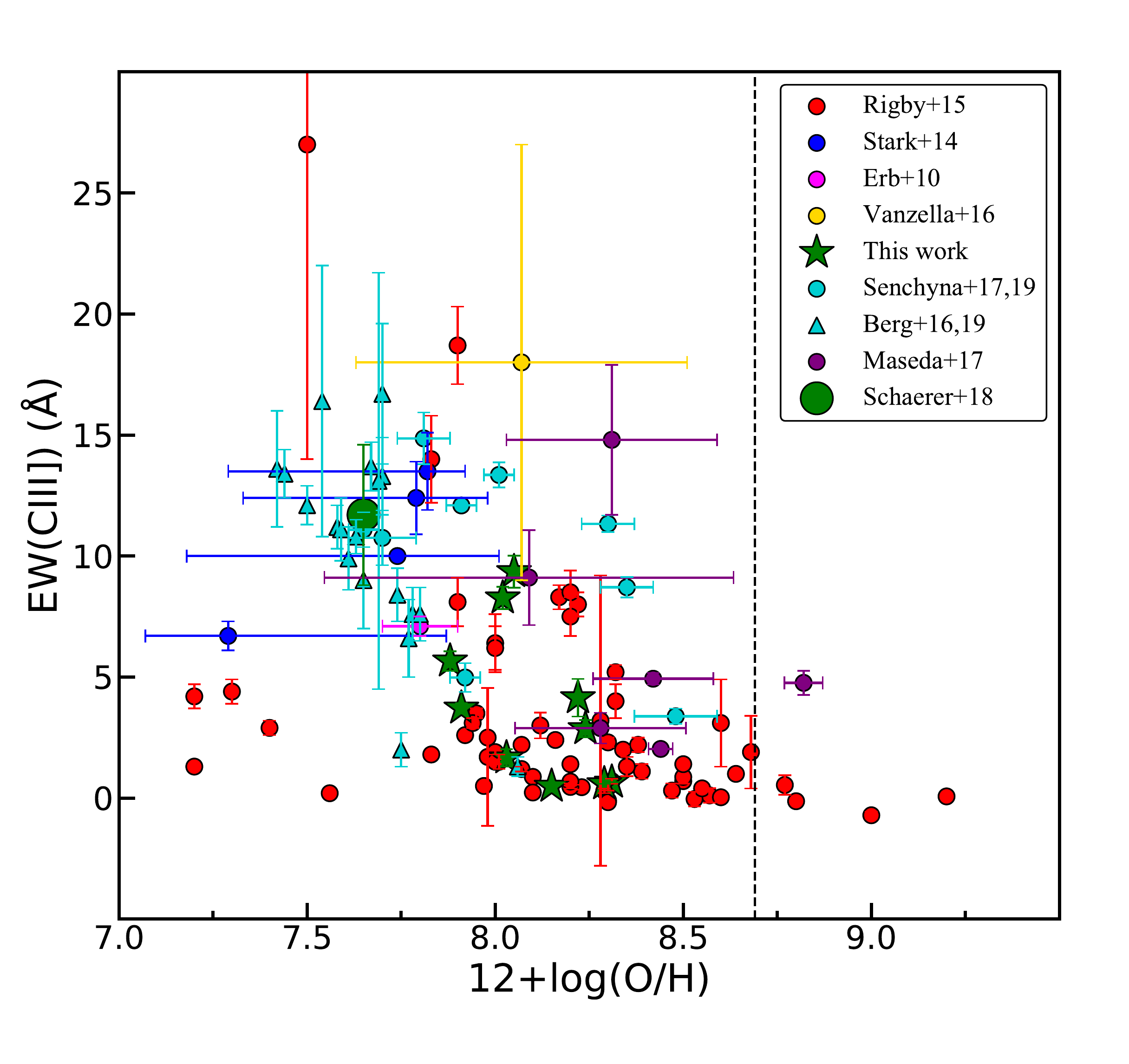}
\caption{: (left) The empirical relation between the rest-frame equivalent widths of Ly$\alpha$ and CIII], and (Right) rest-frame equivalent width of CIII] as a function 
of gas-phase metallicity observed for star-forming galaxies at low and high redshifts. The dashed line in the left panel shows the linear regression fit to the data with 
slope = 0.085$\pm$0.013 and the RMS scatter in EW(CIII) from the fit = 5.13. The vertical dashed line in the right panel indicates the solar metallicity, 12+log(O/H) = 8.69.
The observed data include GPs from this work, along with the LyC-emitting GP galaxy J1154+2443 (Schaerer+18), low redshift He II emitters and extremely metal-poor
galaxies (Senchyna+17, 19), blue compact dwarf galaxies (Berg+16, 19) and compilation of star-forming galaxies (Rigby+15), and high redshift ($1.5<z<7.8$) galaxies 
from the literature (Shapley+03, Erb+10, Stark+14, Vanzella+16, Maseda+17, Ding+17, Hutchison+19, Berg+18, and F\`{e}vre+19)}
\end{figure}

In Figure 4 ({\it left}), we present the observed relation between EW(CIII]) and EW(Ly$\alpha$) for the GPs along with other samples at low and high redshift from 
 the literature. The GPs mostly lie along with the high redshift $z=2-6$ galaxies, with EW(CIII]) that spans the range $\sim 2-10$\AA~ and with EW(Ly$\alpha$) 
$\sim 0-175$\AA~.  The Pearson correlation coefficient suggests a strong linear relationship between EW(CIII]) and EW(Ly$\alpha$) for star-forming galaxies in
Figure 4, with $r_{P} = 0.61$, and a low probability ($p = 7.80 \times 10^{-8}$) that the two quantities are uncorrelated. However, there is no strong correlation that 
is evident within the GPs sample alone. By definition, the GPs selection isolates strong emission line galaxies, and the scatter in Ly$\alpha$ EW among the GPs is 
likely driven by the variations in Ly$\alpha$ optical depth due to the neutral gas. Star-forming galaxies that are non-GPs are not necessarily extreme emission line 
objects, and their weak Ly$\alpha$ emission may indicate that the flux in the emission lines are intrinsically low. One of the two most deviant points in Figure 4 is 
J1457+2232 which has the highest EW(CIII]) among the GPs sample, but the measured EW(Ly$\alpha$) is weak or absent. Jaskot \& Oey (2014) propose that the 
dominant broad absorption feature of the Ly$\alpha$ profile in this galaxy implies the presence of a high column density of neutral gas along the line of sight, and the 
weak Ly$\alpha$ emission probably escapes via scattering, perhaps in a bipolar outflow. J1219+1526, on the other hand, has the highest EW(Ly$\alpha$) = 174$\pm$9\AA~ 
among the GPs sample , and rest-frame EW(CIII]) = 5.66$\pm$0.48\AA~, which is lower than expected from the empirical EW(CIII])-EW(Ly${\alpha}$) relation. There are 
various factors that can impact the CIII] line strengths and are discussed in detail in section 3.5.

\begin{deluxetable*}{ccCcccc}[b!]
\tablecaption{Rest-frame Equivalent Widths of the UV and Optical Emission lines \label{tab:mathmode}}
\tablecolumns{7}
\tablenum{3}
\tablewidth{0pt}
\tablehead{
\colhead{Name} &
\colhead{Ly${\alpha}$$\lambda$1216} &
\colhead{CIII]$\lambda$1909} & \colhead{[OIII]$\lambda$4363} &  \colhead{[OIII]$\lambda$5007} & \colhead{H${\alpha}$$\lambda$6563} \\
\colhead{} & \colhead{(\AA)} &
\colhead{(\AA)} & \colhead{(\AA)} & \colhead{(\AA)} & \colhead{(\AA)}
}
\startdata
J030321$-$075923 & 6$\pm$1&3.72$\pm$0.24 & 11.4$\pm$0.5 & 826$\pm$10& 697$\pm$12\\
J081552$+$215623& 68$\pm$4 & 8.27$\pm$0.53 & 18.5$\pm$1.5 &1365$\pm$29 & 898$\pm$16\\
J091113$+$183108 & 52$\pm$4 & 0.49$^{\dagger}$ &  2.3$\pm$0.5 &274$\pm$3 & 422$\pm$7\\
J105330$+$523752& 8$\pm$1 & 0.66$^{\dagger}$ &  2.3$\pm$0.3 &350$\pm$3 & 401$\pm$4\\
J113303$+$651341 & 35$\pm$2 & 1.66$\pm$0.44 & 5.0$\pm$0.6 &394$\pm$5 & 305$\pm$6\\
J113722$+$352426 & 28$\pm$2 & 0.59$^{\dagger}$ &  4.3$\pm$0.3 &582$\pm$4 & 575$\pm$6\\
J121903$+$152608 & 174$\pm$9 & 5.66$\pm$0.48 &  21.5$\pm$0.7 &1488$\pm$16 & 1266$\pm$21\\
J124423$+$021540 & 34$\pm$2 & 2.87$\pm$0.44 &  8.0$\pm$0.4 & 985$\pm$10 & 841$\pm$11\\
J124834$+$123402 & 96$\pm$6 & 4.14$\pm$0.98 &  7.4$\pm$0.7 & 842$\pm$12 & 743$\pm$19\\
J145735$+$223201 & -4$\pm$1 & 9.35$\pm$0.76 &  19.3$\pm$0.7 &1433$\pm$14 & 1000$\pm$14\\
\enddata

\tablecomments{The Ly$\alpha$ equivalent widths are based on $HST$/COS measurements (Henry et al. 2015; Jaskot\& Oey 2014), the CIII]$\lambda$1909 equivalent 
widths are measured from $HST$/STIS, and the equivalent widths of the optical emission lines are measured from SDSS spectra. The 1-$\sigma$ upper limits are indicated 
by the  $^{\dagger}$ symbol.
}
\end{deluxetable*}

Among the non-GP glaxies, the noticeably deviant points from the empirical correlation between EW(CIII]) and EW(Ly$\alpha$) are the highest redshift galaxies, 
EGS-zs8-1 at $z=7.73$ with high EW(CIII]) $=$ 22$\pm$2\AA~ and weak EW(Ly$\alpha$) = 21$\pm$4\AA~ (Stark et al. 2017), and z7-GND-42912 at $z=7.50$
with EW(CIII]) $=$ 16.23$\pm$2.32\AA~ and EW(Ly$\alpha$) = 33.2$\pm$3.2\AA~ (Hutchison et al. 2019). Both galaxies exhibits very red [3.6]-[4.5] color in the 
Spitzer infrared bands which results from the large rest-frame equivalent width of the [OIII]+H$\beta$ in the 4.5$\mu$m band, a selection criterion that is comparable 
to the GP galaxies selection at low redshift. For EGS-zs8-1, Stark et al. (2017) find that the velocity offset of the Ly$\alpha$ emission line, $\Delta v_{Ly\alpha}$ = 
340 - 520 km/s, implies that the Ly$\alpha$ profile is modulated by the presence of dense neutral gas close to the systemic redshift of the galaxy. The high column 
density of neutral gas is responsible for the observed low EW(Ly$\alpha$). The case of the GP galaxy J1457+2232 is similar, as the Ly$\alpha$ profile shows a large 
velocity separation $\sim$ 750 km/s between the emission peaks, which suggests a high column density along the line of sight. EGS-zs8-1,  z7-GND-42912, and 
J1457+2232 have higher EW(CIII]) than expected for their EW(Ly$\alpha$) from the EW(CIII]) $-$ EW(Ly$\alpha$) empirical relation. Such galaxies illustrate the 
utility of the  CIII] emission line as an alternate diagnostic for the spectroscopic redshifts and ISM conditions, when Ly$\alpha$ is absorbed by the neutral gas in 
the ISM and IGM.

\newpage

\subsection{CIII] and Rest-Optical Emission lines}

The SDSS spectra of GPs provide a suite of rest-frame optical emission lines, which are useful to derive electron temperatures ($T_{e}$), and to estimate the 
gas-phase metallicities or nebular oxygen abundances (12+log(O/H)).  The nebular abundances used in our analysis are based on the direct method using 
$T_{e}$ ([O III]) calculated from the [O III] line fluxes from SDSS spectra.  In Figure 4 ({\it right}), we present EW(CIII]) versus metallicity for the GPs, along with 
various samples of CIII]-emitters at low and high redshifts from the literature (Rigby et al. 2015; Maseda et al. 2017; Senchyna et al. 2017, 2019; Berg et al. 2016, 
2019; Stark et al. 2014; Vanzella et al. 2016; Nakajima et al. 2018b). Previous studies have highlighted the significant trend of increasing EW(CIII]) with decreasing 
metallicity, reaching values $\geq$ 5\AA~ only at 12+log(O/H) $<$ 8.4 or 0.5 Z$_\odot$ (Rigby et al. 2015; Maseda et al. 2017; Nakajima et al. 2018a; Senchyna 
et al. 2017). Using photoionization models it has been shown that low metallicities create ISM conditions that are necessary to produce significant CIII] emission
(JR16; Gutkin et al. 2016; Feltre et al. 2016; Nakajima et al. 2018a; Byler et al. 2018). At low metallicities, stars have higher effective temperatures with weaker 
stellar winds, and their harder ionizing SED boosts the supply of C$^{+}$ ionizing photons. Also, since the ionized nebulae predominantly cool via the forbidden 
emission of metal lines, the low metallicity nebulae tend to have higher $T_{e}$ favoring high collisional excitation rates and stronger CIII] emission. At a fixed 
C/O ratio or carbon abundance, the photoionization models require a metallicity threshold of Z$\sim$ 0.006 for the CIII] to be strong with EWs in excess of 3\AA. 
The GP galaxies have metallicities $Z/Z_{\odot}$ $\leq$ 0.4 (or $Z\leq0.006$), and their EW(CIII]) shows a large spread at a given metallicity, indicating that the 
CIII] emission is also influenced by other factors, such as, ionization parameter, age of the stellar population, and the C/O ratio (JR16, Nakajima et al. 2018a).

\begin{figure}
\plottwo{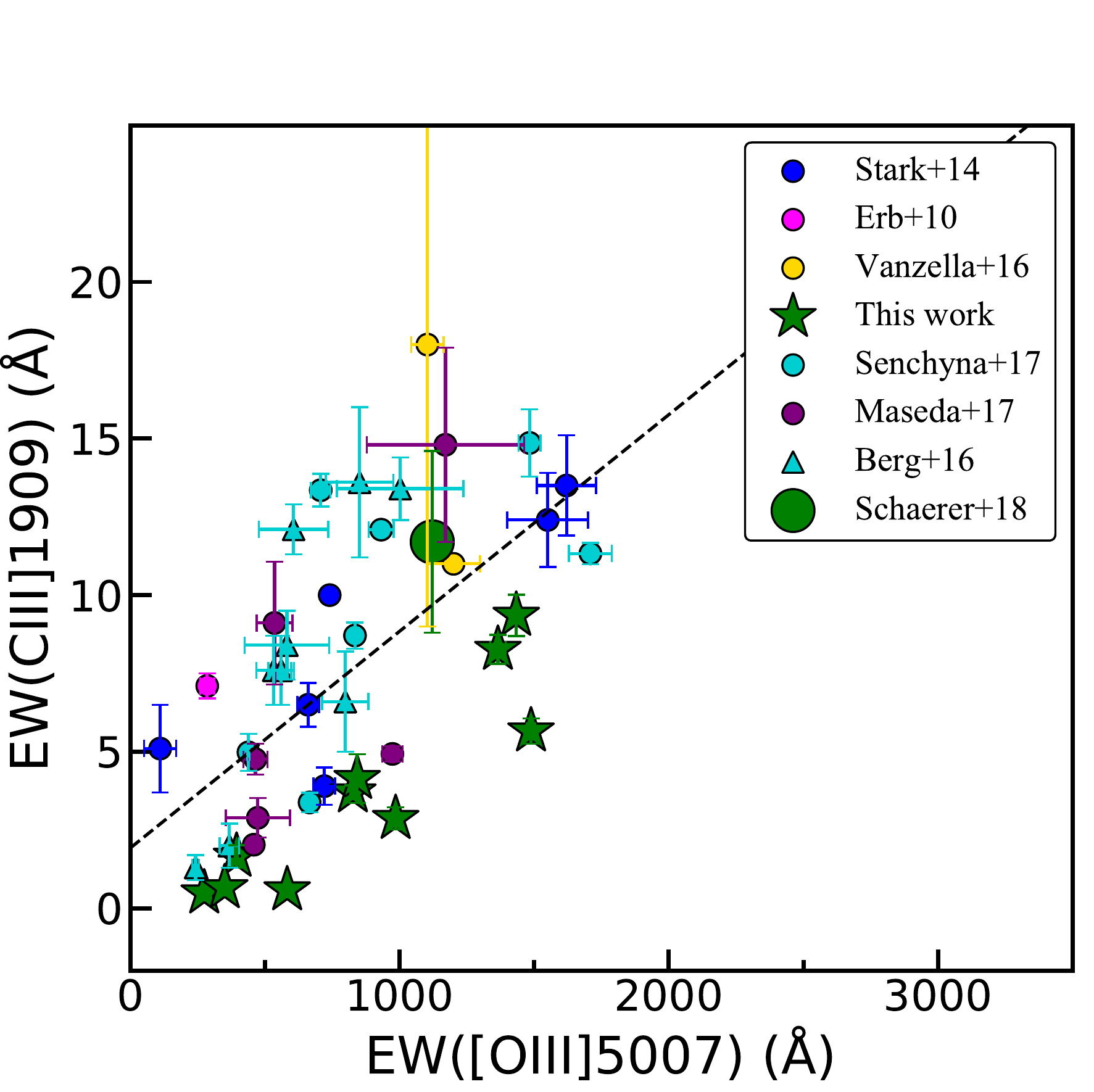}{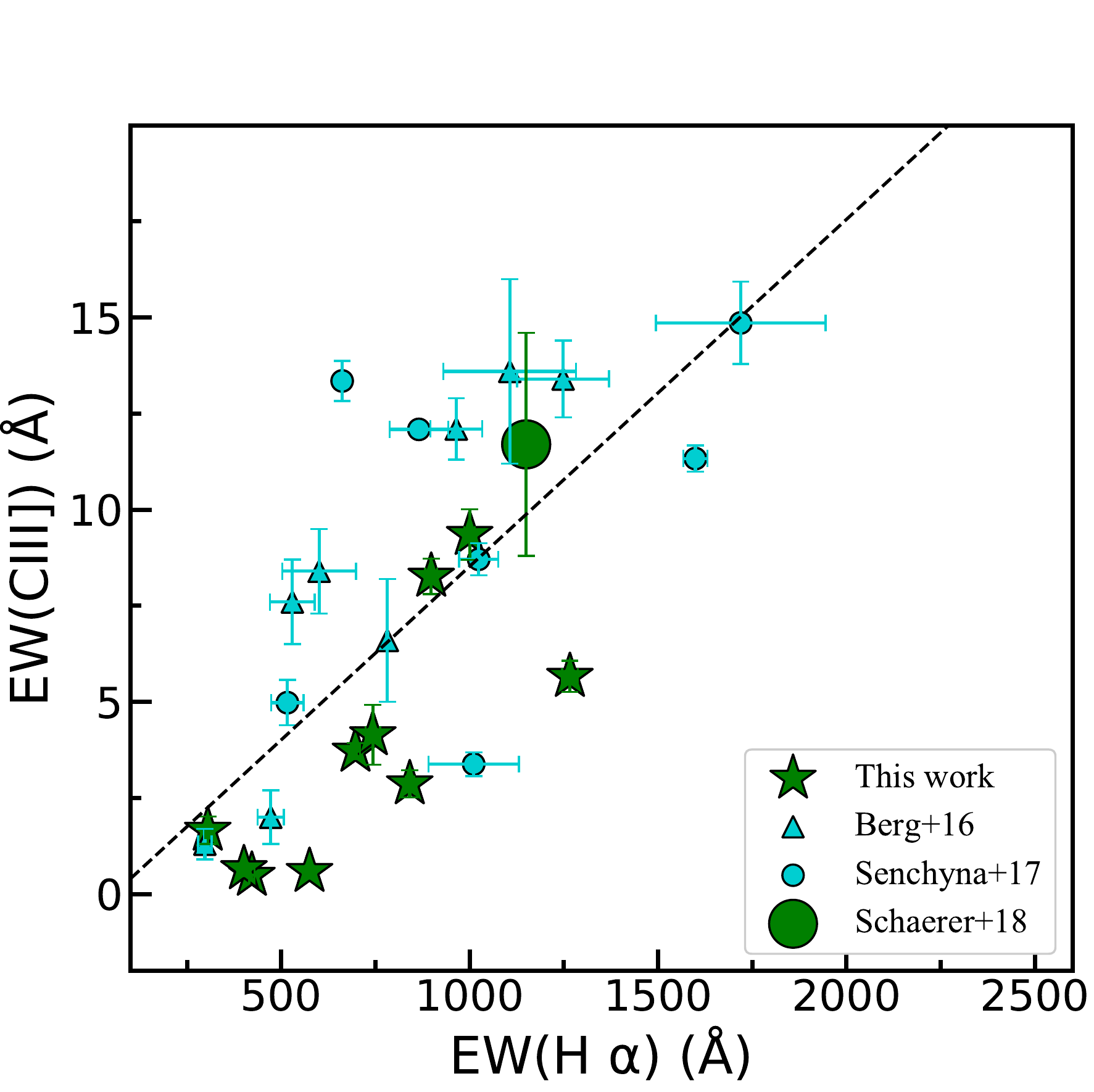}
\caption{(left) The relation between EW(CIII]) and EW([OIII]), and (right) the relation between EW(CIII]) and EW(H$\alpha$) for the GPs and other
CIII]-emitters from the literature.  The dashed line in the panels shows the linear regression fit to the data with slope = 0.007$\pm$0.001 and RMS scatter = 3.63
for EW(CIII]) vs EW([OIII]), and slope = 0.009$\pm$0.002 and RMS scatter = 2.56 for EW(CIII]) vs EW(H$\alpha$). The points shown are: high-z galaxies ($2<z<7$) 
from Stark et al, 2014, 2015 (blue); BX418 from Erb et al. 2010 (magenta); GP sample from this work (green); [OIII]-emitters at $z\sim$3 from Vanzella et al. 2016, 
de Barros et al. 2016 (gold); CIII]-emitters at $1.5<z<4$ from Maseda et al. 2017 (purple); HeII-emitters from Senchyna et al. 2017 (turquoise)and the dwarf galaxies 
from Berg et al. 2016 (turquoise triangle). }
\end{figure}

High nebular temperatures, high ionization parameters, and hard ionizing radiation from metal-poor, young stellar populations with ages $<$ 5 Myrs enhance 
the CIII] $\lambda$1909\AA~ emission. The same physical conditions that produce strong CIII] emission, also favor the emission of strong forbidden lines from 
collisionally-excited species in the optical wavelengths.  The relation between EW(CIII]) and equivalent width of [OIII]$\lambda$5007\AA~ (hereafter, EW([OIII])) 
provides insight into the frequency of CIII] emission among the [OIII]-emitters, and how the relative emission-line fluxes inform us about the ISM properties. In 
Figure 5, we show the empirical relation between EW(CIII]) and EW([OIII]) for the GPs along with other star-forming galaxies for which measurements are available
in the literature. The sample presented in Figure 5 is inhomogeneous, in the sense that some of the measurements correspond to individual compact star-forming 
regions, while the others are integrated measures over the entire galaxies. Also, the apertures used for measuring the rest-UV and rest-optical fluxes are not matched 
in all cases. However, it is clear that there is a strong trend between the CIII] and [OIII] EWs. The EW(CIII])-EW([OIII]) relation is almost linear, since  both CIII] and 
[OIII] are collisionally-excited lines which are favored by higher nebular temperatures and higher ionization parameters. The Spearman rank correlation statistic between 
EW(CIII]) and EW([OIII])  indicates a strong correlation with $r_{Sp} = 0.65$ ($p = 3.02 \times 10^{-6}$), and the Pearson correlation statistic suggests a strong linear 
correlation with $r_{P} = 0.62$ ($p = 1.41 \times 10^{-5}$). As seen from the photoionization models (JR16; Figure 8), the relation between the two EWs exhibits a 
large spread that reflects the range in ionization parameter, nebular temperature, and the dependence on metallicity. Extreme values of EW([OIII]) $>$ 2000\AA, and 
EW(CIII]) $>$ 10\AA~ require both high ionization parameters (log$U$ $>$ -2) and young stellar populations with ages less than 3 Myrs. The energy requirements for 
the excitation of these two emission lines are different, although similar physical conditions tend to favor them. Both emission lines require an ionizing spectrum that 
has high energy photons, since the ionization energy for C$^{+2}$ is 24.4 eV and for O$^{+2}$ is 35.1 eV. However, because of the temperature dependence of the 
collisional excitation rates, the CIII] emission exhibits a greater sensitivity to metallicity. This may also explain the offset of the GPs in our sample from the other 
low-redshift samples. The BCDs (Berg et al. 2016) and He II emitters (Senchyna et al. 2017) mostly have lower metallicities with 12+log(O/H) $<$ 8.0 for the strong 
CIII]-emitters. The location of the low-metallicity GP galaxy J1154+2443 (Schaerer et al. 2018) which lies along with BCDs and He II emitters in Figure 5, also seem 
to suggest that the offset for the GPs in this work may be due to their relatively higher metallicities.

The EW(CIII]) also shows a trend with EW(H$\alpha$) as expected, because the nebular emission lines both depend on the amount of ionizing radiation. The Spearman 
rank correlation statistic between EW(CIII]) and EW(H$\alpha$)  indicates a strong correlation with $r_{Sp} = 0.72$ ($p = 2.85 \times 10^{-5}$), and the Pearson correlation 
statistic suggests a strong linear correlation with $r_{P} = 0.71$ ($p = 4.75 \times 10^{-5}$). The EW(CIII]) versus EW(H$\alpha$) relation is tighter than the EW(CIII]) versus 
EW(Ly$\alpha$) relation (Figure 4) which is  significantly modified by radiative transfer effects as the Ly$\alpha$ is resonantly scattered. Even the GPs which show large 
deviations in the EW(CIII]) vs EW(Ly$\alpha$) relation, follow a tight correlation with EW(H$\alpha$). Since H$\alpha$ is a well-calibrated SFR indicator (Kennicutt 1998), 
it is encouraging to consider the possibility of using the CIII] emission as a diagnostic for the ionizing flux and SFR at high-$z$. CIII] emission depends on the ionization 
parameter and ionizing flux as does the H$\alpha$ emission, but the dependence of CIII] on metallicity and density makes it a weaker diagnostic for SFR. However, when 
H$\alpha$ emission is redshifted beyond the NIR wavelengths at $z>7$, the CIII] emission may serve as a useful diagnostic for estimating the production rate of ionizing 
photons (Chevallard et al. 2018; Schaerer et al. 2018).

\subsection{Estimate of the C/O ratios}

The elemental carbon abundance is one of the parameters that influences the CIII] emission. Previous studies have used the CIII]$\lambda$1909\AA~ and OIII]$\lambda$1663\AA~ 
emission line doublets to derive the C/O ratios in star-forming galaxies (Garnett et al. 1995; Erb et al. 2010; Berg et al. 2016, 2019). Since we do not detect the OIII]$\lambda$1663\AA~ 
doublet, we used the dust-corrected CIII] and optical [OIII] doublets to derive the C/O ratios. Following the equation in Izotov \& Thuan (1999):
\begin{equation}
\frac{C^{2+}}{O^{2+}} = 0.093~ \text{exp} \left(\frac{4.656}{t}\right) \frac{I(C{III]}\lambda1906+\lambda1909)}{I({[OIII]}]\lambda4959+\lambda5007)}
\end{equation}
where $t=T_{e}/10^{4}$ and 
\begin{equation}
\frac{C}{O} = \text{ICF}\left(\frac{C}{O}\right) \frac{C^{2+}}{O^{2+}}
\end{equation}

The correction factor ICF(C/O) was derived from the {\it CLOUDY} photoionization models for log$U$ = -2 and -3. The measured C/O ratios are presented in 
Table 4, and the range of observed values are consistent with the C/O ratio = 0.20 assumed in the JR16 models. We also used the CIII] fluxes along with the 
OIII]$\lambda$1663\AA~ upper limits from the $HST$/STIS spectra to compute the lower limits on the observed C/O using the equations from Erb et al. (2010). 
Taking advantage of the tight correlation between OIII]$\lambda$1663\AA~ and [OIII]$\lambda$4363\AA~ emission (JR16), we also used the observed 
[OIII]$\lambda$4363\AA~ fluxes to get the predicted OIII]$\lambda$1663\AA~ fluxes from the photoionization models. The C/O ratio derived based on the 
predicted OIII] fluxes and observed CIII] are also presented in Table 4. The C/O ratio measured using the observed [OIII] and predicted OIII] fluxes are 
consistent, and range from log(C/O) $\sim$ -0.6 to -1.1, similar to the range of C/O ratios found in high redshift galaxies (Amor{\'i}n et al. 2017; Stark et al. 2014; 
Erb et al. 2010) and in local BCDs (Berg et al. 2016).

\subsection{ISM Conditions and CIII] Detectability}

\begin{figure}
\plottwo{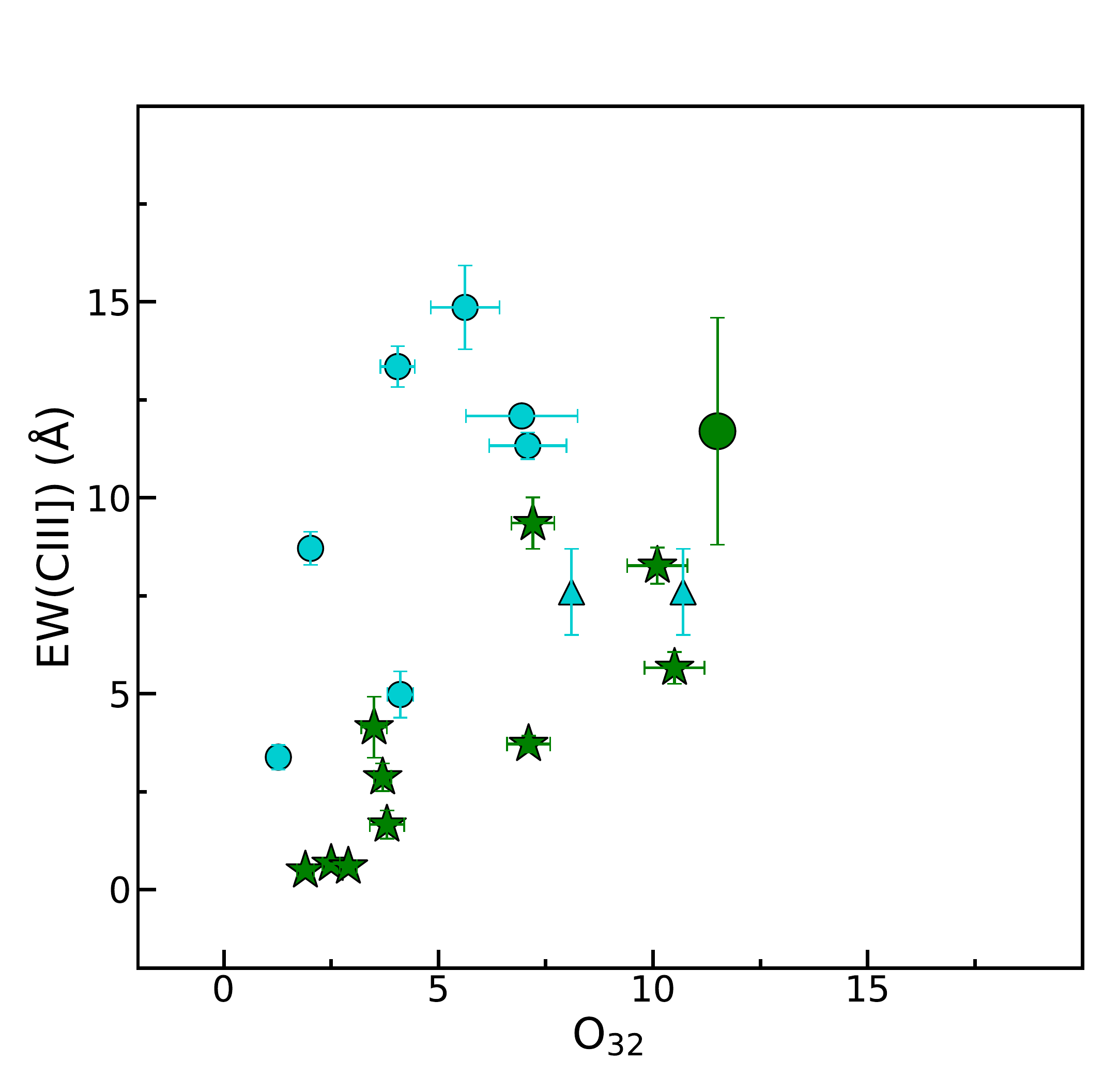}{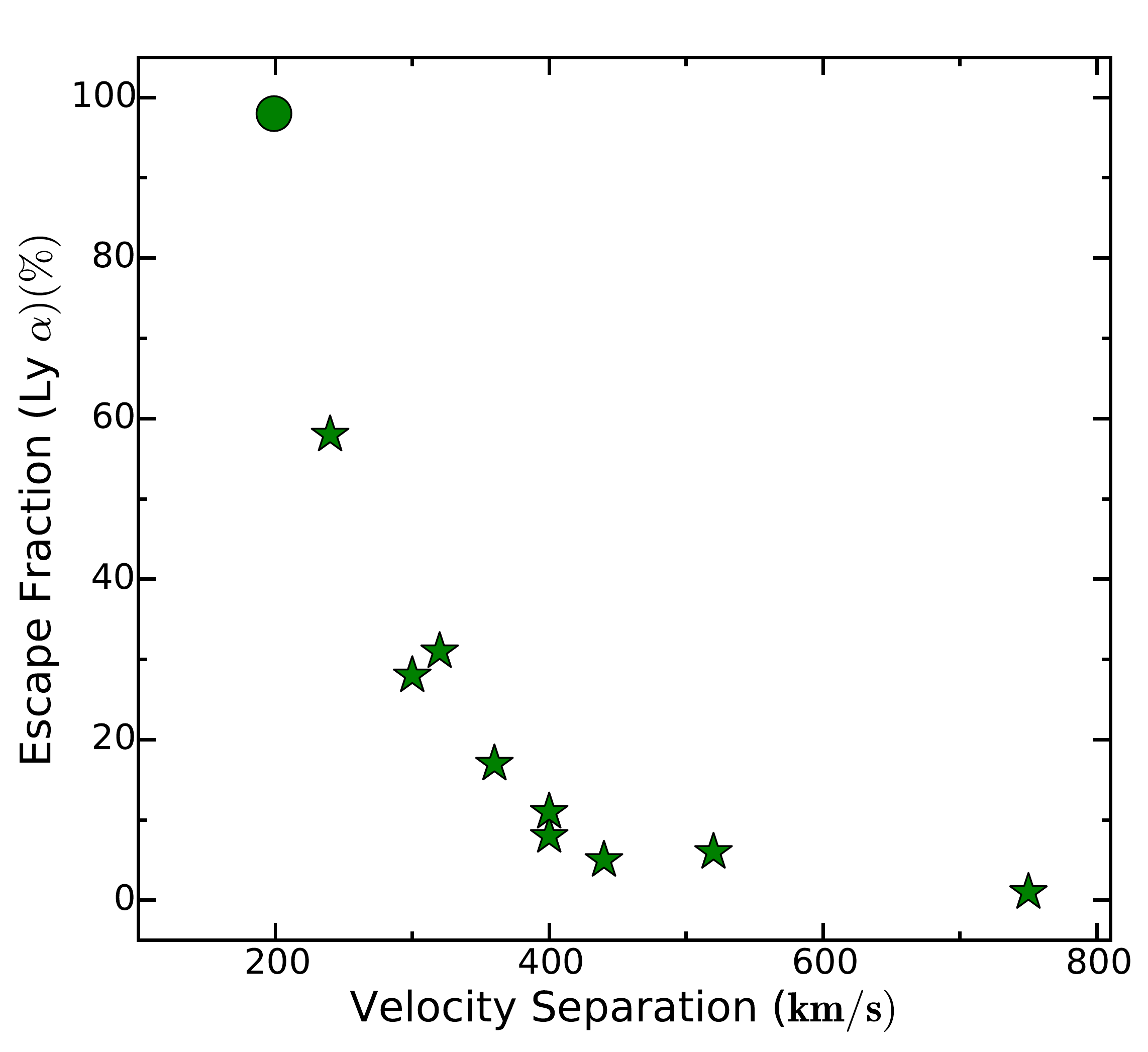}
\plottwo{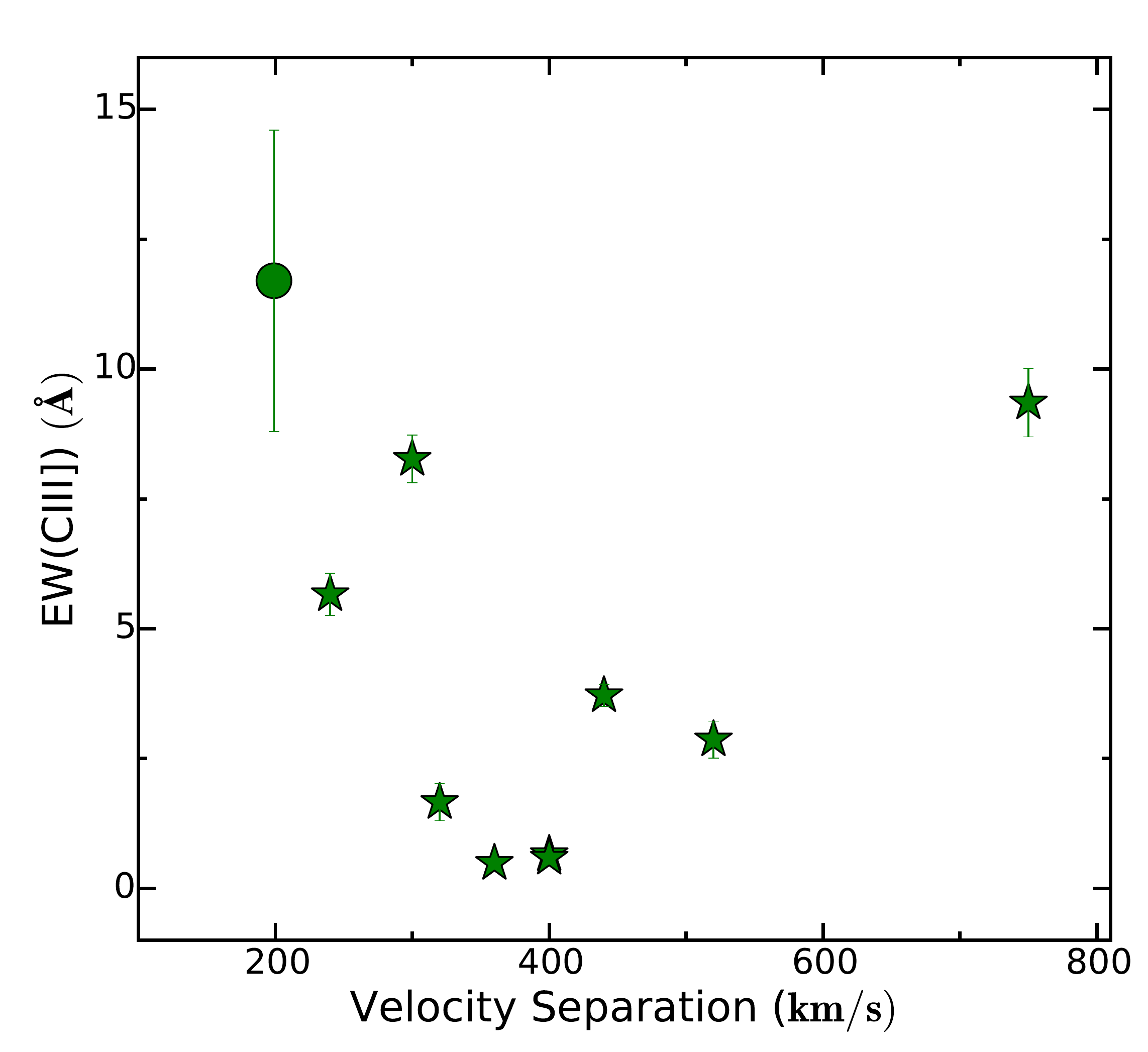}{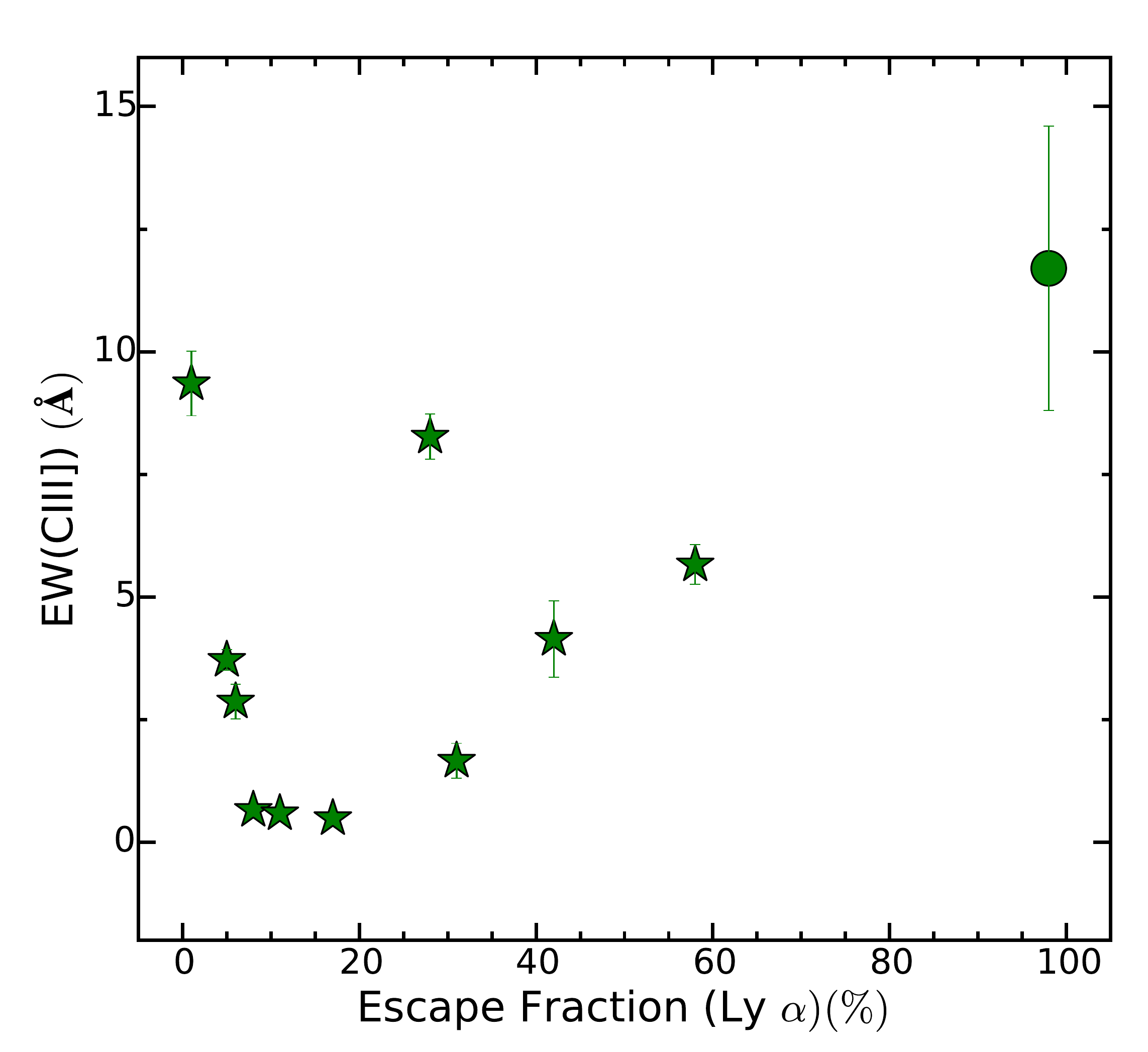}
\caption{The relation between EW(CIII]) and the O$_{32}$ = [OIII]$\lambda$5007/[OII]$\lambda$3727 ratio (upper left), and EW(CIII]) versus the velocity separation 
between the red and blue peaks of the Ly${\alpha}$ emission profile (lower left). The EW(CIII]) versus O$_{32}$ ratio includes BCDs (Berg et al. 2016; turquoise triangles), 
and HeII emitters (Senchyna et al. 2017; turquoise circles) at $z\sim 0$. The Ly${\alpha}$ velocity profile measurements are from Jaskot \& Oey (2014) and Henry 
et al. (2015). The escape fraction of  Ly${\alpha}$ emission as a function of the velocity separation (upper right), and EW(CIII]) (lower right). Larger 
velocity separation implies a larger optical depth and lower Ly${\alpha}$ escape fractions. The GPs from this work are shown as green stars, and J1154+2443 
(Schaerer et al. 2018) is shown as green circle.}
\end{figure}

In this section, we examine how the ionization parameter, and optical depth of the ISM affect the detectability of CIII] emission. In Figure 6, we show the relation 
between EW(CIII]) and the O$_{32}$ ratio. The [OIII] and [OII] lines originate from different ionization levels of oxygen, and hence the O$_{32}$ 
ratio serves as a proxy for the ionization parameter. A high value for O$_{32}$ indicates a high ionization parameter, and for star-forming galaxies this could imply that 
the nebular gas is powered by the ionizing radiation from massive stars of very young ages. Recent studies have suggested that O$_{32}$ is a useful diagnostic of 
Lyman continuum leakage (Izotov et al. 2016a, 2018a; Jaskot \& Oey 2013; Nakajima \& Ouchi 2014). Star-forming galaxies at low redshifts ($z<0.3$) selected from 
the SDSS based on their high O$_{32}$ $\geq$ 5 are mostly found to be Ly$\alpha$ emitters, with few exceptions (Jaskot et al. 2017; McKinney et al. 2019). In a 
sample of 11 GP galaxies with O$_{32}$ $\geq$ 5 targeted with $HST$/COS FUV spectroscopy to look for escaping ionizing radiation, all of them showed evidence for 
LyC leakage with escape fractions (f$_{esc}$) in the range 2-72\% (Izotov et al. 2016a,b; 2018a,b). In general, the f$_{esc}$(LyC)  is found to be higher for GPs with 
higher O$_{32}$, although the relation between the two quantities shows considerable scatter (Izotov et al. 2016a, 2018a). High O$_{32}$ has been proposed to be a 
diagnostic for density-bounded HII regions, which are potential LyC leaker candidates (Guseva et al. 2004; Jaskot \& Oey 2013; Nakajima \& Ouchi 2014; Izotov et al. 
2018a). It is, therefore, interesting that EW(CIII]) also shows a positive trend with O$_{32}$, and hence LyC leakers should be strong CIII]-emitters, because the high 
ionization parameter would favor CIII] emission. However,  the observed strength of the CIII]-emission can be reduced due to other factors, such as the decrease in 
absorbed ionizing flux at low optical depth, the transition of C$^{+2}$ to C$^{+3}$ for log$U$ $>$ -2, and a low elemental carbon abundance. The EW(CIII]) versus 
O$_{32}$ diagram shows a linear relation for the GPs when CIII] emission is detected, with CIII] strength increasing as the O$_{32}$ ratio increases, and EW(CIII]) 
$\geq$ 3\AA~ for O$_{32}$ $\geq$ 5. For similar O$_{32}$ values, even the He II emitters that overlap with GPs in their metallicities have significantly higher 
EW(CIII]), which may be due to contributions from additional ionizing sources (such as, shocks or X-ray binaries). 

\begin{deluxetable*}{cccCCcccc}[b!]
\tablecaption{Elemental Abundances and ISM properties of Green Pea Galaxies \label{tab:mathmode}}
\tablecolumns{9}
\tablenum{4}
\tablewidth{0pt}
\tablehead{
\colhead{Name} &
\colhead{O$_{32}$} & \colhead{12+log(O/H)} & \colhead{log(C/O)} & \colhead{log(C/O)$^\dagger$} &  \colhead{log(C/O)$^{e}$} &
\colhead{f$_{esc}$(Ly$\alpha$)} & \colhead{V$_{sep}$(Ly$\alpha$)}\\
\colhead{} & \colhead{} &  \colhead{} &  \colhead{} &  \colhead{} & \colhead{} &
\colhead{} & \colhead{(km/s)}
}
\startdata
J030321$-$075923  &    7.1$\pm$0.5 &  7.91 & -1.247 (-1.002) &   -1.014 (-0.769)  &  -1.105 (-0.859)  &  0.05  & 460 \\
J081552$+$215623 &  10.1$\pm$0.7 &  8.02 & -0.860 (-0.618) &-0.580 (-0.338)  &  -0.683 (-0.441)  &  0.28  & 260 \\
J091113$+$183108  &   1.9$\pm$0.2  &  8.15 &  $........$  & $........$            &      $........$          &  0.17  & 370 \\
J105330$+$523752 &    2.5$\pm$0.2  &  8.31 &   $........$ & $........$            &     $........$           &  0.08   & 410 \\
J113303$+$651341 &    3.8$\pm$0.4  &  8.03 &  -0.994 (-0.751) &-1.329 (-1.087)  &  -0.742 (-0.499)  &  0.31  & 330 \\
J113722$+$352426 &    2.9$\pm$0.2  &  8.29 &  $........$   & $........$            &     $........$           &  0.11  & 550 \\
J121903$+$152608 &  10.5$\pm$0.7  &  7.88 &  -1.134 (-0.888) & -0.882 (-0.636)  &  -0.994 (-0.748)  &  0.58  & 270\\
J124423$+$021540 &    3.7$\pm$0.2  &  8.24 &  -1.109 (-0.869) &-1.237 (-0.997)  &  -0.901 (-0.661)  &  0.06  & 530 \\
J124834$+$123402 &    3.5$\pm$0.3  &  8.22 &  -0.665 (-0.423) &-1.041 (-0.799)  &  -0.442 (-0.199)  &  0.42  & $...$ \\
J145735$+$223201 &    7.2$\pm$0.5  &  8.05 &  -0.950 (-0.707) &-0.592 (-0.349)  &  -0.756 (-0.513)  &  0.01  & 750\\
\enddata

\tablecomments{The O$_{32}$ and nebular oxygen abundances are based on the SDSS optical spectra. The log(C/O) is computed using the CIII]$\lambda$1909 
line fluxes from $HST$/STIS and [OIII]$\lambda$5007 line fluxes from SDSS spectra. The log(C/O)$^{\dagger}$ values are computed from the observed 
CIII]$\lambda$1909 line fluxes and upper limits for OIII]$\lambda$1663 line fluxes, using ICF from CLOUDY models for ionization parameter log$U$ = -2. The C/O 
abundances derived for the ICF corresponding to  log$U$ = -3 are included in parentheses. The expected C/O abundances, log(C/O)$^{e}$, are computed using 
the observed CIII]$\lambda$1909 fluxes and the OIII]$\lambda$1663 fluxes predicted based on the observed [OIII]$\lambda$4363 emission line in the SDSS 
spectrum. The f$_{esc}$(Ly$\alpha$) are calculated from the $HST$/COS spectra as detailed in Jaskot et al. (2017), and velocity separation between the two peaks 
of the Ly$\alpha$ emission line are from Jaskot \& Oey (2014) and Henry et al. (2015).
}
\end{deluxetable*}

The velocity separation between the peaks of the double-peaked Ly$\alpha$ emission profile shows a tight relation with the escape fraction of Ly$\alpha$ and 
Lyman continuum emission (Izotov et al. 2018a; Verhamme et al. 2017; Verhamme et al. 2015; Henry et al 2015). According to the radiative transfer models from 
Verhamme et al. (2015), the velocity separation is strongly correlated with the neutral hydrogen column density.  At low column densities, the Ly$\alpha$ photons 
experience less scattering in the surrounding neutral medium, and are observed closer to the systemic velocity. The galaxies which have small separations ($\leq$ 
300 km s$^{-1}$) have low column densities (N$_{HI}$ $\leq$ 10$^{18}$ cm$^{-2}$), while those with larger separations ($\geq$ 600 km s$^{-1}$) have higher 
column densities (N$_{HI}$ $\geq$ 10$^{20}$ cm$^{-2}$). We used the velocity separations of the Ly$\alpha$ profile peaks  measured from the  $HST$/COS spectra
for the GPs galaxies in our sample, to examine the ISM conditions that allow the detectability of the LyC escape. As shown in Figure 6, GPs with small velocity 
separations have higher escape fractions for Ly$\alpha$ emission, and are also potential LyC leakers. Since the EW(CIII]) correlates with EW(Ly$\alpha$) for
star-forming galaxies, we examine how the EW(CIII]) relates to velocity separation. There is no obvious correlation overall between EW(CIII]) and the peak velocity 
separation of Ly$\alpha$ profile for the GP sample. 

Among the sample of GPs in this work, J0815+2156 and J1219+1526 have small velocity separation and have high EW(CIII]). The narrow Ly$\alpha$ profiles J0815+2156, 
and J1219+1526 indicate that the Ly$\alpha$ radiation escapes more easily with less resonant scattering. Both of these GPs are suggested to be possible LyC leakage 
candidates by Jaskot \& Oey (2014). Henry et al. (2015) also note that J1133+6513 and J1219+1526 are good candidates for LyC leakers based on the narrow velocity 
separations (see Table 3). J1219+1526 has a high Ly$\alpha$ escape fraction (f$_{esc}$ = 58\%), large EW(Ly$\alpha$) = 174$\pm$9\AA~ and EW(CIII]) = 5.7$\pm$0.48\AA~. 
J0815+2156 also has strong Ly$\alpha$ emission (f$_{esc}$ = 28\%), with EW(Ly$\alpha$) = 68$\pm$4\AA~ and EW(CIII]) = 8.27$\pm$0.53\AA~. J1133+6513 has low 
equivalent widths, EW(Ly$\alpha$) = 35$\pm$2\AA~ and EW(CIII]) = 1.67$\pm$0.44\AA~, even though the velocity separation of the Ly$\alpha$ emission peaks is small 
(330 km s$^{-1}$) and f$_{esc}$ = 31\%. Henry et al. (2015) argue that the low EWs for Ly$\alpha$ and H$\alpha$ emission of this GP are consistent with higher LyC leakage, 
because the lack of any prominent IS absorption lines supports that the system is optically thin along the LOS. The small observed EW(CIII]) is also consistent with the high 
escape fraction of ionizing radiation in optically thin, density-bounded systems. However, the low C III] values can also be accommodated by models with high optical depth 
but older ages for the stellar population (JR16). J1133+6513 has a relatively high escape fraction, and the low EW(H${\alpha}$) suggests a low intrinsic EW(Ly$\alpha$) 
before scattering, which may reflect an older age for the current starburst, or a continuous star formation history. 

J0303-0759 and J1457+2232 have low EW(Ly$\alpha$) and the lowest Ly$\alpha$ escape fractions (f$_{esc}$ $\leq$ 5\% for Ly$\alpha$) among the GPs sample, but have high 
observed EW(CIII]). Their neutral column densities are likely high based on the large velocity separations, 460 km s$^{-1}$  and 750 km s$^{-1}$ for J0303-0759 and J1457+2232 
respectively. While the velocity separation for GP0303-0759 is very similar to the range of velocities seen for the LyC leakers (Izotov et al. 2016a, 2018a), the escape 
fraction for Ly$\alpha$ is low $\sim$ 5\%, and it is likely that f$_{esc}$(LyC) is also low. The high EWs of the optical emission lines ([OIII]$\lambda$4363, and H${\alpha}$) 
along with the high O$_{32}$ ratio suggest that the intrinsic EW(Ly$\alpha$) should be larger than observed for both of these GPs. Jaskot \& Oey (2014) used the 
CII$\lambda$1334\AA~ and SiII$\lambda$1260\AA~ IS absorption lines to infer the optical depth and geometry of these systems. Both galaxies show strong IS absorption 
lines which confirm their high neutral column densities consistent with the observed weak Ly$\alpha$ emission. Both J0303-0759 and J1457+2232 are good local analogs that 
emphasize the utility of the strong CIII]$\lambda$1909~ emission in optically thick systems. J0815+2156, J1219+1526, and J1457+2232 have comparable O$_{32}$ 
(indicating high log$U$), low 12+log(O/H),  large optical emission line EWs, and exhibit the highest EW(CIII]) among the GPs in our sample, while their Ly$\alpha$ emission 
suggests low optical depths in  J0815+2156 and J1219+1526, and a high optical depth in J1457+2232.

\section{Comparing the CIII] Emission to Photoionization Model Predictions} \label{sec:style}

\begin{figure}
\plotone{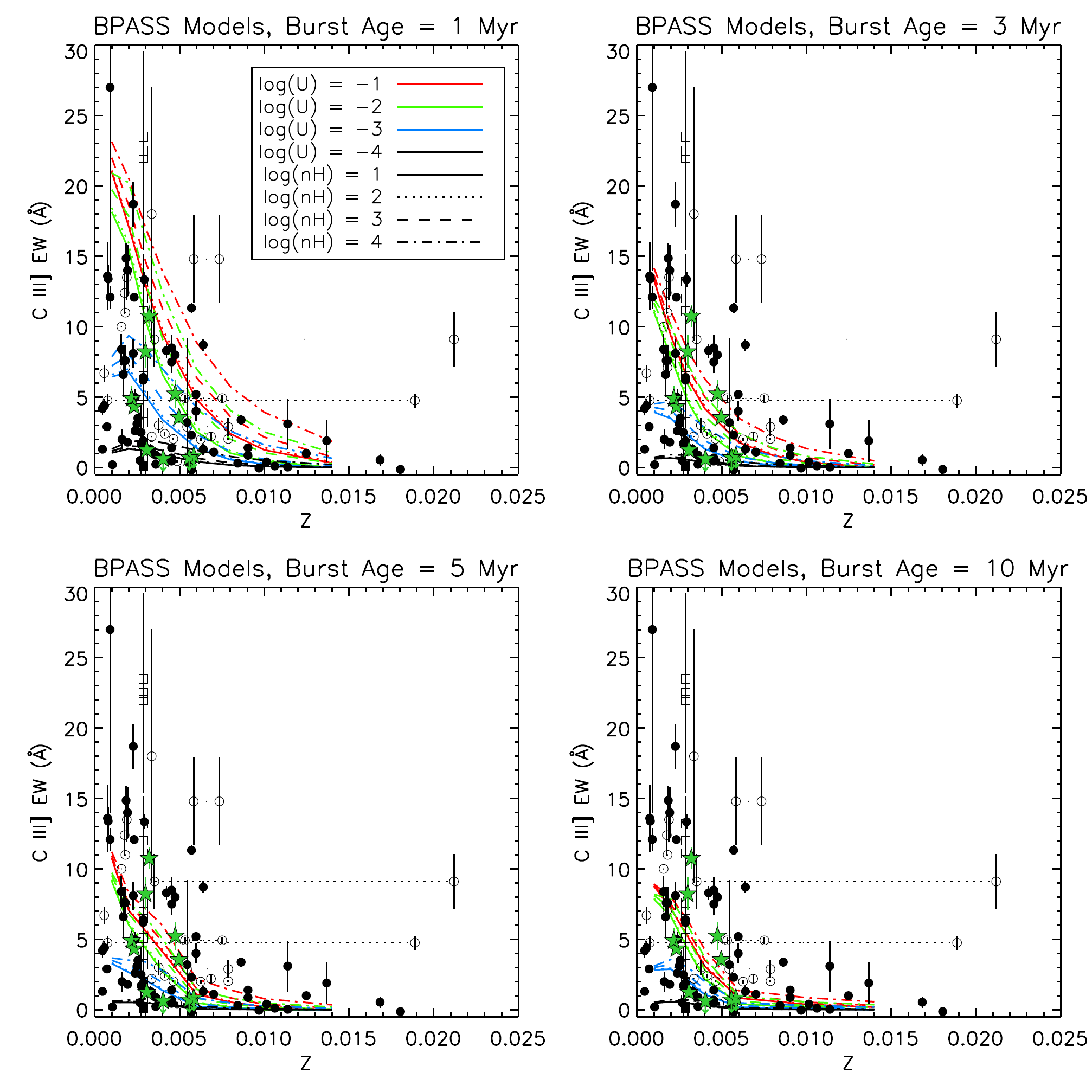}
\caption{The predicted CIII] EW as a function of metallicity from Jaskot \& Ravindranath (2016). Each panel shows a different instantaneous burst age. The models assume log $n_{H}$ =2, and C/O ratio = 0.2 consistent with the values for low-metallicity emission line galaxies, and a low dust-to-metal ratio = 0.1. The fiducial model considers optically thick, filled spherical geometry and does not include effect of shocks. The green star symbols are the GPs from this work, filled black dots are the low redshift galaxies (Rigby et al. 2015, Berg et al. 2016, Senchyna et al. 2017), and open circles are the high redshift samples (Erb et al. 2010; Stark et al. 2014; Le F{\`e}vre et al. 2019; Vanzella et al. 2016, Maseda et al. 2017). Maseda et al. (2017) report two values for metallicity of every galaxy based on the mapping of R$_{23}$ ratio to gas-phase metallicity, and they are shown connected by dotted lines in each panel. \label{fig:fig4}}
\end{figure}

In JR16, we used CLOUDY photoionization models (Ferland et al. 2013) to explore the CIII] EWs and line ratios as a function of starburst age, metallicity, and ionization 
parameter. The models also considered a range of C/O ratios, dust content, gas densities, nebular geometries and optical depths. One of the main conclusions was that 
only the Binary Population and Spectral Synthesis (BPASS; Eldridge \& Stanway 2009; Eldridge et al. 2017) models that incorporate the effects of binary star interactions 
are able to reproduce the highest CIII] EWs ($\geq$ 15-20\AA~) found in high redshift galaxies, and sustain high EW(CIII]) values ($>$ 5\AA~) over longer timescales 
beyond 3 Myrs. The GPs offer a valuable test for the predictions from the photoionization models because the optical SDSS spectra are available for all galaxies in the 
sample and include many diagnostic optical emission lines that offer strong constraints on the metallicity, and ionization parameter. This allows to explore the CIII] EWs 
as a function of age, for the metallicities and ionization parameters determined from the optical emission lines. 

\begin{figure}
\plotone{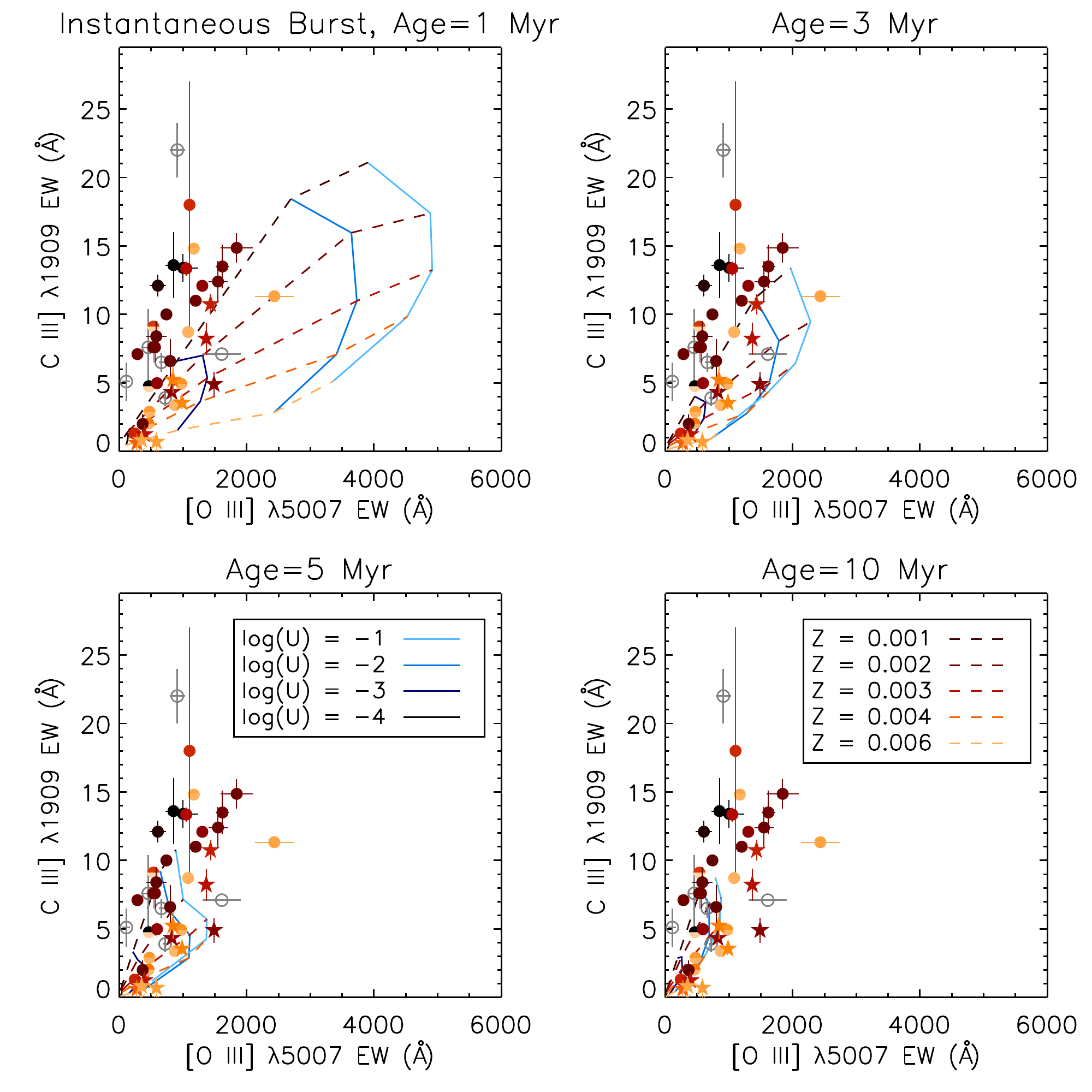}
\caption{The observed EWs of CIII]$\lambda$1909\AA~ and [OIII]$\lambda$5007\AA~ emission lines along with the predictions from CLOUDY models (Jaskot \& 
Ravindranath 2016). The solid lines indicate different values of the ionization parameter, $U$, and dashed lines indicate different metallicity $Z$. The GPs are shown as 
solid star symbols. The data points for low and high redshift galaxies from the literature (see Figure 5) are shown as filled circles. The open circles indicate that the EWs 
are for [OIII]+H$\beta$ as inferred from SED fits to the broad-band photometry, and don't have metallicity estimates. All the other data points are colors-coded by metallicity.
}
\end{figure}

\subsection{Dependence on Metallicity and Ionization Parameter}

In Figure 7, we show the predicted EW(CIII]) as a function of metallicity for the instantaneous burst models with different burst ages (Figure 15 from JR16) along with the 
measured EWs for the GPs. The EW(CIII]) measurements from the literature for high-$z$ (Stark et al. 2014, de Barros et al. 2016, Vanzella et al. 2016, Erb et al 2010, 
Maseda et al. 2017, Le F\`{e}vre et al. 2019), and low-$z$ galaxies (Rigby et al. 2015, Senchyna et al. 2017, Berg et al. 2016) are also shown. The BPASS models are 
able to fully reproduce the range of CIII] EWs observed for GPs, and for high CIII] EWs $>$ 5\AA~, the models require very young burst ages ($<$ 3 Myrs) and high 
ionization parameters (logU $\sim$ $>$ -2) even at low metallicities. The older burst ages are not able to reproduce the high CIII] EWs $>$ 5\AA~ even for high ionization
parameter values (log$U$ $\sim$ -1) at the observed metallicities of the GPs. The BPASS models with the youngest ages $\sim$ 1-3 Myr are also required to accommodate 
the high EW(CIII]) values observed for the He II emitters and BCDs with lower metallicities ($<$ 1/3 Z$_{\odot}$), and for most of the CIII]-emitters at $z>2$. The lower 
EW(CIII]) values can be accommodated by the models with log$U$ $\sim$ -2. Although the GPs in our sample only span a narrow range of metallicities, it is evident that 
their observed EWs follow the expected trend of EW(CIII]) with metallicity from the CLOUDY models. In addition to metallicity and ionization parameter, the C/O ratio 
also affects the observed EW(CIII]). The models presented in Figure 7 assume a fixed C/O = 0.2 which is consistent with the C/O values inferred for our GP sample (Section 3.4).

In Figure 8, the observed EWs of the collisionally-excited emission lines, CIII]$\lambda$1909\AA~ and [OIII]$\lambda$5007\AA~ presented in Figure 5 are compared to 
the predictions from the photoionization models (Figure 8 of JR16). Among the $z>1$ galaxies for which the EWs have been measured,  the $1.8 <z< 2.5$ galaxies from 
Maseda et al. (2017) have metallicities that overlap with the GPs, and have comparable EW(CIII]) and EW([OIII]). The one exception is the galaxy UDF10-164 from Maseda 
et al. (2017), with high CIII] EW = 14.80$\pm$3.10\AA~, but lower [OIII] EW = 1170$\pm$292\AA~ compared to the predicted value $\geq$ 2000\AA~ expected from the 
models for its metallicity (Z$\geq$ 0.004). In Figure 8, this galaxy lies along with the galaxies that have much lower metallicities (Z$\leq$ 0.003).

The observed EWs for low and high redshift non-GP galaxies are clearly offset from the photoionization model grids in the EW(CIII]) versus EW([OIII]) diagram. As noted 
in JR16, the EW([OIII]) is the likely source of the discrepancy, because the models do not include the redder continuum emission from an older population. So, the predicted 
[OIII] EWs is higher than what is observed, while the EW(CIII]) remains almost unaffected. However, this effect alone may not account for all the offsets seen for the measured 
EWs, in particular, for the galaxies that have higher EW(CIII]) than predicted by the models, and are only accommodated by models with the youngest ages ($\sim$ 1-2 Myr) 
and highest ionization parameters. The He II emitters from Senchyna et al. (2017) have metallicities similar to the GPs, but possibly have additional source of nebular heating, 
such as, shocks or WR stars that can explain their higher EW(CIII]). The dwarf galaxies from Berg et al. (2016) have very low nebular oxygen abundances, with 12+log(O/H) 
$<$ 8.0, ranging from $\sim$ 1/5 Z$_{\odot}$ to $\sim$ 1/20 Z$_{\odot}$, which are lower than the average metallicities of the GPs sample in this study. In Figure 5 and 8, 
these dwarf galaxies have consistently higher EW(CIII]) and low EW([OIII]) than the models. On the SDSS images the dwarf galaxies show bright star-forming regions and 
diffuse continuum from the underlying galaxy, which may partly explain the observed EW([OIII]) being lower than the model predictions compared to the GPs. In addition, the 
dwarf galaxies show a range in C/O ratios ranging from 0.15 to 0.50 while the photoionization models use C/O=0.20. The low-metallicity GP, J1154+2443, shows a similar
offset as the blue compact dwarf galaxies, which suggests that the physical conditions in the ionized gas may be different from the input model parameters. Berg et al. (2016) 
infer electron temperatures that are higher ($\sim$ 15,200 - 19,600 K) for the BCDs, compared to the GPs ($\sim$ 14100 - 15500 K;  Jaskot \& Oey 2013). For three of the 
dwarf galaxies, the ionization parameters derived from the UV spectra are $-2.15 <$ log U $< -1.5$ consistent with the high ionization parameters required by the models to 
reproduce the higher EW(CIII]). The comparisons with model predictions in Figure 7 and 8 show that the observed range and trends of EW(CIII]) for GPs can be explained by 
a combination of stellar ages, metallicities, and ionization parameters. While older ages ($>$ 3 Myrs) fail to reproduce the highest observed EWs, the younger ages can produce 
the observed range of EW(CIII]), with the highest EWs occuring for low $Z$ and for high $U$. 

\subsection{Dependence on Optical Depth}

The high LyC escape fractions observed for GPs may indicate that these galaxies are density-bounded systems (Guseva et al. 2004; Jaskot \& Oey 2013; Nakajima \& Ouchi 
2014; Izotov et al. 2018a). Galaxies with highly concentrated star formation as in the compact GPs, have high surface density of star formation, and the feedback from such 
systems can be very effective in clearing out pathways that allow the escape of LyC and Ly$\alpha$ (Heckman et al. 2011; Verhamme et al. 2017). The high LyC escape 
fractions also have implications for the observed emission line ratios. The density-bounded nebulae are optically thin, and compared to the radiation-bounded nebulae, they 
have lower column densities of surrounding gas in the outer layers where the low ionization lines originate (Pellegrini et al. 2012; Jaskot \& Oey 2013; Zackrisson et al. 2013). 
In JR16, we found that predicted CIII] EWs from the photoionization models are lower for density-bounded nebulae, and suggested that the CIII] could be a possible diagnostic 
for optical depth. The transition from radiation-bounded to density-bounded nebulae is equivalent to truncating the nebular gas at different outer radii within the Stromgren 
sphere. To characterize the effect of varying optical depth in the models, we followed Stasinska et al. (2015) and used the $f_{H\beta}$ parameter which is the ratio of the total 
H${\beta}$ produced inside the nebular radius to the total integrated H${\beta}$ in a radiation-bounded Stromgren sphere. In this parametrization, $f_{H\beta}$ =1 for optically 
thick radiation-bounded nebulae, and $f_{H\beta}$ $<$ 1 for optically thin density-bounded nebulae. The photoionization models show that the CIII] flux declines with decreasing 
$f_{H\beta}$, and the effect is strongest for the models with the highest ionization parameters. For high values of $U$, the CIII] originates at larger radii because the higher 
ionization CIV emission dominates in the inner regions of the nebulae closer to the ionizing source.

\begin{figure}
\plotone{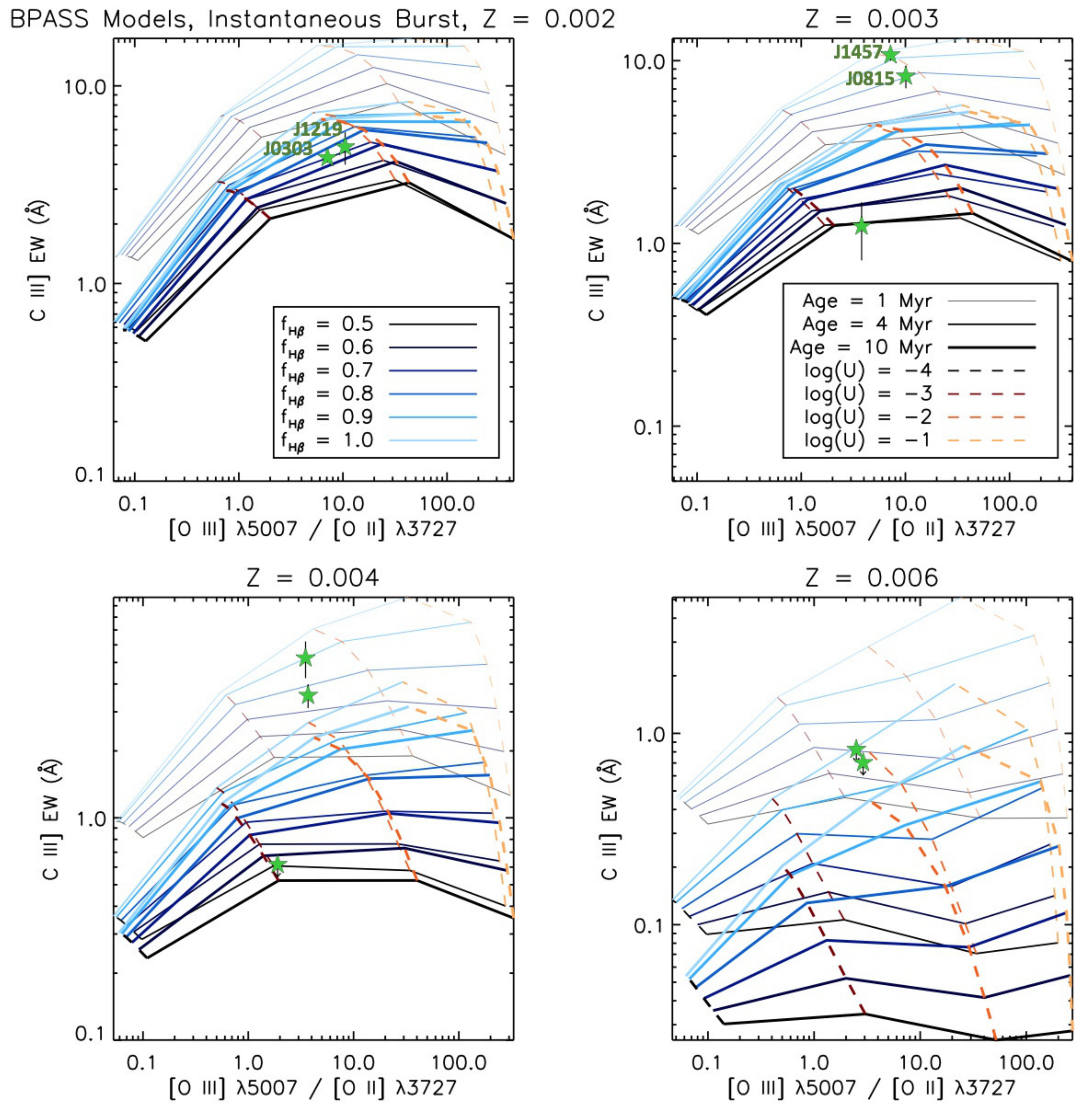}
\caption{: The predicted C III] EWs vs. O$_{32}$ ratio as a function of ionization parameter, $U$, (dashed orange lines), optical depth, $f_{H\beta}$, (solid blue lines), and 
instantaneous burst age (grid thickness). The green star symbols show the observed values for the green pea galaxies on model grids (JR16) corresponding to their 
metallicity. For a given value of O$_{32}$ , the optically thin systems tend to have smaller EW(CIII]) due to escape of ionizing photons, making the combination of these
two parameters a diagnostic for potential LyC leakers. The CLOUDY models are generated for an assumed C/O = 0.2. However, a higher C/O ratio or younger ages can 
also lead to high values of observed EW(CIII]).}
\end{figure}

In Figure 9, the model grids for CIII] EWs versus O$_{32}$ are shown as a function of age, ionization parameter, and optical depth for four different metallicities. At a given 
metallicity, the O$_{32}$ ratio depends on the ionization parameters, but there is a weak dependence on optical depth. The CIII] EW also depends on both the ionization 
parameter $U$, and the optical depth $f_{H\beta}$ at a given metallicity. Since the O$_{32}$ ratio serves as a proxy for $U$, the JR16 models predict that galaxies with 
low EW(CIII]) for a given O$_{32}$ tend to be optically thin with high escape fractions for LyC. Both age and metallicity also affect the scaling of CIII] EWs with optical depth. 
In each panel, we show the location of the GPs on the diagnostic grids corresponding to the metallicities derived from the optical spectra. The model grids in Figure 9 are 
based on an instantaneous burst scenario and assume a C/O ratio = 0.2. 

J0815+2156 and J1219+1526 are GPs with similar O$_{32}$ ratios $\sim$ 10 and are likely optically thin based on the velocity separation measured for the Ly$\alpha$ profiles. 
Both galaxies also have high CIII] EWs, as expected given their high ionization parameters and low metallicites. These GPs, similar to J1154+2443 from Schaerer et al. (2018), 
are high EW(CIII]) galaxies which have high f$_{esc}$(Ly$\alpha$) $>$ 25\% and are expected to be optically thin to the LyC. The strong CIII] emission in all these GPs is 
primarily driven by the very young ages, low metallicity, and high ionization parameter which may be a common property among the LyC leakers, and the effect of optical depth 
($f_{H\beta}$) may be relatively small. Some of the LyC leaker candidates have such high CIII] EWs that can only be reproduced with starburst ages of only 1-2 Myrs, suggesting 
that the LyC leakers are very young, or have harder ionizing spectra than the non-leakers. 

J0303-0759 and J1457+2232 also have low metallicities and comparable high O$_{32}$ values $\sim$ 7.2, but their Ly$\alpha$ profiles suggest they are optically thick with 
f$_{esc}$(Ly$\alpha$) $\sim$ 0. Both J0303-0759 and J1457+2232 have high EW(CIII]) $>$ 3\AA~.  Interestingly, the CIII] EWs of J0815+2156 and J1219+1526 are slightly 
lower than the optically thick galaxy J1457+2232, in spite of their high O$_{32}$ values. It is plausible that there is a slight suppression of the CIII] EW due to the lower optical 
depth to LyC in J0815+2156 and J1219+1526, although the dominant factors that influence the CIII] strength appears to be the age and ionization parameter. J0303-0759, on 
the other hand has lower EW(CIII]) than the other three GPs, although it has low metallicity ($Z \leq 0.003$ ) and high O$_{32}$=7.1. The EWs of the optical nebular lines 
[OIII]$\lambda$5007\AA~ and H${\alpha}$ are also relatively lower for this galaxy, which suggests a lower intrinsic ionizing flux and possibly an older burst age for the stellar 
population. For the remaining GPs in our sample, with low CIII] EW and low O$_{32}$ $<$ 5, the combined effects of an older average starburst age and higher metallicities 
influence the EW(CIII]), and do not offer sufficient constraints on the optical depth. Since EW(CIII]) depends on various factors, the model grids presented in Figure 9 only 
serves as a diagnostic for the effect of optical depth for CIII]-emitters with comparable ages, ionization parameters, and metallicities. The analysis of a larger sample is required 
to isolate the influence of LyC escape on the strength of the CIII] emission.

\section{Discussion}

\subsection{Interpreting CIII] emission in low-metallicity, star-forming galaxies}

The semi-forbidden CIII] nebular line is one of the most prominent emission features in the rest-UV spectra of low-metallicity, star-forming galaxies. CIII] emission appears to 
be ubiquitous in $z>2-6$ low-metallicity galaxies selected by various criteria, including the $z>2$ Lyman-break galaxies (Shapley et al. 2003, Steidel et al. 2016), Lyman-alpha 
emitters at $z\sim3$ (Vanzella et al. 2016, de Barros et al. 2016, Erb et al. 2010), gravitationally-lensed low-mass galaxies at z$>$6 (Stark et al. 2017, Stark et al. 2015a), and 
UV luminosity-selected star forming galaxies at $1.5\leq z\leq 4$ (Maseda et al. 2017, Le F{\`e}vre et al. 2019). As discussed in the previous sections, the interpretation of the CIII] 
emission depends on various parameters, including metallicity, shape of the ionizing spectrum, starburst age, LyC optical depth, and dust extinction. The high fraction of CIII]-emitters 
among the local GPs (8/11 including J1154+2443) and the IRAC color-selected galaxies at high redshifts, shows that low-metallicity galaxies that are selected by their strong 
[OIII] emission line are likely to also show CIII] emission. The higher effective temperatures of low-metallicity ionized nebulae favor the collisionally-excited [OIII] and CIII] 
emission lines. In the presence of hard ionizing radiation with high ionization parameters (log$U$ $\geq$ -2) powered by very young ($\sim$ 1 Myr) metal-poor massive stars
(Figure 7), or a weak AGN (Le F{\`e}vre et al. 2019), the CIII] emission can be very strong with EW(CIII]) $>$ 20\AA~.  In the inner regions of such highly ionized nebulae, CIII] 
emission can be lowered by the transition to triply ionized carbon resulting in CIV emission. At metallicities $Z<1/5 Z_{\odot}$, the CIV emission line has been observed in local 
star-forming dwarf galaxies (Senchyna et al. 2017; Berg et al. 2016). Strong CIV emission frequently detected in $z>6$ galaxies suggests that such hard ionizing SEDs may 
be common for star-forming galaxies in the reionization epoch (Stark et al. 2015b; Schmidt et al. 2017; Mainali et al. 2017).

The burst ages and star formation history can affect the observed emission line EWs. As shown in JR16, for an instantaneous burst, the high CIII] EWs $\sim$ 5\AA~ at
young ages $<$ 3 Myrs are primarily driven by the ionization parameter and metallicity. The same is true for EW(CIII]) $>$ 10\AA~ and a continuous star formation history. 
In both star formation scenarios, for a given metallicity and ionization parameter the EW(CIII]) declines quickly with age initially due to the decrease in the production rate of 
high energy photons. However, in the case of continuous star formation an equilibrium is reached between the birth and death of massive stars beyond $\sim$ 20 Myrs, 
such that the CIII] nebular emission remains approximately constant with age and only the increasing non-ionizing UV continuum at 1909\AA~ from the stellar population 
lowers the EW(CIII]). The EWs of the optical lines are lowered by the continuum from the current star formation, in addition to the contribution to the optical continuum from 
the underlying older stellar population. Therefore, when comparing the CIII] EWs against EW([OIII]) or EW(H${\alpha}$), the contribution to the continuum from the older 
stellar component has to be considered, although the photoionization models used here do not include multiple stellar populations (Figures 5, 8).  For the CIII]-emitters at 
$z=2$, Stark et al. (2014) propose photoionization models with two-component stellar population, one young ($\leq$ 3 Myrs) and an older stellar population to provide a 
better fit to their SEDs. Multi-component star formation histories with a recent burst ($<$ 10 Myr) that powers the nebular emission and an older few 100 Myr stellar 
population that contributes to the UV-optical continuum have been proposed to consistently explain the observed UV spectra and IRAC (rest-optical) colors of $z = 6-7$ 
galaxies (Stark et al. 2015a). The GPs and EELGs at lower redshifts also show evidence for multiple stellar populations. Amor{\'i}n et al. (2012) find that the star formation 
history of GPs indicates the presence of an evolved stellar component with age between 10$^{8}$ yr and several Gyrs. For example, in the case of J1133+6513 with 
EW(CIII])=1.67\AA~ and EW([OIII]) = 394\AA~, which is in common with our sample, they note that the presence of the broad Mg I $\lambda\lambda$ 5167, 5173 absorption 
feature confirms the presence of old stars. The high EWs of the nebular lines, however, are powered by the ionizing continuum from the young stellar component, which 
may only be $< $ 20\% of the total mass. The high SFR ($>$ 4 - 60 M$_{\odot}$/yr) and short mass doubling time of $\leq$ 1 Gyr, imply that the GPs are currently 
experiencing a powerful starburst phase which dominates their UV-optical continuum. Izotov et al. (2011) find that the GPs and similar luminous compact star-forming 
galaxies have M$_{young}$/M$_{total}$ $\sim$ 0.03 - 0.05 on average, for a galaxy with total stellar mass of M$_{total}$ = 10$^{9}$ M$_{\odot}$.  Although there is a 
higher fraction of young stars in the lower mass galaxies, the range in the fraction of young stars for a given galaxy mass can be large based on their star formation 
histories. Among the GPs in this work, four of the galaxies (J0303-0759, J0815+2156, J1219+1526, and J1457+2232) have O$_{32}$ $>$ 5, low metallicity $Z\leq 0.003$, 
and their large optical emission line EWs are consistent with a very young starburst ($<$ 3Myrs). The lower CIII] and optical line EWs of the other GPs may result from 
older burst ages, lower ionization parameters, or high metallicities, and the effect of each individual parameter cannot be entirely disentangled using the present data.

The same physical conditions that produce strong intrinsic LyC and Ly$\alpha$ emission also favor strong C III] emission at low metallicities. However, while LyC and Ly$\alpha$ 
are absorbed or scattered by neutral gas, C III] can escape unimpeded. C III] is therefore a potentially useful probe for systems where LyC and Ly$\alpha$ are suppressed 
due to a surrounding neutral ISM or IGM. For instance, J1457+2232 and J0303-0759 in our GP sample are optically thick with low escape fractions for Ly$\alpha$ and 
possibly to LyC and have high EW(CIII]). Many GPs have lower optical depths, allowing the escape of the ioninzing continuum, and the LyC leakage can lower the observed 
EW(CIII]). At fixed age, metallicity, and ionization parameter, a lower EW (C III]) can be interpreted as arising from a star-forming region that is optically thin to the LyC. As 
discussed in section 4.2, the trends seen for the EW(CIII]) with optical depth for a subset of the current GPs sample appear to be consistent with this interpretation and 
should be revisited using a larger sample of CIII]-emitters.

The overall carbon abundances and dust content in the ionized nebular regions are also expected to influence the CIII] EWs. Since carbon acts as a coolant in ionized regions, 
the lower carbon abundance in the low metallicity galaxies results in higher nebular temperatures and increases the CIII] collisional excitation rates.  As shown in JR16, the CIII] 
emission does not scale linearly with the C/O ratio, but the higher nebular temperatures at low C/O ratios can partially compensate for the lower carbon abundance and result in 
strong CIII] emission. The LyC emitting GP galaxy J1154+2443 with a  high EW(CIII]) = 11.7$\pm$2.9\AA~, has low C/O $\sim$ 0.13 (Schaerer et al. 2018) compared to most 
GPs in our sample. The enhanced CIII] emission in this low-metallicity galaxy with 12+log(O/H) $\sim$ 7.6 is consistent with the JR16 models that predict strongest CIII] emission 
at $Z<0.002$ because of the higher electron temperature even if there are fewer carbon atoms. The dust content in ionized nebulae also affects the emergent CIII] EWs, through 
its dependence on the dust extinction, and the role of photoelectric heating versus cooling via the forbidden lines. The photoelectric heating can enhance the CIII] emission, but 
as the dust content increases the extinction begins to be dominant and lowers the EW(CIII]).  However, low-metallicity star-forming galaxies are relatively dust-poor systems. The 
GPs in our sample have very low nebular extinction, with E(B-V) = $0.03-0.2$ as inferred from the Balmer decrement.

\subsection{Constraints on the ionizing sources and inputs to the Photoionization models}

Although a large number of CIII] observations have become available in the recent years, the nature of the ionizing sources that provide the high ionization parameters to account 
for the CIII] emission strengths are not fully understood. The CIII] emission requires high energy photons ($>$ 24.4 eV) which can be provided by young, massive stars in low 
metallicity SFGs, or by AGNs (Feltre et al 2016; Gutkin et al. 2016; Nakajima et al. 2018a). In the case of SFGs, the effects of binary stellar evolution have to be incorporated to 
successfully reproduce the high CIII] EWs measured in star-forming galaxies consistent with their ages (JR16, Nakajima et al. 2018a). The X-ray observations of GPs reported 
by Svoboda et al. (2019) show that some GPs have X-ray luminosities that are a factor of 6 higher than expected for star-forming galaxies, which may be attributed to a hidden AGN, 
Ultra-luminous X-ray sources, or a higher fraction of high-mass X-ray binaries. However, the optical emission line ratios of GPs are compatible with star-forming galaxies in the BPT 
diagram (Baldwin et al. 1981), and are inconsistent with an AGN contribution although some of extreme GPs can lie close to the maximum line for starburst (Jaskot \& Oey, 2013). 
While the emission lines from GPs in our sample can be accommodated by models with ages $\gtrsim$ 3 Myrs, and ionization parameter log$U$ $\gtrsim$ -2, some of the strong 
CIII]-emitters in the literature require extremely young ages ($\sim$ 1 Myr) and very high ionization parameters (log$U$ $\sim$ -1). The role of other exotic ionizing sources, such as 
very massive stars with masses $>$ 100 M$_{\odot}$ (Smith et al. 2016), and contribution from low-luminosity AGNs ( Le F{\`e}vre et al. 2019; Nakajima et al. 2018a) cannot be ruled 
out for the most extreme CIII] emitters. 

The CIII] emission line in combination with other UV and optical emission lines forms an important diagnostic which helps to reveal the ionizing sources that are responsible for 
the nebular emission (Feltre et al. 2016; Byler et al. 2018). In addition to photoionization, the contribution of shocks can enhance the flux of CIII] emission, and low-ionization optical 
emission lines. Among the six GPs in Jaskot \& Oey (2013), five of them showed HeII $\lambda$4686 emission which may be produced by WR stars or shocks. The radiative shocks 
originating from SNe explosions can generate He II emission even without the WR stars, and are most efficient in dense, low-metallicity systems which are the characteristic of GPs 
(Guseva et al. 2000; Thuan \& Izotov 2005). In order to understand the ionizing spectrum and optical depth effects, the contribution from shocks should be subtracted. Although GPs 
are young enough to host large numbers of ionizing O stars or WR stars, SNe and stellar winds from an ongoing or previous burst of star formation can reshape their ISM. In JR16, 
we showed that the presence of shock always contributes to an increase in the observed CIII] flux, and proposed a diagnostic involving CIII]/HeII versus C IV/ HeII to distinguish 
nebular emission from purely photoionized gas and shock-ionized gas. However, we only detect weak HeII $\lambda$1640\AA~ in the composite STIS spectrum from this study. 
Future spectroscopic observations of GPs with higher spectral resolution and high signal-to-noise to measure various shock diagnostics in the UV-optical wavelengths will be required 
to calibrate the shock diagnostics. 

The UV spectra of GPs can constrain the input ISM parameters used in the photoionization models, such as C/O ratios and electron densities. The sample of $\sim$ 8 GPs with detectable 
CIII] emission (including J1154+2443 from Schaerer et al. 2018), show that the C/O ratios range from $\sim$ 0.08 to 0.35 and are similar to the other CIII]-emitters locally (Garnett et al. 1995; 
Berg et al. 2016, 2019; Senchyna et al. 2017), and at $z\geq2$ (Amor{\'i}n et al. 2017; Stark et al. 2014, Erb et al. 2010, Shapley et al. 2003). The variation of CIII] emission over the same 
range of C/O ratios has been explored in JR16, and found to have a significant effect on the EW(CIII]). The electron densities, $n_{e} \sim$ 100 cm$^{-3}$, used in the photoionization models 
(JR16), are similar to that seen in typical HII regions. However, various measurements of electron densities from the [CIII]$\lambda$1906 + CIII]$\lambda$1908 doublet give $n_{e}$ values 
that are about two orders of magnitude higher than derived using optical diagnostics (e.g., [O II] or [S II] doublets) for galaxies at low and high redshifts (James et al. 2014, 2018; Bayliss et al. 
2014; Berg et al. 2018). A possible reason for the different $n_{e}$ values is that the density-sensitive doublets in the optical are tracing the low density regions compared to the CIII] doublet 
which originates in higher density regions closer to the ionizing source (James et al. 2018; Berg et al. 2018). The existing $HST$/STIS observations of GPs do not resolve the CIII] doublet 
lines, and high resolution UV spectra of GPs would be required to constrain the electron densities used for the model predictions of CIII] emission.

\subsection{CIII] Emission Diagnostics to Explore the Galaxies in the Reionization Epoch}

One of the key goals of the {\it James Webb Space Telescope (JWST)} and upcoming large 20-m class ground-based telescopes is to reveal the physical properties of the galaxies responsible 
for the reionization of the Universe, and to quantify their contribution of ionizing photons. The relative contributions of star-forming galaxies and AGNs to the total ionizing budget for reionizing 
the Universe are still debated (Madau \& Haardt 2015; Finkelstein 2016; Matsuoka et al. 2018; Hassan et al. 2018). The UV luminosity function from deep surveys suggest that star-forming 
galaxies are the primary agents for reionization (Finkelstein et al. 2016), but recent faint AGN surveys suggest that AGNs may have also contributed to the transition (Madau \& Haardt 2015). 
The spectroscopic capabilities of the JWST instruments will play a critical role in identifying AGNs and starbursts in the reionization epoch. While the commonly employed optical nebular 
diagnostics of the BPT-diagram (Baldwin et al. 1981) will be redshifted to the mid-IR wavelengths at z $>$8, the bulk of the JWST spectroscopic data at NIR wavelengths obtained using NIRSpec 
and slitless spectroscopy with NIRISS and NIRCam will provide access to the rest-UV emission lines (eg; Lyman-$\alpha$, CIV, OIII], He II, and CIII]).  For the gravitationally-lensed galaxies 
at $z>8$, the NIRSpec IFU, NIRISS and NIRCam grism modes will provide spatially-resolved emission-line maps over physical scales of few tens to hundreds of parsecs. 

The star-forming galaxies in the reionization epoch are compact, have high SSFRs and low-metallicities similar to the GPs and are likely to show strong CIII] and nebular UV emission lines.
The young ages of the starbursts and low-metallicities of the ISM would favor high CIII] EWs (Figure 5, 8), making CIII] one of the most easily detected emission features in the NIR spectra 
of $z>6$ galaxies. Since the observability of Ly$\alpha$ emission drops at $z>6$ due to the increased IGM absorption, the CIII] emission could be used as an alternate probe of the ionizing 
flux and SFR based on the empirical relation between CIII] and H$\alpha$ emission lines (Figure 5). The CIII] emission line in combination with other nebular lines, such as, NV$\lambda$1240\AA~, 
CIV$\lambda$1548,1550\AA~, HeII$\lambda$1640\AA~, OIII]$\lambda$1661, 1666\AA~, and Si III]1883,1892\AA~, can be used to reveal the nature of the ionizing sources, and to derive the 
nebular abundances (Feltre et al. 2016; Gutkin et al. 2016, JR16; Nakajima et al. 2018a; Byler et al. 2018). The CIV/CIII] ratio that uses UV emission lines which originate from different ionization 
states of carbon can constrain the ionization parameter, similar to the O$_{32}$ optical emission line ratio. The CIII] doublet lines and SiIII] doublet lines can both be used for deriving the electron 
densities (JR16, Gutkin et al. 2016; Byler et al. 2018), while the commonly used CIII]/OIII] can be used to determine the elemental carbon abundances (Berg et al. 2016, 2018; Garnett et al. 
1995). Feltre et al. (2016) have shown using photoionization models that the AGNs and star-forming galaxies separate out well in the CIV/CIII] versus CIV/HeII, and CIII]/HeII versus CIV/HeII 
diagnostic diagrams. The HeII line serves as a key diagnostic for hard ionizing radiation with high energy photons ($>$ 54.4 eV), since the enhanced HeII emission would lower the CIII]/HeII ratio. The 
models which include shock contribution can produce emission line ratios that overlaps with the AGNs in the CIII] /HeII versus CIV/HeII diagnostic diagram, but the pure photoionization models 
occupy an entirely different part of the diagram (JR16). As seen from the composite spectrum of CIII]-emitters among GPs, the HeII emission line can be strong (section 2.3) enough to be 
observable in the low-metallicity high redshift galaxies with JWST. 

The semi-forbidden CIII] nebular emission doublet serves as an important diagnostic spectral line to infer the physical properties of reionizers. However, extensive calibration by applying the 
UV diagnostics to low-redshift galaxies is necessary for them to be used effectively to interpret sources in the reionization epoch. Currently, the number of galaxies in the low-metallicity regime 
with the required wavelength coverage to test the diagnostics based on rest-UV emission line ratios at $z\geq 2$ is very sparse. Only in recent years have the rest-UV spectroscopic data 
become available for local analogs of the high-$z$ galaxies, which include GPs, BCDs, and EELGs at lower redshifts. In this work, we have explored the conditions that favor the CIII] emission 
and examined diagnostics involving the CIII] and optical emission lines for the GP galaxies. We have shown that the GPs with high O$_{32}$ also have high EW(CIII]) because the high ionization 
parameters favor CIII] emission, but there is also a weak dependence on the optical depth to Lyman continuum (Figure 9). CIII] emission tends to be weaker for optically-thin density-bounded 
nebulae compared to radiation-bounded nebulae for similar values of the ionization parameter. If these trends are calibrated for a sample of confirmed LyC leaking GPs, the dependence of CIII] 
EW on the optical depths can be used to quantify the Lyman continuum escape from star-forming galaxies in the reionization epoch. Future work will require deep UV spectroscopy to detect 
the weaker UV spectral lines for a larger sample of GPs and other local analogs that extend these analyses to lower metallicities and a wide range of Ly$\alpha$ emission profiles. Detailed 
calibrations of the rest-UV nebular diagnostics in combination with the commonly used optical emission line diagnostics is a crucial step to be able to characterize and interpret the spectroscopic 
observations of the sources of reionization obtained with JWST.

\section{Summary}

We have analyzed the {\it HST}/STIS NUV spectra for a sample of ten GP galaxies at redshifts 0.1$\leq z\leq 0.3$, to explore the semi-forbidden CIII]$\lambda$1909\AA~ nebular emission line 
in low metallicity star-forming galaxies with 7.8$\leq$12+log(O/H)$\leq$8.4. We selected galaxies that have archival {\it HST}/COS FUV spectroscopic observations which includes  the Ly${\alpha}$ 
emission. The UV spectra were used along with the optical spectra from SDSS to constrain the metallicity and ionization parameter, to examine the correlations between the UV and optical nebular 
lines, and to compare the CIII] emission properties with predictions from the photoionization models. We summarize the results below:

(1) CIII] emission is detected in 7/10 GP galaxies confirming that CIII] emission is almost ubiquitous in low-metallicity galaxies. The composite spectrum of the CIII]-emitters shows an emission 
feature at the location of HeII $\lambda$1640. The composite spectrum of the CIII] non-emitters shows strong interstellar absorption lines, particularly CIV absorption with broad blue-shifted 
profile wing, likely from the presence of strong outflows.

(2) The observed CIII] EWs of GPs are in the range 2-10\AA~, consistent with the predictions from the photoionization models of JR16 which used constraints on model inputs (such as, metallicity 
and ionization parameter) from the optical SDSS spectra. The GPs have CIII] EWs that overlap with the range of values seen in $z>2$ star-forming galaxies, but do not reach the high EW(CIII]) 
values ($>$15\AA~) seen in some $z>2$ star-forming galaxies, local He II emitters and BCDs that have very low metallicities with 12+log(O/H) $\lesssim$ 7.5. At very low metallicities, the lack 
of metals that act as coolants can considerably enhance the CIII] emission even when the abundance of carbon atoms is low.

(3) Although the ensemble of star-forming galaxies appear to follow the empirical relation between EW(CIII]) and EW(Ly$\alpha$), the GPs do not seem to closely follow this relation. GPs are 
strong emission line galaxies by definition, and the observed trend in EW(CIII]) versus EW(Ly$\alpha$) diagram is primarily driven by the variation in the Ly$\alpha$ optical depth. For non-GP 
galaxies, a weak Ly$\alpha$ emission may indicate that the emission lines are intrinsically weak due to a weak ionizing spectrum.  J1457+2232 has the strongest CIII] emission in our sample 
with EW(CIII]) = 9.35$\pm$0.76\AA~, but is offset from the EW(CIII]) versus EW(Ly$\alpha$) relation. This galaxy has a broad absorption in the Ly$\alpha$ profile at the systemic velocity, and 
has profile peaks with large velocity separation, $\Delta v$ = 750 km/s indicating a high neutral gas column along the line of sight. At high redshift, EGS-zs8-1 at z=7.73 (Stark et al. 2017) is a
galaxy in the reionization epoch with substantial neutral gas present at the Ly$\alpha$ line center leading to a low EW(Ly$\alpha$) = 21$\pm$4\AA~, but has high EW(CIII]) = 22$\pm$2\AA~. 
Such examples highlight the utility of the CIII] emission as a key nebular diagnostic when Ly$\alpha$ is attenuated by the ISM or IGM.

(4) CIII] emission does not show any obvious relation between velocity separation of the Ly$\alpha$ profiles and escape fraction of the Ly$\alpha$.  For a given flux of hard ionizing radiation powered 
by young massive stars of fixed low-metallicity, a narrow velocity separations for the Ly$\alpha$ profile implies low neutral hydrogen column densities and high Ly$\alpha$ escape fraction. In such 
optically thin systems, the intrinsic and observed EW(Ly$\alpha$) and EW(CIII]) will be high for a high ionizing flux. However, such systems are likely to have higher LyC leakage which may lower the 
observed EW(CIII]). In the case of optically thick systems the Ly$\alpha$ profile has broad velocity separation and Ly$\alpha$ emission may be weak with low EW(Ly$\alpha$) or even absent. The
CIII] EW remains unaffected compared to the optically thin case.

(5) The CIII] emission in GPs correlates with the [OIII]$\lambda$5007\AA~, and H${\alpha}$ optical emission lines which are also powered by the ionization by massive stars with young ages
The presence of strong CIII] and [OIII] emission confirms the presence of hard ionizing radiation required to produce the high-ionization nebular lines. The  CIII] EW in GPs also correlates strongly 
with the O$_{32}$ ratio, which is a proxy for the ionization parameter.

(6) The observed EW(CIII]) and EW([OIII]) values for the GP galaxies lie within the predictions from the model grids. However, most of the non-GP galaxies have higher EW(CIII]) for a given EW([OIII]). 
Unlike most of the GPs, and other EELGs which have very strong nebular lines with EW([OIII]) $\gtrsim$ 800\AA~ that dominate their spectrum, most of the star-forming galaxies selected using different 
criteria have a significant older stellar population that contributes to the continuum, which may explain the offset from the models. Other factors that can influence the CIII] and [OIII] EWs include the C/O
ratios, nebular temperatures, and electron densities which are different compared to the GPs.

(7) We compared the properties of the CIII] emission for the GPs to the predictions from the JR16 photoionization models that use BPASS input SEDs and a grid of model parameters (eg; metallicities, 
ages, log$U$, and optical depth). The observed range of EW(CIII]) are consistent with the model predictions, and require young stellar ages ($\lesssim$ 3-5Myrs), high ionization parameters (log$U$ 
$\geq$ -2), and low-metallicities (Z$\leq$ 0.006). 

(8) GPs are very likely to be LyC emitters, and the leakage of LyC radiation from the nebular region can reduce the CIII] emission. The JR16 models predict that for density-bounded regions with 
similar metallicities, ages, and O$_{32}$ ratios, the EW(CIII]) will be lower when the optical depth to LyC is lower. At very young stellar ages ($<$ 3 Myrs), high ionization parameters (log$U$ $>$ -2), 
and low metallicities (Z$<$0.15Z$_{\odot}$) the EW(CIII]) values are high ($>$ 3\AA~), mainly due to the high ionizing flux and high nebular temperatures, and the effect of optical depth is not the 
dominant factor. Four of the GPs in our sample with high O$_{32}$ $>$ 5 are found to have EW(CIII]) $>$ 3\AA~. J0815+2156 and J1219+1526 are two of the most promising candidates for LyC 
escape from our sample and much like the LyC leaking GP galaxy, J1154+2443 (Schaerer et al. 2018), they have high C III] EWs consistent with extremely young ages and high $U$. These could 
be common properties of LyC-leaking GPs. Interestingly, J0815+2156 and J1219+1526 are candidate LyC emitters based on their Ly$\alpha$ profiles and have lower EW(CIII]) than J1457+2232 
which has lower O$_{32}$ and comparable metallicity, but is likely optically thick to the LyC.

In this work, we have focused on the semi-forbidden CIII] and trends involving CIII] EWs and other UV-optical lines. Since CIII] is one of the strongest and predominant nebular emission lines that will 
be used to reveal the nature of ionizing sources at the reionization epoch, it is important to calibrate line ratios involving CIII] and other UV lines for a larger sample of galaxies with a broad range of nebular
properties. Such calibrations based on low redshift samples where both the UV and the commonly-used optical diagnostics can be combined, will be crucial to interpret the spectra of $z>7$ galaxies that 
will be observed by JWST and future large telescopes.
\acknowledgments

A.E.J and J.T acknowledge support provided by NASA through grant {\it HST}-GO-14134 from Space Telescope Science Institute, which is operated by the Association of Universities for Research in Astronomy (AURA) under NASA contract NAS-5-26555. We thank the referee for careful reading of the manuscript and for the useful comments.




\begin{thebibliography}{}

\bibitem[Amor{\'i}n et al. (2012)]{2012ApJ...749..185A} Amor{\'i}n, R., P{\'e}rez-Montero, E., Vilchez, J.M., Papaderos, P.  \ 2012, \apj, 749, 185
\bibitem[Amor{\'i}n et al. (2017)]{2017NatAs...1E..52A} Amor{\'i}n, R., Fontana, A., P{\'e}rez-Montero, E., Castellano, M., et al.\ 2017, Nature, 1, 52
\bibitem[Asplund et al. (2009)]{2009ARA&A..47..481A} Asplund, M., Grevesse, N., Sauval, A.J., \& Scott, P. \ 2009, ARAA, 47, 481
\bibitem[Baldwin et al. (1981)]{1981PASP...93....5B} Baldwin, J.A., Phillips, M.M., \& Terlevich, R. \ 1981, PASP, 93, 5
\bibitem[Bayliss et al. (2014)]{2014ApJ...790..144B} Bayliss, M.B., Rigby, J.R., Sharon, K., Wuyts, E., Florian, M., Gladders, M. et al. \ 2014, \apj, 790, 144
\bibitem[Berg et al. (2016)]{2016ApJ...827..126B} Berg, D. A., Skillman, E. D., Henry, R.B.C., Erb, D.,K., Carigi, L. \ 2016, \apj, 827, 126
\bibitem[Berg et al. (2018)]{2018ApJ...859..164B} Berg, D. A., Erb, D. K., Auger, M.W., Pettini, M., Brammer, G. \ 2018, \apj, 859, 164
\bibitem[Berg et al. (2019)]{2019ApJ...874..93B} Berg, D. A., Erb, D. K., Henry, R.B.C., Skillman, E. D., McQuinn, K.B.W  \ 2019, \apj, 874, 93
\bibitem[Byler et al. (2018)]{2018ApJ...863...14B} Byler, N., Dalcanton, J.J., Conroy, C., Johnson, B.D., Levesque, E.M., Berg, D. \ 2018, \apj, 863, 14
\bibitem[Cardamone et al. (2009)]{2009MNRAS.399.1191C} Cardamone, C., Schawinski, K., Sarzi, M., Bamford, S. P., Bennert, N., Urry, C.M., et al. \ 2009, \mnras, 399, 1191
\bibitem[Cardelli, Clayton, \& Mathias (1989)]{1989ApJ...345..245C} Cardelli, J.A., Clayton, G.C., \& Mathis, J.S. \ 1989, \apj, 345, 245
\bibitem[Chevallard et al. (2018)]{2018MNRAS.479.3264C} Chevallard, J., Charlot, S., Senchyna, P., Stark, D.P., Vidal-Garci{\'a}, A., Feltre, A. et al. \ 2018, \mnras, 479, 3264
\bibitem[de Barros et al. (2016)]{2016A&A...585A..51D} de Barros, S., Vanzella, E., Amo{\'r}in, R., Castellano, M., Siana, B., et al. \ 2016, A\&A, 585, 51
\bibitem[Ding et al. (2017)]{2017ApJ...838L..22D} Ding, J., Cai, Z., Fan, X., Stark, D.P., Bian, F., Jiang, L., McGreer, I.D., Robertson, B. E., Siana, B. \ 2017, \apj, 838, 22
\bibitem[Eldridge \& Stanway (2009)]{2009MNRAS.400.1019E} Eldridge, J.J. \& Stanway, E. \ 2009, \mnras, 400, 1019
\bibitem[Eldridge et al. (2017)]{2017PASA...34...58E} Eldridge, J.J., Stanway, E.R., Xiao, L., McClelland, L.A.S., \ 2017, PASA, 34, 58
\bibitem[Erb et al. (2010)]{2010ApJ...719.1168E} Erb, D.K., Pettini, M., Shapley, A.E., Steidel, C.C, Law, D.R., Reddy, N.A. \ 2010, \apj, 719, 1168
\bibitem[Ferland et al. (2013)]{2013RMxAA..49..137F} Ferland, G.J., Porter, R.L., van Hoof, P.A.M., Williams, R.J.R., Abel, N.P., \ 2013, RMxAA, 49, 137
\bibitem[Feltre et al. (2016)]{2016MNRAS.456.3354F} Feltre, A., Charlot, S., \& Gutkin, J. \ 2016, \mnras, 456 3354
\bibitem[Finkelstein (2016)]{2016PASA...33...37F} Finkelstein, S.L. \ 2016, PASA, 33, 37
\bibitem[Fitzpatrick (1999)]{1999PASP..111...63F} Fitzpatrick, E.L. \ 1999, PASP, 111, 63
\bibitem[Fosbury et al. (2003)]{2003ApJ...596..797F} Fosbury, R.A.E., Villar-Mart{\'i}n, M., Humphrey, A., Lombardi, M., Rosati, P. et al. \ 2003, \apj, 596, 797
\bibitem[Garnett et al. (1995)]{1995ApJ...443...64G} Garnett, D.R., Skillman, E.D., Dufour, R.J., Piembert, M., Torres-Piembert, S., et al. \ 1995, \apj, 443, 64
\bibitem[Guseva et al. (2000)]{2000ApJ...531..776G} Guseva, N.G., Izotov, Y. I., Thuan, T.X. \ 2000, \apj, 531, 776
\bibitem[Guseva et al. (2004)]{2004A&A...421..519G} Guseva, N.G., Papaderos, P., Izotov, Y. I., Noeske, K. G., Fricke, K. J. \ 2004, A\&A, 421.519
\bibitem[Gutkin et al. (2016)]{2016MNRAS.462.1757G} Gutkin, J, Charlot, S., Bruzual, G \ 2016, \mnras, 462, 1757
\bibitem[Hassan et al. (2018)]{2018MNRAS.473..227H} Hassan, S., Dav{\'e}, R., Mitra, S., Finlator, K., et al. \ 2018, \mnras, 473, 227
\bibitem[Heckman et al. (2011)]{2011ApJ...730....5H} Heckman, T.M., Borthakur, S., Overzier, R., Kauffmann, G., Basu-Zych, A., et al. \ 2011, \apj, 730, 5
\bibitem[Henry et al. (2015)]{2015ApJ...809...19H} Henry, A., Scarlata, C., Martin, C. L., Erb, D. K. \ 2015, \apj, 809, 19
\bibitem[Hutchison et al. (2019)]{2019ApJ...879...70H} Hutchison, T.A., Papovich, C., Finkelstein, S. L., Dickinson, M.E., et al. \ 2019, \apj, 879, 70
\bibitem[Izotov \& Thuan (1999)]{1999ApJ...511..639I} Izotov, Y. I., \& Thuan, T.X. \ 1999, \apj, 511, 639
\bibitem[Izotov et al. (2011)]{2011ApJ...728..161I} Izotov, Y.I., Guseva, N. G., \& Thuan, T. X., \ 2011, \apj, 728, 161
\bibitem[Izotov et al. (2016a)]{2016MNRAS.461.3683I} Izotov, Y. I., Schaerer, D., Thuan, T.X., Worseck. G., Guseva, N.G., Orlitov{\'a}, I., Verhamme, A. \ 2016a, \mnras, 461, 3683
\bibitem[Izotov et al. (2016b)]{2016Natur.529..178I} Izotov, Y. I., Orlitov{\'a}, I., Schaerer, D., Thuan, T.X., Verhamme, A., Guseva, N.G., Worseck. G. \ 2016b, Nature, 529, 178
\bibitem[Izotov et al. (2017)]{2017MNRAS.467.4118I} Izotov, Y. I., Schaerer, D., Guseva, N.G., Fricke, K.J., Henkel, C., Schaerer, D. \ 2017, \mnras, 467, 4118
\bibitem[Izotov et al. (2018a)]{2018MNRAS.478.4851I} Izotov, Y. I., Worseck, G., Schaerer, D., Guseva, N. G., Thuan, T. X., Fricke, K.J., Verhamme, A., Orlitov{\'a}, I. \ 2018a, \mnras, 478, 4851
\bibitem[Izotov et al. (2018b)]{2018MNRAS.474.4514I} Izotov, Y. I., Schaerer, D., Worseck, G., Guseva, N. G., Thuan, T. X., Verhamme, A., Orlitov{\'a}, I., Fricke, K.J. \ 2018b, \mnras, 474, 4514
\bibitem[James et al. (2014)]{2014MNRAS.440.1794J} James, B.L., Pettini, M., Christensen, L., Auger, M.W., Becker, G.D., et al. \ 2014, \mnras, 440, 1794
\bibitem[James et al. (2018)]{2018MNRAS.476.1726J} James, B.L., Auger, M., Pettini, M., Stark, D.P., Belokurov, V., Carniani, S. \ 2018, \mnras, 476, 1726
\bibitem[Jaskot \& Oey (2013)]{2013ApJ...766...91J} Jaskot, A.E., \& Oey, M.S.\ 2013, \apj, 766, 91 
\bibitem[Jaskot \& Oey (2014)]{2014ApJ...791L..19J} Jaskot, A.E., \& Oey, M.S.\ 2014, \apj, 791, 19
\bibitem[Jaskot \& Ravindranath(2016)]{2016ApJ...833..136J} Jaskot, A.E., \& Ravindranath, S.\ 2016, \apj, 833, 136 
\bibitem[Jaskot et al. (2017)]{2017ApJ...851L...9J} Jaskot, A.E., Oey, M.S., Scarlata, C., Dowd, T. 2017, ApJL, 851, 9
\bibitem[Kennicutt (1998)]{1998ARA&A..36..189K} Kennicutt, R.C., \ 1998, ARAA, 36, 189
\bibitem[Kewley \& Dopita (2002)]{2002ApJS..142...35K} Kewley, L.J. \& Dopita, M. A. \ 2002, \apjs, 142, 35
\bibitem[Konno et al. (2014)]{2014ApJ...797...16K} Konno, A., Ouchi, M., Ono, Y., Shimasaku, K., Shibuya, T., Furusawa, H., et al. \ 2014, \apj, 797, 16
\bibitem[Laporte et al. (2017)]{2017ApJ....851...40L} Laporte, N., Nakajima, K., Ellis, R.S., Zitrin, A., Stark, D.P., Mainali, R., Roberts-Borsani, G.W., \ 2017, \apj, 851, 40
\bibitem[Leitherer et al. (2011)]{2011AJ....141...37L} Leitherer, C., Tremonti, C.A., Heckman, T.M., Calzetti, D. \ 2011, AJ, 141, 37
\bibitem[Le F{\`e}vre et al. (2019)]{2019A&A...625A..51L} Le F{\`e}vre, O., Lemaux, B.C., Nakajima, K., Schaerer, D., Talia, M., Zamorani, G. et al. \ 2019, A\&A, 625, 51
\bibitem[Madau \& Haardt (2015)]{2015ApJ...813L...8M} Madau, P., \& Haardt, F. \ 2015, \apj, 813, 8
\bibitem[Mainali et al. (2017)]{2017ApJ...836L..14M} Mainali, R., Kollmeier, J.A., Stark, D.P., Simcoe, R.A., et al. \ 2017, \apj, 836, 14
\bibitem[Matsuoka et al. (2018)]{2018ApJ...869..150M} Matsuoka, Y., Strauss, M.A., Kashikawa, N., Onoue, M. et al. \ 2018, \apj, 869, 150
\bibitem[Mason et al. (2018)]{2018ApJ...856....2M} Mason, C. A., Treu, T., Dijkstra, M., Mesinger, A., Trent, M., et al. \ 2018, \apj, 856, 2
\bibitem[Maseda et al. (2014)]{2014ApJ...791...17M }Maseda, M.V., van der Wel, A., Rix, H.-W., da Cunha, E., Pacifici, C., et al. \ 2014, \apj, 791, 17
\bibitem[Maseda et al. (2017)]{2017A&A...608A...4M} Maseda, M.V., Brinchmann, J., Franx, M., Bacon, R., Bouwens, R.J., Schmidt, K. B., Boogard, L. A., et al. \ 2017, A\&A, 608, 4
\bibitem[McKinney et al. (2019)]{2018ApJ...856....2M} McKinney, J. H., Jaskot, A.E., Oey, M.S., Yun, M.S., Dowd, T., Lowenthal, J.D \ 2019, \apj, 874, 52
\bibitem[Nakajima \& Ouchi (2014)]{2014MNRAS.442..900N} Nakajima, K., \& Ouchi, M. \ 2014, \mnras, 442, 900
\bibitem[Nakajima et al. (2016)]{2016ApJ...831L...9N} Nakajima, K., Ellis, R. S., Iwata, I., Inoue, A.K., Kusakabe, H., et al. \ 2016, \apj, 831, 9
\bibitem[Nakajima et al. (2018a)]{2018A&A...612A..94N} Nakajima, K., Schaerer, D., Le F{\`e}vre, O., Amo{\'r}in, R., et al.\ 2018, A\&A, 612, 94
\bibitem[Nakajima et al. (2018b)]{2018MNRAS.477.2098N} Nakajima, K., Fletcher, T., Ellis, R. S., Robertson, B., Iwata, I. \ 2018, \mnras, 477, 2098
\bibitem[Pellegrini et al. (2012)] {2012ApJ...755...40P} Pellegrini, E.W., Oey, M.S., Winkler, P.F., Points, S.D., Smith, R.C., et al. \ 2012, \apj, 755, 40
\bibitem[Rigby et al. (2015)]{2015ApJ...814L...6R} Rigby, J.R., Bayliss, M.B., Gladders, M.B., Sharon, K., et al.\ 2015, \apj, 814, 6
\bibitem[Roberts-Borsani et al. (2016)]{2016ApJ...823..143R} Roberts-Borsani, G.W., Bouwens, R. J., Oesch, P. A., Labbe, I., Smit, R. et al. \ 2016, \apj, 823, 143
\bibitem[Schaerer et al. (2018)]{2018A&A...616L..14S} Schaerer, D., Izotov, Y.I., Nakajima, K., Worseck, G., Chisholm, J., Verhamme, A. et al. \ 2018, A\&A, 616, L14
\bibitem[Schmidt et al. (2017)]{2017ApJ...839...17S} Schmidt, K.B., Huang, K.-H., Treu, T., Hoag, A., et al. \ 2017, \apj, 839, 17
\bibitem[Senchyna et al. (2017)]{2017MNRAS.472.2608S} Senchyna, P., Stark, D. P., Vidal-Garc{\'i}a, A., Chevallard, J., Charlot, S. \ 2017, \mnras, 472, 2608
\bibitem[Senchyna et al. (2019)]{2019MNRAS.488.3492S} Senchyna, P., Stark, D. P., Chevallard, J., Charlot, S., Vidal-Garc{\'i}a, A.,  \ 2019, \mnras, 488, 3492
\bibitem[Shapley et al. (2003)]{2003ApJ...588...65S} Shapley, A.E., Steidel, C.C., Pettini, M., \& Adelberger, K.L. \ 2003, \apj, 588, 65
\bibitem[Shapley et al. (2015)]{2015ApJ...801...88S} Shapley, A.E., Reddy, N.A., Kriek, M., Freeman, W.R., Sanders, R.L., et al. \ 2015, \apj, 801, 88
\bibitem[Shirazi \& Brinchmann (2012)]{2012MNRAS.421.1043S} Shirazi, M., \& Brinchmann, J. \ 2012, \mnras, 421, 1043
\bibitem[Schlafly \& Finkbeiner (2011)]{2011ApJ...737..103S} Schlafly, E.F., \& Finkbeiner, D.P. \ 2011, \apj, 737, 103
\bibitem[Smit et al. (2014)]{2014ApJ...784...58S} Smit, R., Bouwens, R.J., Labb{\'e}, I., Zheng, W., Bradley, L., et al. \ 2014, \apj, 784, 58
\bibitem[Smit et al. (2015)]{2015ApJ...801..122S} Smit, R., Bouwens, R.J., Franx, M., Oesch, P.A., Ashby, M.L.N, et al. \ 2015, \apj, 801, 122
\bibitem[Smith et al. (2017)]{ 2016ApJ...823...38S} Smith, L., Crowther, P.A., Calzetti, D., \& Sidoli, F.  \ 2016, \apj, 823, 38
\bibitem[Stark et al. (2017)]{2017MNRAS.464..469S} Stark, D.P., Ellis, R.S., Charlot, S., Chevallard, J., et al.\ 2017, \mnras, 464, 469
\bibitem[Stark (2016)]{2016ARA&A..54..761S} Stark, D.P. \ 2016, ARAA, 54, 761
\bibitem[Stark et al. (2015a)]{2015MNRAS.450.1846S} Stark, D.P., Richard, J., Charlot, S., Cl{\'e}ment, B., Ellis.R., Siana, B., et al.\ 2015, \mnras, 450, 1846
\bibitem[Stark et al. (2015b)]{2015MNRAS.454.1393S} Stark, D.P., Walth, G., Charlot, S., Cl{\'e}ment, B., Feltre, A., et al. \ 2015, \mnras, 454, 1393
\bibitem[Stark et al. (2014)]{2014MNRAS.445.3200S} Stark, D.P., Richard, J., Siana, B., Charlot, S., Freeman, W.R., Gutkin, J., et al.\ 2014, \mnras, 445, 3200
\bibitem[Stasinska et al. (2015)]{2015A&A...576A..83S} Stasinska, G., Izotov, Y., Morisset, C., \& Guseva, N. \ 2015, A\&A, 576, 83
\bibitem[Steidel et al. (2014)]{2014ApJ...795..165S} Steidel, C.C., Rudie, G.C., Strom, A.L., Pettini, M., Reddy, N.A., et al. \ 2014, \apj, 795, 165 
\bibitem[Steidel et al. (2016)]{2016ApJ...826..159S} Steidel, C.C., Strom, A.L., Pettini, M., Rudie, G. C., Reddy, N.A., Trainor, R.F. \ 2016, \apj, 826, 159
\bibitem[Svoboda et al. (2019)]{2019ApJ...880..144S} Svoboda, J., Douna, V., Orlitova, I., \& Ehle, M. \ 2019, \apj, 880, 144 
\bibitem[Thuan \& Izotov (2005)]{2005ApJS..161..240T} Thuan, T.X., \& Izotov, Y.I. \ 2005, \apjs, 161, 240
\bibitem[Tilvi et al. (2014)]{2014ApJ...794....5T} Tilvi, V., Papovich, C., Finkelstein, S.L., Long, J., Song, M. et al. \ 2014, \apj, 794, 5
\bibitem[van der Wel et al. (2011)]{2011ApJ...742..111V} van der Wel, A., Straughn, A.N., Rix, H.-W., Finkelstein, S.L., Koekemoer, A.M., Weiner, B.J., et al. \ 2011, \apj, 742, 111
\bibitem[Vanzella et al. (2016)]{2016ApJ...821L..27V} Vanzella, E., de Barros, S., Cupani, G., Karman, W., Gronke, M., Balestra, I. et al. \ 2016, \apj, 821, 27
\bibitem[Verhamme et al. (2015)]{2015A&A...578A...7V} Verhamme, A., Orlitov{\'a}, I., Schaerer, D., Hayes, M. \ 2015, A\&A, 578, 7
\bibitem[Verhamme et al. (2017)]{2017A&A...597A..13V} Verhamme, A., Orlitov{\'a}, I., Schaerer, D., Izotov, Y., Worseck, G., Thuan, T.X., Guseva, N. \ 2017, A\&A, 597, 13
\bibitem[Zackrisson et al. (2013)]{2013ApJ...777...39Z} Zackrisson, E., Inoue, A. K., \& Jenson, H. \ 2013, \mnras, 777, 39


\end{thebibliography}
\end{document}